\documentclass[pdflatex,sn-mathphys-num]{sn-jnl}

\usepackage{graphicx}
\usepackage{multirow}
\usepackage{amsmath,amssymb,amsfonts}
\usepackage{amsthm}
\usepackage{mathrsfs}
\usepackage[title]{appendix}
\usepackage{xcolor}
\usepackage{textcomp}
\usepackage{manyfoot}
\usepackage{booktabs}
\usepackage{listings}
\usepackage{threeparttable}
\usepackage{makecell}
\usepackage{placeins}

\usepackage{etoolbox}
\usepackage{subcaption}
\DeclareCaptionFont{manuscript}{\fontsize{8}{9.5}\selectfont}
\hypersetup{
  pdftitle={MW-Nowcast: Six-hour ensemble nowcasting of extreme precipitation},
  pdfauthor={Ning Wang, Zuliang Fang, Weixin Jin, Zhongjian Lv, Shuang Qin, Pengcheng Zhao, Siqi Xiang, Jiang Bian, Haoyi Xiong, Nan Guan, Bin Zhang, Liangjie Zhang, Denvy Deng, Qi Zhang, Matt Corey, Jitu Keshri, Sridhar Iyer, Hongyu Sun, Kit Thambiratnam, Jonathan Weyn, Richard E. Turner, Haiyu Dong}
}

\makeatletter
\patchcmd{\@maketitle}{\vskip21pt}{\vskip6pt}{}{\PackageError{ms-nowcast-layout}{Title spacing patch failed}{Check the sn-jnl class definition.}}
\patchcmd{\@maketitle}{\vskip24pt}{\vskip12pt}{}{\PackageError{ms-nowcast-layout}{Contact spacing patch failed}{Check the sn-jnl class definition.}}
\patchcmd{\@maketitle}{\vskip24pt}{\vskip12pt}{}{\PackageError{ms-nowcast-layout}{Abstract spacing patch failed}{Check the sn-jnl class definition.}}
\makeatother

\newcommand{\ours}{\emph{MW-Nowcast}}
\newcommand{\mX}{\mathbf{X}}

\begin{document}

\title[MW-Nowcast]{\ours{}: Six-hour ensemble nowcasting of extreme precipitation}

\author[1,2]{\fnm{Ning} \sur{Wang}}\equalcont{These authors contributed equally to this work.}
\author[1]{\fnm{Zuliang} \sur{Fang}}\equalcont{These authors contributed equally to this work.}
\author[1]{\fnm{Weixin} \sur{Jin}}\equalcont{These authors contributed equally to this work.}
\author[1]{\fnm{Zhongjian} \sur{Lv}}
\author[1]{\fnm{Shuang} \sur{Qin}}
\author[1]{\fnm{Pengcheng} \sur{Zhao}}
\author[1]{\fnm{Siqi} \sur{Xiang}}
\author[1]{\fnm{Jiang} \sur{Bian}}
\author[1]{\fnm{Haoyi} \sur{Xiong}}
\author[2]{\fnm{Nan} \sur{Guan}}
\author[1]{\fnm{Bin} \sur{Zhang}}
\author[1]{\fnm{Liangjie} \sur{Zhang}}
\author[1]{\fnm{Denvy} \sur{Deng}}
\author[1]{\fnm{Qi} \sur{Zhang}}
\author[1]{\fnm{Matt} \sur{Corey}}
\author[1]{\fnm{Jitu} \sur{Keshri}}
\author[1]{\fnm{Sridhar} \sur{Iyer}}
\author[1]{\fnm{Hongyu} \sur{Sun}}
\author[1]{\fnm{Kit} \sur{Thambiratnam}}
\author[1]{\fnm{Jonathan} \sur{Weyn}}
\author[3]{\fnm{Richard E.} \sur{Turner}}
\author*[1]{\fnm{Haiyu} \sur{Dong}}\email{haiyu.dong@microsoft.com}

\affil[1]{\orgname{Microsoft Corporation}}
\affil[2]{\orgname{City University of Hong Kong}}
\affil[3]{\orgname{University of Cambridge}}

\abstract{
Extending reliable nowcasting of extreme precipitation could provide critical additional time for warnings and emergency response during high-impact events such as flash floods.
Radar-based generative machine-learning models have enabled skilful hyperlocal precipitation nowcasting, but accurate prediction of intense precipitation remains confined to the first few hours. Because storm-scale structure is predictable for longer than individual cells, a natural strategy is to predict that structure while generatively modelling only the uncertain local growth, decay, reorganisation and initiation of storms.
Here we present Microsoft Weather Nowcast (\ours{}), a six-hour ensemble radar nowcasting model that jointly learns a deterministic predictor to capture organised precipitation structure shared across ensemble members, and a generator to produce diverse local residuals around this shared prediction.
Across independent test data from the United States, Europe and China, \ours{} achieves higher detection skill than leading methods for heavy and extreme precipitation throughout the 6 h horizon.
For the most intense rainfall, \ours{} doubles the available warning time across all three regions, delivering 6 h forecasts with skill previously limited to 3 h for the leading generative baseline. A cost--loss decision analysis shows that \ours{} retains substantial value for a broad range of applications even at 4--6 h, where alternative methods offer little benefit. These additional hours can give forecasters and emergency managers the time to warn and act before extreme rainfall strikes, helping to protect lives and property.
}

\maketitle

\section{Introduction}

Precipitation nowcasting aims to provide weather forecasts for the next few hours, translating available information into guidance that is local and early enough to support immediate proactive mitigation. The World Meteorological Organisation frames nowcasting as local weather prediction from the present to up to 6 h ahead~\cite{guidelines}. Within this window, predicting hazardous precipitation reliably can broaden the range of possible responses, including issuing warnings, positioning emergency resources, and managing drainage systems, roads, aviation and energy operations. This need is especially acute for extreme precipitation, because assessing its potential impacts requires forecasts that extend far enough to characterize the likely duration of intense rainfall over affected locations~\cite{wmo_ffgs}.

Operational nowcasting requires forecasts to be updated frequently as new observations arrive and to resolve rapidly-evolving precipitation at convective scales. Numerical weather prediction provides physically-grounded forecasts and can assimilate recent radar observations. However, its computational cost makes updates at radar cadence challenging~\cite{nwp_challenge,sun2014_nwp_nowcasting}. These constraints have motivated methods that infer future precipitation directly from recent radar sequences. Advection-based statistical methods such as STEPS provide fast and interpretable ensemble nowcasts by combining the extrapolation of recent radar echoes with stochastic, scale-dependent evolution~\cite{steps,pysteps}. However, their reliance on transported recent echoes limits their ability to represent local growth, decay, reorganisation and initiation, especially as lead time extends beyond 3 h.

Over the last decade, machine-learning methods have become the gold standard for radar nowcasting, retaining the speed and radar-native character of extrapolation while learning nonlinear precipitation evolution from data. Early machine-learning approaches based on deep learning generally formulated this task as supervised sequence prediction, training deterministic networks with grid-cell losses on observed future radar sequences~\cite{convlstm,predrnn,earthformer}. They improved the representation of nonlinear spatiotemporal evolution beyond extrapolation, but forecasts often became smooth at longer lead times, with weakened intense cores and limited uncertainty information. Deep Generative Models of Radar (DGMR) marked an important step towards probabilistic nowcasting by learning a conditional generative model for entire future precipitation fields, producing sharper and more useful ensemble forecasts~\cite{dgmr}. NowcastNet extended this direction by conditioning the generative network on an advection-informed deterministic evolution network~\cite{nowcastnet}. It was also strongly preferred by professional meteorologists in 3 h case evaluations.

Despite these advances, DGMR and NowcastNet reported results for forecast horizons of 90 min and 3 h, respectively, while the full 0--6 h nowcasting window remains less explored for extreme precipitation~\cite{dgmr,nowcastnet}. At longer lead times, future precipitation evolution remains partly constrained by the observed radar history, whereas its conditional distribution becomes increasingly broad. Existing generative systems typically sample complete future radar sequences, even when a deterministic prediction is supplied as conditioning information. In this formulation, the generative component must simultaneously preserve the storm-scale structure supported by recent observations and represent uncertainty in local cell evolution. Coupling these roles may make it harder to maintain coherent storm placement and organisation while generating sufficient ensemble variability at longer lead times. Inspired by scale-dependent precipitation predictability and stochastic representations of unresolved weather variability, a promising direction is to make the allocation between predictable dynamics and stochastic variability itself learnable~\cite{germann2002_predictability,predictability,Palmer_2000}.

We show that long-lead-time ensemble nowcasting benefits from an explicit, jointly-learned separation between a deterministic forecast shared across members and member-dependent probabilistic residuals. We developed Microsoft Weather Nowcast (\ours{}), a radar nowcasting model that couples two decoders, one producing the deterministic prediction and the other modelling the probabilistic residual relative to it. The two are trained jointly rather than as a fixed two-stage cascade, so that the allocation between shared storm-scale structure and local evolution adapts throughout training. Across radar networks in the United States (US), Europe (EU) and China (CN), \ours{} extends skilful detection of heavy and extreme precipitation from the previously demonstrated 3 h to the full 6 h operational window while preserving the multiscale structure of the observations. In two recent high-impact events, the storm that spawned an EF4 tornado in Mississippi and the July 2025 extreme rainfall over Beijing, \ours{} kept the intense rain cores over the affected areas for longer than baselines, which weakened or displaced them. A cost--loss analysis shows that this advantage translates into value at 4--6 h lead times across the full range of protective decisions. Within these hours, authorities can issue warnings, manage drainage and transport systems, and position emergency resources before flash floods develop.

\section{\ours{}}

\begin{figure}
  \centering
  \includegraphics[width=0.8\linewidth]{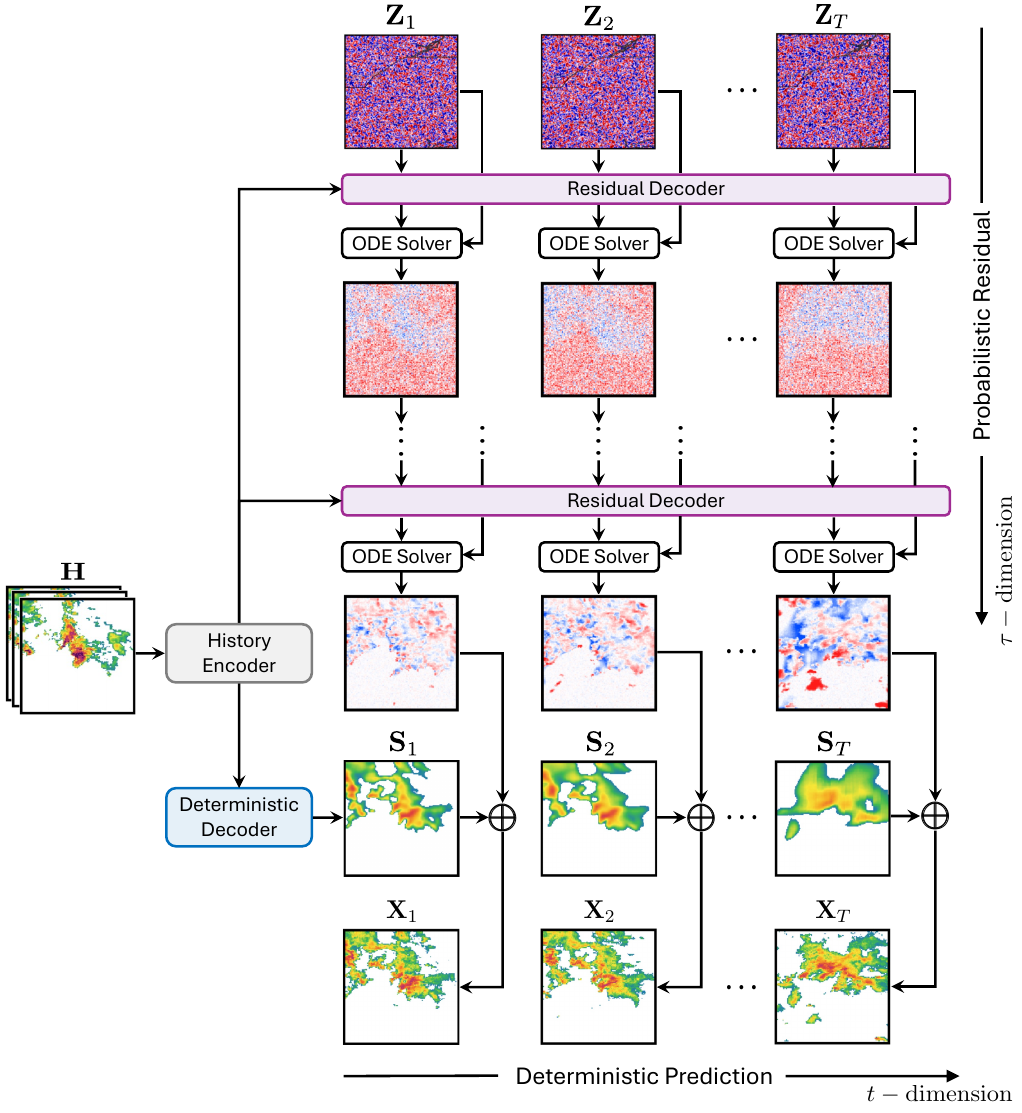}
  \caption{\textbf{\ours{} for 6 h ensemble precipitation nowcasting.} A history encoder provides the deterministic and residual decoders with a shared representation of the observed radar sequence $\mathbf{H}$. The deterministic decoder produces the common full-horizon forecast $\mathbf{S}_{1:T}$, while the flow-matching residual decoder transforms Gaussian noise $\mathbf{Z}^{(k)}$ into sampled residuals through an ODE solver. At each lead time $t$, the sampled residual is added to $\mathbf{S}_t$ to obtain $\mathbf{X}_t^{(k)}$; different noise draws yield the forecast ensemble.}
  \label{fig:main}
\end{figure}

\ours{} is a 6 h ensemble radar nowcasting framework that predicts the future radar sequence $\mX_{1:T}=(\mX_1,\ldots,\mX_T)$ from an observed radar history $\mathbf{H}=(\mX_{-T_0+1},\ldots,\mX_0)$, with $T_0=9$ and $T=36$ at 10 min intervals, by combining a deterministic prediction with sampled probabilistic residuals. As shown in Fig.~\ref{fig:main}, the model first encodes the radar history with a shared encoder parameterised by $\psi$, then branches into a deterministic decoder and a flow-matching residual decoder. We denote the full-horizon mappings by $\mathcal{S}_{\psi,\theta}$ for deterministic prediction and $\mathcal{R}_{\psi,\phi}$ for residual sampling. Both mappings include the shared encoder; ``deterministic decoder'' and ``residual decoder'' refer only to the trainable modules after the encoder. Stochasticity is introduced by initialising the residual flow with a Gaussian noise tensor $\mathbf{Z}^{(k)}$ for each ensemble member. The $k$-th ensemble member can therefore be written as
\begin{equation}
  \mX_{1:T}^{(k)}
  =
  \mathbf{S}_{1:T}
  +
  \mathbf{R}_{1:T}^{(k)},
  \label{eq:forecast_factorization}
\end{equation}

where $\mathbf{S}_{1:T}:=\mathcal{S}_{\psi,\theta}(\mathbf{H})$ is the full-horizon deterministic prediction, and $\mathbf{R}_{1:T}^{(k)}:=\mathcal{R}_{\psi,\phi}(\mathbf{Z}^{(k)},\mathbf{H})$ is the probabilistic residual generated for the $k$-th ensemble member. The noise tensor satisfies $\mathbf{Z}^{(k)}\in\mathbb{R}^{T\times S\times S}$ with $Z_{t,i,j}^{(k)}\overset{\mathrm{i.i.d.}}{\sim}\mathcal{N}(0,1)$, independently across ensemble member $k$, physical lead-time index $t$ and spatial indices $(i,j)$.
The residual decoder is conditioned on the shared radar-history representation.

During training, the shared encoder, deterministic decoder and residual decoder are optimised as one model. The encoder is trained to provide a radar-history representation that supports both the deterministic decoder and the residual decoder. Through joint optimisation, the deterministic decoder and residual decoder co-adapt to allocate future radar evolution between a common predictable component and conditional residuals designed to represent plausible local variability over the 6 h horizon. In the Supplementary Information, we show that joint optimisation yields stronger long-lead-time high-threshold skill than training the deterministic decoder before the residual decoder, a cascade-style strategy analogous to residual correction in generative downscaling~\cite{corrdiff}.

Training combines two objectives. The first penalises deviations between the deterministic prediction and the observed future sequence. The second is a probabilistic-residual flow-matching loss, with the target dynamically recomputed as
\begin{equation}
  \mathbf{R}^{\star}_{1:T}
  =
  \mX_{1:T}-\mathbf{S}_{1:T},
  \label{eq:residual_target}
\end{equation}
so the distribution of learned probabilistic residuals changes as the deterministic prediction is updated. The residual decoder is implemented with flow matching. Unlike GAN-based nowcasting, which encourages realism through a discriminator, flow matching trains the residual decoder to predict a velocity field at intermediate states along a probability path from Gaussian noise to the target residual distribution~\cite{cfm}. Recent diffusion and flow-matching nowcasting models have improved sample diversity and uncertainty quantification compared to GAN-based systems~\cite{prediff,ldcast, flowcast}.

During sampling, we integrate the learned probability-flow ODE to transform each independent noise tensor into a plausible residual sequence conditioned on the observed radar history. Adding this residual sequence to the deterministic prediction yields an ensemble member with detailed local structure, rather than an average over possible futures.

\section{Experimental Setup}

We train and evaluate \ours{} on three regional radar datasets covering the US, EU and CN domains. All datasets are converted to a common 6 h nowcasting task defined above over a 256 km $\times$ 256 km domain at 2~km resolution. The US dataset pairs NOAA/NCEI Storm Events reports with MRMS Seamless Hybrid Scan Reflectivity radar composites~\cite{noaa_storm_events,zhang2016mrms}; the EU dataset pairs European Severe Weather Database records with OPERA radar mosaics~\cite{eswd,kock2000opera}; and the CN dataset is constructed from China Meteorological Administration composite reflectivity data using radar-native, peak-anchored strong-precipitation sampling.
Training sets contain 130,000--175,000 training sequences per region. Each region is evaluated on 4,000 test sequences drawn from a strictly later observation period, selected under matching criteria to focus the benchmark on strong-precipitation and severe convective regimes (\nameref{sec:methods}).

We compare \ours{} against three representative baselines spanning operational extrapolation, deterministic deep prediction and probabilistic ensemble nowcasting. We use the STEPS implementation in PySTEPS as a widely used advection-based ensemble nowcasting baseline~\cite{steps,pysteps}. SimVPv2 is a recent deterministic spatiotemporal prediction model that has served as a strong, state-of-the-art baseline across prediction benchmarks~\cite{simvpv2}. NowcastNet is a leading generative ensemble nowcasting model that showed strong performance for extreme precipitation in both US and CN evaluations~\cite{nowcastnet}. We retrain SimVPv2 and NowcastNet using their official model implementations and evaluate all baselines under the same protocol.

Evaluation combines threshold-based, probabilistic, structural, and decision-analytic diagnostics. Threshold-based verification uses rain-rate exceedances at 16, 32 and 64 mm h$^{-1}$ via the Critical Success Index within a $5\times5$ grid-cell neighbourhood (CSIN)~\cite{csi, nowcastnet}. All ensemble methods are evaluated with four members and aggregated using Probability Matched Mean (PMM) wherever a single field is required~\cite{ebert2001_pmm}. Probabilistic forecasts are evaluated using the Continuous Ranked Probability Score (CRPS) and the Continuous Ranked Probability Skill Score (CRPSS), which measures how well the ensemble distribution agrees with the observations~\cite{crps,muller2005debiased}. Power Spectral Density (PSD) is used separately as a threshold-free diagnostic of spatial structure~\cite{psd}. Practical decision utility is quantified using Relative Economic Value (REV) across cost--loss ratios~\cite{richardson2000,dgmr}.

\section{Precipitation Cases}

We present two high-impact precipitation cases from contrasting storm regimes: a rapidly-evolving tornadic convective system in the United States and persistent mountainous heavy rainfall in China. We assess how well each method retains the precipitation system’s organisation and embedded intense cores as lead time increases, and how \ours{} allocates the forecast between shared, predictable storm-scale structure and uncertain local evolution across ensemble members. Additional cases from all three domains are presented in Extended Data Figs.~\ref{fig:extended-us-event182747}--\ref{fig:extended-eu-175810}.

\begin{figure*}[!t]
  \centering{%
    \includegraphics[width=\textwidth]{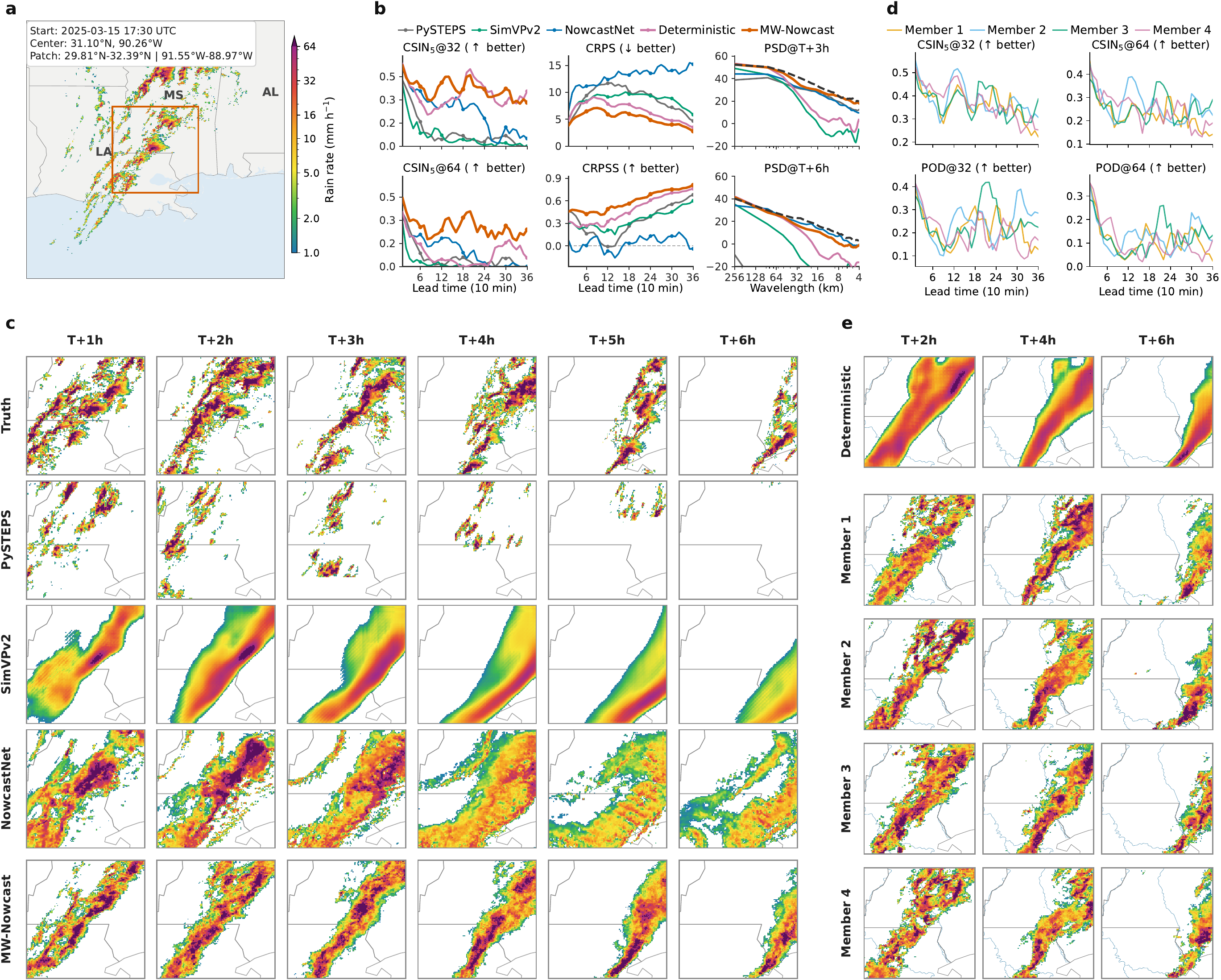}%
  }
  \caption{\textbf{Case study of a US tornado event over the 6 h forecast horizon.} The case shows a convective storm over Walthall County, Mississippi, on 15 March 2025 that spawned an EF4 tornado; the forecast is initialised at 17:30 UTC.
  \textbf{a}, Geographic and radar context, with the forecast domain outlined in orange.
  \textbf{b}, Forecast diagnostics: CSIN in a $5\times5$ grid-cell neighbourhood at rain-rate thresholds of 32 and 64 mm h$^{-1}$, CRPS, CRPSS, and radially averaged PSD at lead times of 3 and 6 h. ``Deterministic'' denotes the output of the deterministic decoder of \ours{}.
  \textbf{c}, Observed and predicted rain-rate fields at hourly lead times from 1 to 6 h. Rows show observations, PySTEPS, SimVPv2, NowcastNet and \ours{}.
  \textbf{d}, CSIN and POD at thresholds of 32 and 64 mm h$^{-1}$ for the four \ours{} ensemble members and their PMM summary.
  \textbf{e}, The deterministic prediction and four ensemble members at lead times of 2, 4 and 6 h.}
  \label{fig:case-us-tornado}
\end{figure*}

The convective storm over Walthall County, Mississippi, on 15 March 2025 (Fig.~\ref{fig:case-us-tornado}a) spawned an EF4 tornado. The event was documented by an NWS Storm Survey, with 3 fatalities, 8 injuries and approximately \$5.0 million in reported property damage. We initialise the forecast at 17:30 UTC. In the lead-time diagnostics (Fig.~\ref{fig:case-us-tornado}b), \ours{} maintains higher CSIN than the baselines at 32 and 64 mm h$^{-1}$ over much of the final three hours, with the separation clearest at 64 mm h$^{-1}$. It also better preserves multiscale spatial variability and achieves higher probabilistic skill than the ensemble baselines over most lead times.

The forecast fields show how \ours{} overcomes the characteristic failure modes of existing baselines (Fig.~\ref{fig:case-us-tornado}c). Throughout the forecast window, a southwest--northeast--oriented organised precipitation band remains identifiable, while multiple embedded intense cores undergo local intensification, weakening, fragmentation and reorganisation. \ours{} preserves rain-band continuity, concentrated heavy-rain cores and the overall spatial extent of the system at longer lead times, with member-wise scores confirming that the late-lead heavy-precipitation signal is carried by individual ensemble members (Fig.~\ref{fig:case-us-tornado}d). By contrast, PySTEPS advects part of the early motion, but its intensity fades progressively within the domain. SimVPv2 keeps the band continuous, but smooths the embedded convective cores into a broad, uniform region of heavy rain. NowcastNet produces sharper local structures at early and middle leads, but its later forecasts substantially underestimate heavy-core intensity while spreading light-to-moderate rain over an excessively broad area. These are precisely the modes that the design of \ours{} targets. Its deterministic predictor learns the predictable storm-scale structure from data rather than transporting recent echoes, and its generator adds the small-scale detail as sampled residuals, which keeps the forecast sharp while remaining anchored to that structure rather than generating the entire field.

Comparing the deterministic prediction of \ours{} with its ensemble members reveals this scale-dependent structure of forecast uncertainty (Fig.~\ref{fig:case-us-tornado}e). The deterministic branch captures the rain-band location, orientation and overall organisation shared across the ensemble, while the residual branch generates member-dependent variations in the locations, intensities and morphologies of the embedded convective cores. Rather than merely adding spatial detail to the deterministic forecast, the residual branch represents multiple possible pathways of local convective intensification, weakening and reorganisation while preserving the storm-scale evolution supported by the observed history.

\begin{figure*}[!t]
  \centering
  \includegraphics[width=\textwidth,keepaspectratio]{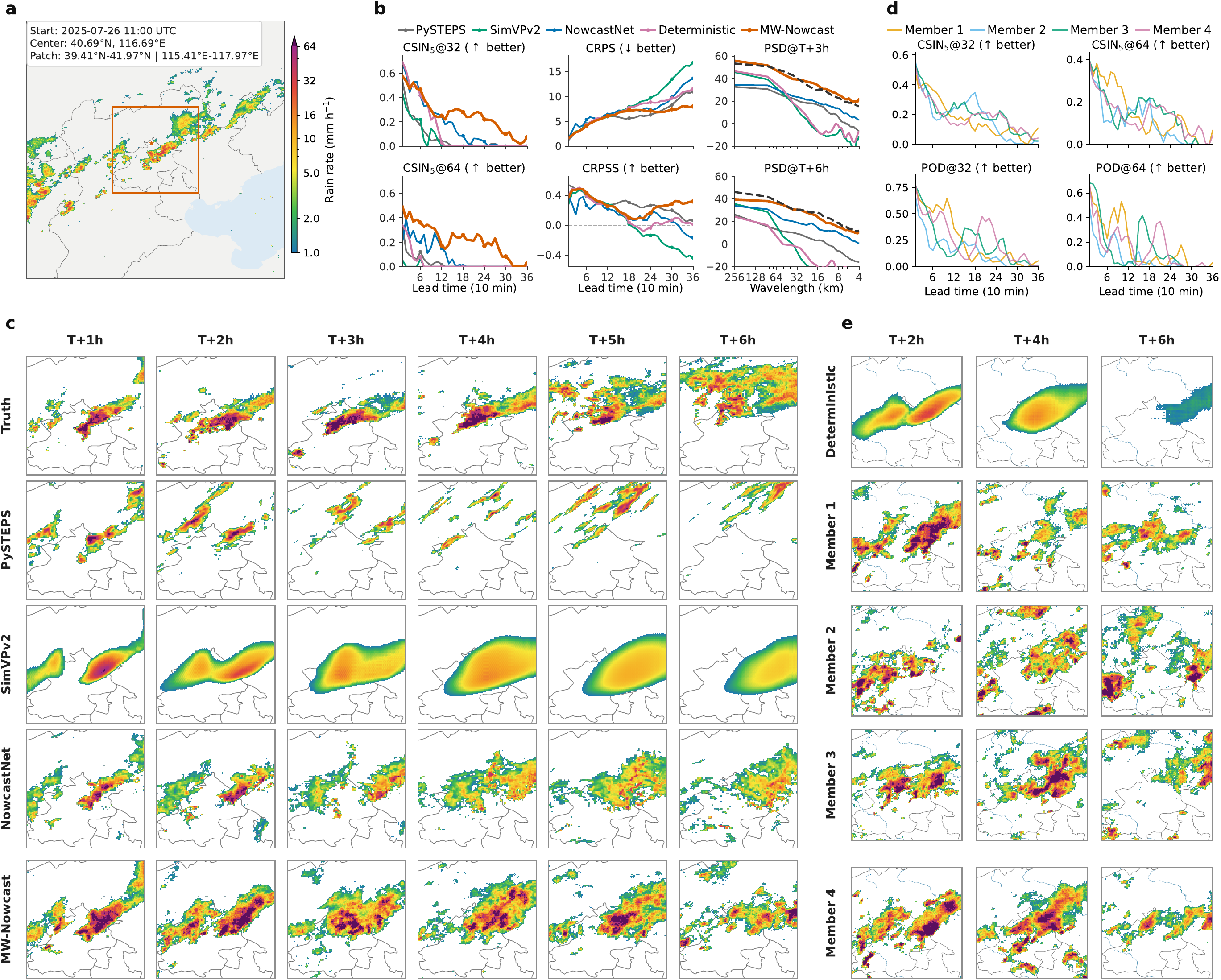}
  \caption{\textbf{Case study of the Beijing Miyun mountainous extreme-rainfall event over the 6 h forecast horizon.}  Extreme rainfall over the northern Miyun district of Beijing and adjacent Hebei during the late-July 2025 episode; forecast initialised at 11:00 UTC on 26 July 2025 (19:00 local time). Panel descriptions are as in Fig.~\ref{fig:case-us-tornado}.}
  \label{fig:case-cn-miyun}
\end{figure*}

\ours{} likewise maintains its advantage in a contrasting convective predictability regime, where organised storm structure progressively scatters and deterministic predictability is rapidly lost (Fig.~\ref{fig:case-cn-miyun}a). The late-July 2025 event over the northern Miyun district of Beijing and adjacent Hebei caused 30 deaths and led to the relocation of 80,332 people. For a forecast initialised at 11:00 UTC (19:00 local time) on 26 July 2025, during the evening-to-night development of the storm, \ours{} retains markedly higher skill than the baselines over the final three hours, especially at 64 mm h$^{-1}$, while achieving the highest probabilistic skill and closest spectral agreement with observations (Fig.~\ref{fig:case-cn-miyun}b). In the forecast fields, the baselines exhibit the same error modes as in the previous case, and NowcastNet, although generative, likewise loses the intense cores at later lead times, whereas \ours{} retains the sustained heavy-rainfall signal over Miyun to about 5 h (Fig.~\ref{fig:case-cn-miyun}c).

The comparison between the deterministic prediction and the ensemble members shows how the allocation learned by \ours{} shifts as predictability is lost (Fig.~\ref{fig:case-cn-miyun}e). Over the first 4 h, the observed rainfall over the Miyun region is relatively concentrated, the deterministic prediction carries a coherent rainfall signal there, and the members largely agree on its location while differing in core detail. As the observed rainfall becomes increasingly scattered, the deterministic prediction fades to a faint patch by 6 h, whereas the members continue to produce intense cores in varied locations, often where the deterministic forecast shows almost nothing. This differs substantially from the Mississippi case, in which the deterministic branch carried most of the forecast throughout (Fig.~\ref{fig:case-us-tornado}e). The diagnostics quantify the shift: deterministic CSIN at 64 mm h$^{-1}$ falls to zero within about 3 h, whereas the ensemble retains skill to about 5 h (Fig.~\ref{fig:case-cn-miyun}b), with member-wise scores confirming that this signal is carried by individual ensemble members at middle lead times (Fig.~\ref{fig:case-cn-miyun}d). By the final hour, with little storm-scale structure left to hold onto, the residual branch still supplies most of the heavy-rainfall signal as diverse scenarios. While the widening spread marks the limit of local predictability in this case, with individual members increasingly diverging from the observed cores towards 6 h, the ensemble still conveys a sustained likelihood of heavy rain over the region, an important signal that supports continued monitoring rather than de-escalation.

\FloatBarrier

\section{Quantitative Evaluation}

\begin{figure*}[htbp]
  \centering
  \includegraphics[width=\textwidth]{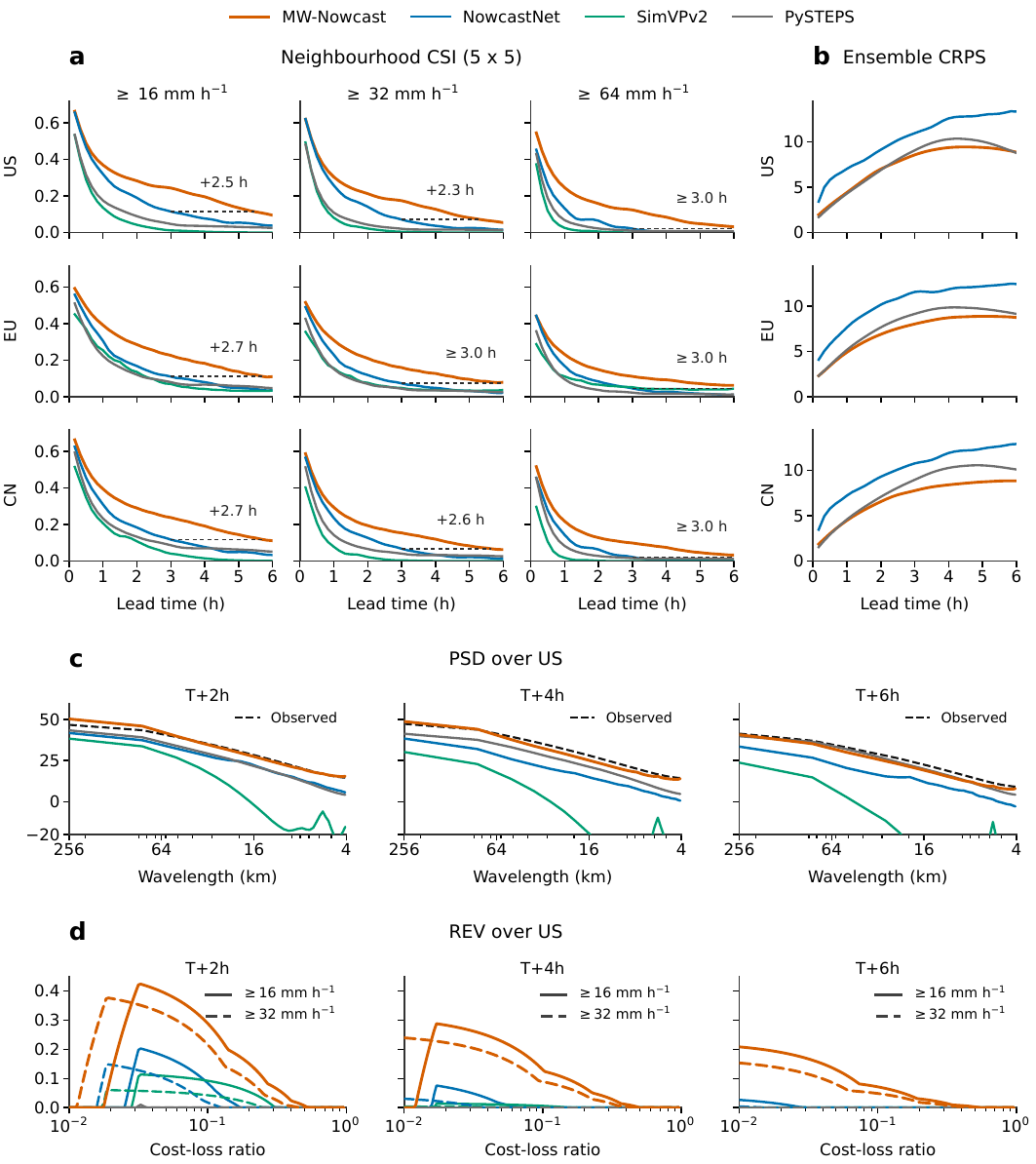}
  \caption{\textbf{Quantitative evaluation of 6 h precipitation nowcasts.} Curves compare \ours{}, NowcastNet, SimVPv2 and PySTEPS. \textbf{a}, Critical Success Index in a $5\times5$ grid-cell neighbourhood over the 6 h forecast horizon for the US, EU and CN test sets at rain-rate thresholds of 16, 32 and 64 mm h$^{-1}$; higher values are better. \textbf{b}, grid-point Continuous Ranked Probability Score (CRPS) of the four-member ensembles for the three regions; lower values are better. \textbf{c}, radially averaged power spectral density (PSD) over the US at lead times of 2, 4 and 6 h, with the observed spectrum shown for reference. \textbf{d}, relative economic value (REV) over the US as a function of cost--loss ratio at lead times of 2, 4 and 6 h for rain-rate exceedances of 16 and 32 mm h$^{-1}$.}
  \label{fig:quantitative-evaluation}
\end{figure*}

Figure~\ref{fig:quantitative-evaluation} summarises verification over the complete US, EU and CN test sets. \ours{} achieved the highest CSIN at every evaluated lead time and threshold across all three regions (Fig.~\ref{fig:quantitative-evaluation}a). Whereas baseline CSIN rapidly approached zero after approximately 3 h at a threshold of 64 mm h$^{-1}$, \ours{} retained measurable skill throughout the 6 h horizon. NowcastNet was originally evaluated over a 3 h forecasting horizon, whereas we retrain it here to predict the full 6 h sequence. Using the 3 h CSIN of this retrained baseline as the reference, \ours{} sustained equivalent skill for an additional 2.3 to more than 3 h across the nine region--threshold combinations, with larger gains for more extreme precipitation. These gains are supported by structured, member-dependent residual adjustments to the common prediction, illustrated by residual sampling and ensemble-spread visualisations (Supplementary Sections~\ref{sec:supp-single-member} and~\ref{sec:supp-ensemble}). Consistently, the PMM ensemble better recovers the observed high-intensity rain-rate tail than the deterministic prediction across all three regions (Extended Data Fig.~\ref{fig:extended-regional-rain-density}). Both generating the complete future sequence without a deterministic branch and training the deterministic and residual models sequentially reduce long-lead high-threshold skill (Supplementary Section~\ref{sec:supp-ablation} and Supplementary Table~\ref{tab:supp-ablation-metrics}). These results support jointly-learned residual generation as a key contributor to sustained extreme-precipitation skill at longer forecast horizons.

We next evaluated the four-member ensembles directly using grid-point continuous ranked probability score (CRPS), which measures the agreement between the predictive distribution and the observation at each location (Fig.~\ref{fig:quantitative-evaluation}b). \ours{} achieved lower CRPS than both NowcastNet and PySTEPS across all three regions, indicating a better balance between individual-member accuracy and ensemble spread. PySTEPS nevertheless outperformed NowcastNet in CRPS. Because domain-averaged grid-point CRPS weights every location equally and does not specifically emphasise intense precipitation, conservative forecasts can remain competitive even when they lose high-threshold skill.

We further assessed scale-dependent precipitation variability in all three regions using radially averaged PSD at lead times of 2, 4 and 6 h. Figure~\ref{fig:quantitative-evaluation}c shows the US results. SimVPv2 progressively lost spectral power at medium and short wavelengths, consistent with increasingly smooth forecasts. By contrast, \ours{} remained close to the observed spectrum throughout the 6 h horizon, outperforming NowcastNet and more closely matching observations than PySTEPS at 2 and 4 h, with comparable agreement at 6 h. Similar spectral behaviour was observed in EU and CN (Supplementary Section~\ref{sec:supp-regional}). These results indicate better retention of precipitation variability across spatial scales.

Beyond standard verification metrics, we assessed whether improved forecast skill could better support protective decision-making using the relative economic value (REV) framework adopted in previous probabilistic forecasting studies~\cite{dgmr,gencast}. REV measures whether forecast-based action can reduce expected expense in a stylised cost--loss decision problem and is normalised such that 1 represents a perfect forecast and 0 represents no improvement over the optimal climatological strategy (\nameref{sec:methods}). DGMR~\cite{dgmr} evaluated REV for precipitation accumulated over a 90 min forecast window, whereas we evaluated the instantaneous value separately at lead times of 2, 4 and 6 h to examine whether forecasts retain decision value at longer lead times and could therefore support earlier protective action. At each lead time, we calculated REV across cost--loss ratios for rain-rate exceedances of 16 and 32 mm h$^{-1}$ in all three regions. Across these regions, \ours{} provided substantially greater REV overall than the baselines at both thresholds and all three lead times (Fig.~\ref{fig:quantitative-evaluation}d and Supplementary Information). The US results in Fig.~\ref{fig:quantitative-evaluation}d show that, at 4 and 6 h, \ours{} retained positive value over a broad range of cost--loss ratios, whereas the baseline values were close to zero. These results suggest that the long-lead-time skill of \ours{} can support useful protective decisions across users with different action costs, extending the time available for preparation throughout the 6 h nowcasting window.

\FloatBarrier

\section{Conclusion}

The value of extreme-precipitation nowcasting depends not only on producing realistic precipitation fields, but also on retaining useful skill far enough ahead to support early warning and emergency response, yet the latter has remained a difficult frontier. We have presented \ours{}, which jointly learns a shared deterministic prediction and conditional probabilistic residuals around it to provide six-hour ensemble radar nowcasting. Controlled experiments show that learning this division end-to-end is essential to sustaining skill at longer lead times. Across radar datasets from the United States, Europe and China, \ours{} delivers skilful ensemble nowcasts of extreme precipitation across the full 6 h window, doubles the available warning time for the most intense rainfall from three hours to six, and retains decision value at 4--6 h, where existing methods offer little.

Future work should extend precipitation nowcasting beyond the information available from radar alone. Combining radar with geostationary satellite imagery and numerical-weather-prediction guidance could provide earlier information about cloud development and environmental conditions, with particular potential for improving forecasts of convective initiation and rapidly evolving storms. Our evaluation spans three regions with dense radar networks, but radar observations could also serve as training targets for models driven by satellite and numerical inputs which are available more widely, bringing nowcasting to regions with sparse or no operational radar coverage. Such approaches could improve both the accuracy and geographical reach of long-lead precipitation nowcasting and extend the decision value demonstrated here to communities with the most limited early-warning capabilities.

\bibliography{references}

\section{Methodology}\label{sec:methods}

We use this section to define the architecture, training objective and sampling procedure with the same notation as the main text. A temporal stack $\mX_{(a,b]}$ denotes the left-open, right-closed sequence $(\mX_{a+1},\ldots,\mX_b)$. The observed history is $\mathbf{H}=(\mX_{-T_0+1},\ldots,\mX_0)$, equivalently $\mathbf{H}=\mX_{(-T_0,\,0]}$, and therefore contains exactly $T_0=9$ frames, including the issue-time frame $\mX_0$. The future target sequence is $\mX_{1:T}=(\mX_1,\ldots,\mX_T)$, equivalently $\mX_{(0,\,T]}$, with $T=36$. The parameterised full-horizon mappings for deterministic prediction and residual sampling are $\mathcal{S}_{\psi,\theta}$ and $\mathcal{R}_{\psi,\phi}$, respectively. Both mappings include the shared encoder, and neither symbol denotes a decoder module alone. Their outputs are $\mathbf{S}_{1:T}:=\mathcal{S}_{\psi,\theta}(\mathbf{H})$ and $\mathbf{R}_{1:T}^{(k)}:=\mathcal{R}_{\psi,\phi}(\mathbf{Z}^{(k)},\mathbf{H})$. Throughout, calligraphic symbols denote mappings, whereas bold symbols denote tensor-valued inputs and outputs. Thus, Eq.~\eqref{eq:forecast_factorization} is the single top-level definition of the model used in both the main text and Methods. For physical lead time $t$, we write $\mX_t^{(k)}:=[\mX_{1:T}^{(k)}]_t$, $\mathbf{S}_t:=[\mathbf{S}_{1:T}]_t$ and $\mathbf{R}_t^{(k)}:=[\mathbf{R}_{1:T}^{(k)}]_t$. We omit the member index $k$ when referring to a generic probabilistic residual.

\subsection{Deterministic-probabilistic formulation}

For the discrete forecasting task, $\mX_t\in\mathbb{R}^{S\times S}$ denotes the reflectivity field at physical lead index $t$, and bold uppercase symbols such as $\mX_{(a,b]}$ denote sequences stacked along the time dimension. The deterministic prediction mapping $\mathcal{S}_{\psi,\theta}$ produces the full $T$-frame output $\mathbf{S}_{1:T}$ in parallel from $\mathbf{H}$, avoiding autoregressive error accumulation. During training, this deterministic prediction defines the probabilistic residual target in Eq.~\eqref{eq:residual_target}. This target is not an observed physical forcing; it is precisely the part of the future sequence not represented by the current deterministic prediction. Because the model components underlying $\mathcal{S}_{\psi,\theta}$ and $\mathcal{R}_{\psi,\phi}$ are trained jointly, the probabilistic residual target evolves as the deterministic decoder improves.

The residual sampling map $\mathcal{R}_{\psi,\phi}$ is defined by a residual decoder trained with conditional flow matching and an ODE solver used at inference. Physical lead time $t\in\{1,\ldots,T\}$ indexes forecast frames, whereas generative time $\tau\in[0,1]$ indexes an internal trajectory from a Gaussian noise tensor to a probabilistic residual. To keep Eqs.~\eqref{eq:fm_path}--\eqref{eq:residual_flow_ode} compact, we write $\mathbf{r}^{\star}:=\mathbf{R}^{\star}_{1:T}$ for the full-sequence probabilistic residual target and $\mathbf{r}(\tau)$ for the full sequence-valued flow state; the physical lead dimension is therefore implicit in both symbols. For a noise tensor $\mathbf{Z}$ drawn as specified in Eq.~\eqref{eq:forecast_factorization} and $\tau\sim\mathcal{U}[0,1]$, the linear probability path is
\begin{equation}
  \mathbf{r}(\tau)
  =
  (1-\tau)\mathbf{Z}+\tau\mathbf{r}^{\star}.
  \label{eq:fm_path}
\end{equation}
Its target velocity is
\begin{equation}
  \mathbf{u}^{\star}
  =
  \frac{d\mathbf{r}(\tau)}{d\tau}
  =
  \mathbf{r}^{\star}-\mathbf{Z}.
  \label{eq:fm_target_velocity}
\end{equation}
The residual decoder, parameterised by $\phi$ and conditioned on the history representation produced by the shared encoder parameterised by $\psi$, predicts this velocity from the current flow state, generative time and observed history:
\begin{equation}
  \mathbf{v}_{\psi,\phi}
  (\mathbf{r}(\tau),\tau,\mathbf{H})
  \approx
  \mathbf{u}^{\star}.
  \label{eq:fm_velocity_network}
\end{equation}

At inference, a probabilistic residual is obtained by solving
\begin{equation}
  \frac{d\mathbf{r}(\tau)}{d\tau}
  =
  \mathbf{v}_{\psi,\phi}
  (\mathbf{r}(\tau),\tau,\mathbf{H}),
  \qquad
  \mathbf{r}(0)=\mathbf{Z},
  \qquad
  \mathbf{R}_{1:T}
  :=
  \mathcal{R}_{\psi,\phi}(\mathbf{Z},\mathbf{H})
  =
  \mathbf{r}(1).
  \label{eq:residual_flow_ode}
\end{equation}
This last identity connects the terminal ODE state directly to the probabilistic residual in Eq.~\eqref{eq:forecast_factorization}. For numerical inference, we discretise generative time as $\tau_n=n/N$, $n=0,\ldots,N$, with $N=20$ and step size $\Delta\tau=1/N$, and write $\mathbf{r}_n:=\mathbf{r}(\tau_n)$. At a discrete solver state, the residual decoder predicts $\mathbf{v}_n:=\mathbf{v}_{\psi,\phi}(\mathbf{r}_n,\tau_n,\mathbf{H})$. The fixed-step Heun integrator advances the flow by
\begin{equation}
  \tilde{\mathbf{r}}_{n+1}
  =
  \mathbf{r}_n+\Delta\tau\,\mathbf{v}_n,
  \qquad
  \mathbf{v}_{n+1}^{\mathrm{p}}
  =
  \mathbf{v}_{\psi,\phi}(\tilde{\mathbf{r}}_{n+1},\tau_{n+1},\mathbf{H}),
  \qquad
  \mathbf{r}_{n+1}
  =
  \mathbf{r}_n+\frac{\Delta\tau}{2}\bigl(\mathbf{v}_n+\mathbf{v}_{n+1}^{\mathrm{p}}\bigr).
  \label{eq:heun_step}
\end{equation}
For ensemble member $k$, $\mathbf{r}_0^{(k)}=\mathbf{Z}^{(k)}$ and $\mathbf{R}_{1:T}^{(k)}=\mathbf{r}_N^{(k)}$, while the deterministic prediction remains shared across members.

\subsection{Model details}

For all experiments, each sample covers a 256 km $\times$ 256 km physical domain, while the model input is represented on a square radar grid with side length $S=128$. The input history has shape $(T_0,S,S,1)=(9,128,128,1)$ and the target future sequence has shape $(T,S,S,1)=(36,128,128,1)$, corresponding to 1.5 h of history and 6 h of prediction at 10 min resolution. A stride-4 patch projection gives feature-grid side length $s=S/4=32$, and the token feature dimension is $c=256$. The Gaussian noise tensor $\mathbf{Z}$ has mathematical shape $(T,S,S)=(36,128,128)$ and implementation shape $(36,128,128,1)$. Its 36 temporal slices are sampled independently.

\begin{figure}
  \centering
  \includegraphics[width=\linewidth]{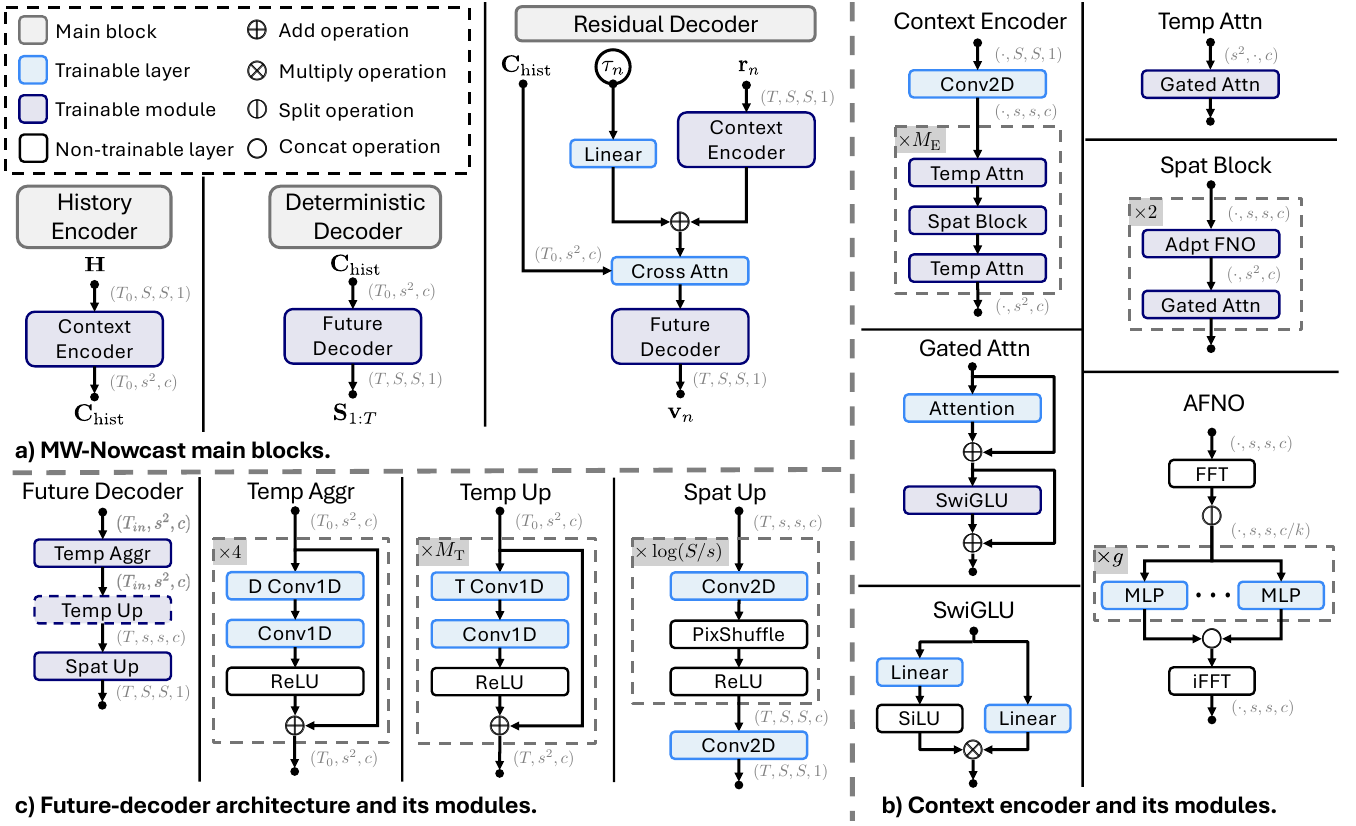}
  \caption{\textbf{Network architecture of \ours{}.} The history encoder maps $\mathbf{H}$ to $\mathbf{C}_{\mathrm{hist}}$, and the deterministic decoder produces $\mathbf{S}_{1:T}$. At discrete ODE state $(\mathbf{r}_n,\tau_n)$, the residual decoder uses the shared history context to predict $\mathbf{v}_n$. In panel c, $T_{\mathrm{in}}=T_0$ for the deterministic decoder and $T_{\mathrm{in}}=T$ for the residual decoder. The dashed temporal-upsample block is applied only in the deterministic decoder and is bypassed as an identity operation in the residual decoder. The Heun solver iterates from $\mathbf{r}_0^{(k)}=\mathbf{Z}^{(k)}$ to $\mathbf{R}_{1:T}^{(k)}=\mathbf{r}_N^{(k)}$. Symbols, shapes and repeated-module counts are defined in Table~\ref{tab:architecture-notation}.}
  \label{fig:network}
\end{figure}

\ours{} contains three trainable modules: a shared history encoder $\mathcal{E}_{\psi}$, a deterministic decoder parameterised by $\theta$, and a residual decoder parameterised by $\phi$. The latter two are the post-encoder modules, whereas $\mathcal{S}_{\psi,\theta}$ and $\mathcal{R}_{\psi,\phi}$ remain reserved for the corresponding end-to-end mappings. The encoder supplies the flattened history context $\mathbf{C}_{\mathrm{hist}}$ to both decoders. The deterministic decoder produces $\mathbf{S}_{1:T}$, while each residual-decoder evaluation maps $(\mathbf{r}_n,\tau_n,\mathbf{C}_{\mathrm{hist}})$ to the probability-flow velocity $\mathbf{v}_n$. Table~\ref{tab:architecture-notation} summarises the symbols, tensor shapes and repeated-module counts used in Fig.~\ref{fig:network}.

\begin{table}[t]
  \centering
  \small
  \setlength{\tabcolsep}{4pt}
  \renewcommand{\arraystretch}{1.12}
  \caption{Architecture and discrete-inference notation.}
  \label{tab:architecture-notation}
  \begin{tabular}{@{}p{0.10\linewidth}p{0.62\linewidth}p{0.18\linewidth}@{}}
    \toprule
    Symbol & Meaning & Value \\
    \midrule
    $T_0,T$ & Numbers of history and future frames & $9,36$ \\
    $K$ & Number of ensemble members & $4$ \\
    $N$ & Number of Heun intervals & $20$ \\
    $S,s,c$ & Radar-grid side, feature-grid side and feature width & $128,32,256$ \\
    $g$ & Number of AFNO channel groups & $8$ \\
    $M_E$ & Number of repeated TSF blocks in each context backbone & $3$ \\
    $M_T$ & Number of temporal-upsample stages in the deterministic decoder & $2$ \\
    $\mathbf{H}$ & Observed radar history & $(9,128,128,1)$ \\
    $\mathbf{C}_{\mathrm{hist}}$ & Flattened history-context tokens & $(9,1024,256)$ \\
    $\mathbf{S}_{1:T}$ & Deterministic prediction & $(36,128,128,1)$ \\
    $\mathbf{Z}^{(k)}$ & Initial noise for ensemble member $k$ & $(36,128,128,1)$ \\
    $\tau_n$ & Discrete generative time at solver state $n$ & $0,0.05,\ldots,1$ \\
    $\mathbf{r}_n$ & Probabilistic-residual flow state at $\tau_n$ & $(36,128,128,1)$ \\
    $\mathbf{v}_n$ & Velocity predicted from $\mathbf{r}_n$ at $\tau_n$ & $(36,128,128,1)$ \\
    $\mathbf{R}_{1:T}^{(k)}$ & Terminal probabilistic residual for ensemble member $k$ & $(36,128,128,1)$ \\
    \bottomrule
  \end{tabular}
\end{table}

The history encoder patchifies each observed frame with a strided $4\times4$ convolution, reducing the spatial grid from $S\times S$ to $s\times s$ and projecting the single input channel to width $c=256$. The flattened history-token sequence has shape $(T_0,s^2,c)=(9,1024,256)$. The encoder applies $M_E=3$ stacked Temporal--Spatial--Frequency (TSF) blocks. Each block uses temporal attention, spatial mixing, AFNO frequency mixing and a second temporal-attention layer. Temporal attention uses 4 heads at width 256; the spatial modules alternate shifted-window and global-token attention, and AFNO uses $g=8$ channel groups. All attention and MLP sublayers use pre-normalisation, SwiGLU activations and residual connections. The output is $\mathbf{C}_{\mathrm{hist}}\in\mathbb{R}^{T_0\times s^2\times c}$.

The deterministic decoder predicts all $T=36$ future frames in parallel. Four 1D temporal convolution blocks with kernel size 3 and dilation rates $1,2,4,8$ aggregate the history tokens at width $c$. The temporal decoder uses $M_T=2$ transposed 1D convolution stages to expand the temporal axis from $T_0=9$ to $T=36$. Two PixelShuffle stages then increase the feature grid from $s=32$ to $S=128$, followed by the convolutional output head. The result is $\mathbf{S}_{1:T}\in\mathbb{R}^{T\times S\times S\times1}$.

At solver state $n$, the residual decoder receives $\mathbf{r}_n\in\mathbb{R}^{T\times S\times S\times1}$, $\tau_n$ and $\mathbf{C}_{\mathrm{hist}}$. A stride-4 projection produces $(T,s^2,c)$ future tokens, and the sinusoidal embedding of $\tau_n$ is projected to width $c$. The residual pathway applies $M_E=3$ TSF blocks with cross-attention to the history context. Four dilated temporal-convolution blocks then mix information across the $T=36$ future positions without changing their number. Because the residual decoder has equal input and output lengths, its temporal-upsample operator is the identity; the same two-stage spatial upsampler reconstructs the radar grid from $s=32$ to $S=128$. The decoder output is $\mathbf{v}_n:=\mathbf{v}_{\psi,\phi}(\mathbf{r}_n,\tau_n,\mathbf{H})\in\mathbb{R}^{T\times S\times S\times1}$.

The shared encoder, deterministic decoder and residual decoder contain approximately 18.6M, 11.2M and 23.4M trainable parameters, respectively. Diversity is introduced exclusively through the frame-wise independent noise tensor $\mathbf{Z}$ in the residual decoder. Consequently, ensemble spread represents conditional uncertainty in local growth, decay and structural reorganisation around a shared deterministic prediction, rather than stochasticity injected into the deterministic decoder.

\FloatBarrier

\subsection{Training objective function}

The three modules are optimised jointly. The deterministic decoder is supervised by an intensity-weighted reconstruction loss, while the residual decoder is supervised by the conditional flow-matching objective. The total loss is
\begin{equation}
  \mathcal{L}
  =
  \lambda_{\mathrm{det}}\mathcal{L}_{\mathrm{det}}
  +
  \mathcal{L}_{\mathrm{FM}},
  \label{eq:training_total_loss}
\end{equation}
where $\lambda_{\mathrm{det}}=0.2$ in all experiments.

\subsubsection{Deterministic loss}
Let $\mathbf{S}_t=[\mathbf{S}_{1:T}]_t$ be the deterministic prediction at physical lead time $t$. The deterministic loss is
\begin{equation}
  \mathcal{L}_{\mathrm{det}}
  =
  \frac{s_{\mathrm{det}}}{T}
  \sum_{t=1}^{T}
  \ell_{\mathrm{det}}(\mathbf{S}_t,\mX_t),
  \label{eq:lpix_total}
\end{equation}
where $s_{\mathrm{det}}=0.5$. The per-frame loss combines weighted MSE and MAE terms:
\begin{equation}
  \ell_{\mathrm{det}}(\mathbf{S},\mX)
  =
  \frac{1}{S^2}\sum_{i=1}^{S}\sum_{j=1}^{S}
  W(\mX)_{i,j}
  \left[
    \lambda_{\mathrm{mse}}(\mathbf{S}_{i,j}-\mX_{i,j})^2
    +
    \lambda_{\mathrm{mae}}\left|\mathbf{S}_{i,j}-\mX_{i,j}\right|
  \right],
  \label{eq:lpix_final}
\end{equation}
with $\lambda_{\mathrm{mse}}=\lambda_{\mathrm{mae}}=1$. The intensity weight is
\begin{equation}
  [W(\mX)]_{i,j}=q_{k_{i,j}},
  \qquad
  k_{i,j}=\sum_{m=1}^{6}\mathbf{1}\{\mX_{i,j}>\theta_m\},
\end{equation}
where $(i,j)$ indexes a grid cell, $\boldsymbol{\theta}=(-0.4,0,0.2667,0.4667,0.6667,0.9)$ and $\mathbf{q}=(1,1,2,4,6,8,10)$ for normalised reflectivity values in $[-1,1]$.

\subsubsection{Probabilistic-residual flow-matching loss}
Using the probabilistic residual target in Eq.~\eqref{eq:residual_target}, the interpolation in Eq.~\eqref{eq:fm_path} and the target velocity in Eq.~\eqref{eq:fm_target_velocity}, we train the residual decoder with
\begin{equation}
  \mathcal{L}_{\mathrm{FM}}
  =
  \mathbb{E}_{(\mathbf{H},\mX_{1:T})\sim p_{\mathrm{data}},\,\tau,\,\mathbf{Z}}
  \left[
    \frac{1}{TS^2}
    \left\|
      \mathbf{v}_{\psi,\phi}(\mathbf{r}(\tau),\tau,\mathbf{H})
      -
      \mathbf{u}^{\star}
    \right\|_2^2
  \right].
  \label{eq:lfm_final}
\end{equation}

We train for $200{,}000$ optimisation steps with a batch size of 8 using AdamW ($\beta_1=0.9$, $\beta_2=0.999$, $\epsilon=10^{-8}$ and weight decay $0.01$). The learning rate is linearly increased from $10^{-5}$ to a peak of $2\times10^{-4}$ over the first 10\% of training and then cosine-decayed to $10^{-5}$ over the remaining steps. Training uses mixed precision and DeepSpeed gradient clipping with a global-norm threshold of 1.0. No data augmentation beyond random sample shuffling is used.

\subsection{Datasets}
\label{sec:dataset-details}
We evaluate all methods on three regional radar datasets from the US, EU and CN domains. Although the three datasets are constructed from different event sources and radar archives, they are converted to a common nowcasting protocol. Each sample contains 9 historical radar frames and 36 future target frames at 10 min cadence, corresponding to a 1.5 h input window and a 6 h forecast horizon. Radar fields are represented on a common spatial grid and extracted over a fixed 256 km $\times$ 256 km physical domain. Models use preprocessed radar reflectivity as input and predict future reflectivity. Reflectivity is then converted to rain rate for verification and precipitation visualisation.

The US dataset is constructed by pairing NOAA/NCEI Storm Events reports with MRMS Seamless Hybrid Scan Reflectivity radar composites~\cite{noaa_storm_events,zhang2016mrms}. The NCEI records provide the occurrence time, location, event type and impact information for severe weather events, while MRMS provides the corresponding high-resolution precipitation structure within the event-centred radar window. We extract radar sequences using the reported event time and location as anchors, and construct the training and evaluation samples after quality control and de-duplication. The training period spans January 2021 to September 2024 and contains 152,408 samples. The evaluation period spans October 2024 to February 2026. For the main evaluation, we use a compact test set of 4,000 samples constructed from the future holdout period; this set preserves spatiotemporal diversity while prioritising high-impact events.

The EU dataset follows a similar event-record-driven strategy, but uses the European Severe Weather Database (ESWD) as the event source and OPERA radar mosaics as the radar source~\cite{eswd,kock2000opera}. ESWD provides event-level records for severe convective weather, including wind, hail, heavy precipitation, tornadoes and lightning. We retain events that fall within the OPERA radar coverage and have a complete nowcasting time window, and then extract the corresponding radar sequences. Because the reported event time and location do not necessarily coincide with a pronounced precipitation echo in the radar field, we apply radar-signal quality control to remove samples with little or no precipitation signal. The training period spans January 2020 to December 2023 and contains 131,761 samples. The evaluation period spans January 2024 to 23 July 2024. The main evaluation uses a compact 4,000-sample test set that balances event diversity and high-impact event coverage.

The CN dataset is rebuilt from China Meteorological Administration composite reflectivity data using a radar-native, peak-anchored sampling strategy. The sampler first identifies spatially organised future strong-echo peaks in the radar sequence by requiring both local window-mean and window-maximum reflectivity to satisfy strong-precipitation criteria, reducing the chance that isolated noisy pixels trigger candidate samples. For each retained peak, the forecast initialisation time is generated by shifting backward from the peak time using sparse lead-to-peak buckets of 60, 120, 180 and 240 min. This peak-anchored design samples near-peak, mid-growth and early-growth stages, allowing the evaluation to probe 6 h forecasts of heavy-precipitation growth, maintenance and reorganisation.

To reduce near-duplicate samples from the same storm system, we apply frame- and peak-level non-maximum suppression and limit both the daily number of samples and the number of issue times retained for each peak. Training samples are additionally required to retain late-label strong-echo activity, ensuring that the 6 h target window contains meaningful heavy-precipitation evolution. The CN training period spans November 2021 to June 2025 and contains 174,753 samples. The evaluation period spans July 2025 to June 2026, and the main evaluation uses a temporally held-out strict compact test set of 4,000 samples. This compact test set requires strong-echo activity in every forecast hour and future centre-region reflectivity maxima of at least 45 dBZ. It is stratified by month, convective-growth stage and lead-to-peak bucket, and uses a 2 h / 200 km time-space diversity rule to reduce near-duplicate samples.

Together, the three regional datasets support a common evaluation goal: testing long-lead radar nowcasting skill for heavy precipitation and high-impact convective weather. The US and EU datasets provide event-record-centred evaluations, while the CN dataset provides a complementary radar-native evaluation of strong-precipitation processes. Despite these differences in event definition, all three datasets are converted to the same input-output protocol, allowing models and baselines to be compared under a consistent technical setting. This design also determines the focus of our evaluation: high rain-rate thresholds, long forecast lead times and the preservation of organised strong-echo structure, rather than average performance over all precipitation intensities.

\begin{table*}[t]
\centering
\caption{\textbf{Regional radar datasets used in this study.} All datasets are converted to a common 9-frame input and 36-frame forecast protocol at 10 min cadence.}
\label{tab:regional-corpora}
\footnotesize
\setlength{\tabcolsep}{4pt}
\begin{tabular}{llll}
\toprule
Region & Construction & Training set & Evaluation set \\
\midrule
US &
\makecell[l]{NOAA/NCEI Storm Events\\+ MRMS SeamlessHSR} &
\makecell[l]{Jan. 2021--Sep. 2024\\152,408 samples} &
\makecell[l]{Oct. 2024--Feb. 2026\\4,000 samples} \\
\midrule
EU &
\makecell[l]{ESWD severe-weather records\\+ OPERA radar mosaics} &
\makecell[l]{Jan. 2020--Dec. 2023\\131,761 samples} &
\makecell[l]{Jan. 2024--23 Jul. 2024\\4,000 samples} \\
\midrule
CN &
\makecell[l]{Radar-native strong-precipitation\\sampling + CMA reflectivity} &
\makecell[l]{Nov. 2021--Jun. 2025\\174,753 samples} &
\makecell[l]{Jul. 2025--Jun. 2026\\4,000 samples} \\
\bottomrule
\end{tabular}
\end{table*}

\subsection{Baselines}
We compare against three representative baselines spanning classical extrapolation, deterministic deep prediction and physics-guided generative nowcasting. All methods use the same 9-frame history length, 36-frame forecast horizon, reflectivity preprocessing and $128\times128$ verification targets over a $256\,\mathrm{km}\times256\,\mathrm{km}$ domain. The learned models receive the corresponding $128\times128$ input histories. As described below, PySTEPS is intentionally supplied with a larger $256\times256$ spatial context before its forecasts are centre-cropped to the common verification domain.

\subsubsection{PySTEPS}
PySTEPS~\cite{pysteps} is used as the classical transport-based ensemble baseline. We apply the STEPS nowcasting method~\cite{steps} through the official PySTEPS implementation, using Lucas--Kanade motion estimation and a second-order autoregressive model. For each sample, PySTEPS receives the 9-frame, $256\times256$ radar history and predicts the subsequent 36 frames. The forecasts are then centre-cropped to the common $128\times128$ verification domain. This setting provides PySTEPS with broader spatial context than the learned models, including upstream precipitation that may subsequently advect into the verification domain. We generate four ensemble members using independent random seeds for probabilistic evaluation.

\subsubsection{SimVPv2}
SimVPv2~\cite{simvpv2} is used as the deterministic deep-learning baseline. We train it on the same training split as \ours{} using the official model implementation, with the input and output lengths adjusted to 9 and 36 frames, respectively, and the input tensor shape fixed at $128\times128$. To provide a strong deterministic baseline for the high-intensity precipitation regime considered here, SimVPv2 is trained with the same per-frame intensity-weighted MSE-plus-MAE reconstruction loss as the deterministic pixel head of \ours{}, including identical intensity-bin weights and equal coefficients for the MSE and MAE terms. This setting places the same emphasis on intense precipitation in both deterministic predictors, allowing their comparison to focus more directly on differences in model architecture. All other hyperparameters follow the default optimiser and scheduler settings of the released code unless adjustment is required by the modified input--output configuration.

\subsubsection{NowcastNet}
NowcastNet \cite{nowcastnet} is used as the physics-guided generative baseline. We retrain it on our dataset using the official released code, with the input/output length adapted to the 9-frame history and 36-frame forecast setting used in this paper. We then evaluate the retrained model on the corresponding chronological held-out split under the same preprocessing and reflectivity-space evaluation protocol as the other methods.

\subsection{Evaluation settings and metrics}
The primary forecast object in this paper is the ensemble itself. For selected figures and thresholded summary comparisons that require a single field, we also report Probability Matched Mean (PMM) \cite{ebert2001_pmm}. For a grid with $S^2$ cells and $K$ ensemble members, let $m_i=K^{-1}\sum_{k=1}^{K}x_i^{(k)}$ be the ensemble mean at cell $i$, and let $r_i$ be the rank of $m_i$ among the $S^2$ mean-field values. If $Q_{\mathrm{ens}}(p)$ denotes the empirical quantile of the pooled ensemble values $\{x_j^{(k)}:j=1,\ldots,S^2;\,k=1,\ldots,K\}$, the PMM field is computed as
\begin{equation}
  \tilde{x}_i = Q_{\mathrm{ens}}\left(\frac{r_i-1/2}{S^2}\right).
\end{equation}
Equivalently, PMM ranks the grid cells of the ensemble mean and replaces them with values drawn from the pooled ensemble-value distribution according to the same rank order. This retains the spatial organisation of the ensemble mean while avoiding some intensity damping caused by arithmetic averaging.

All models operate in reflectivity-space (dBZ). When rain-rate-based thresholds are needed for evaluation, we convert reflectivity to rain rate using the WSR-88D $Z$--$R$ relation \cite{fulton1998wsr88d}
\begin{equation}
  Z = 300R^{1.4}, \qquad Z=10^{\mathrm{dBZ}/10},
\end{equation}
which yields
\begin{equation}
  R=\left(\frac{10^{\mathrm{dBZ}/10}}{300}\right)^{1/1.4}.
\end{equation}
This conversion is applied only in evaluation and figure labelling, not in model training.

\subsubsection{Grid metrics}
\textit{Critical Success Index Neighbourhood (CSIN).}
Following the neighbourhood verification used by NowcastNet~\cite{nowcastnet}, CSIN allows limited spatial displacement when assessing threshold exceedances. We first convert the forecast and observation to rain rate and form binary exceedance masks $\mathbf{1}\{R\geq\rho\}$. Each mask is then dilated by taking the maximum over a centred $w\times w$ grid-cell neighbourhood. Hits ($\mathrm{TP}_w$), false alarms ($\mathrm{FP}_w$) and misses ($\mathrm{FN}_w$) are accumulated from the resulting neighbourhood masks over all grid cells in the test set, giving
\begin{equation}
  \mathrm{CSIN}(\rho,w)
  =
  \frac{\mathrm{TP}_w}
  {\mathrm{TP}_w+\mathrm{FP}_w+\mathrm{FN}_w}.
\end{equation}
We use the $5\times5$ neighbourhood adopted by NowcastNet unless a different window is specified. For the regional quantitative comparison in Figure~\ref{fig:quantitative-evaluation}, we report CSIN at $\rho\in\{16,32,64\}$ mm h$^{-1}$; other case and ensemble figures use $\rho\in\{32,64\}$ mm h$^{-1}$. When a method produces an ensemble, CSIN is reported on the PMM summary field unless otherwise specified in a figure caption. Curves by forecast lead are computed first and, where a single score is needed, averaged over the reported lead times.

\textit{Probability of Detection (POD).}
POD measures the fraction of observed threshold exceedances that are detected by the forecast. Using the original pointwise threshold masks, it is defined as
\begin{equation}
  \mathrm{POD}(\rho)=\frac{\mathrm{TP}}{\mathrm{TP}+\mathrm{FN}}.
\end{equation}

\subsubsection{Probabilistic metrics}
\textit{Continuous Ranked Probability Score (CRPS).}
CRPS \cite{crps} is computed directly from the empirical ensemble distribution rather than from a fitted Gaussian. It measures the distance between the forecast distribution and the observation, with lower values indicating better probabilistic forecasts. For an ensemble forecast $\{x^{(k)}\}_{k=1}^{K}$ and observation $y$, the grid-cell CRPS is
\begin{equation}
  \mathrm{CRPS}(\{x^{(k)}\},y)=\frac{1}{K}\sum_{k=1}^{K}\left|x^{(k)}-y\right| - \frac{1}{2K^2}\sum_{k=1}^{K}\sum_{l=1}^{K}\left|x^{(k)}-x^{(l)}\right|.
\end{equation}
We use $K=4$ members for \ours{}, PySTEPS, and NowcastNet when probabilistic evaluation is reported. CRPS is computed in precipitation-space at the original grid resolution, then averaged over all grid cells and lead times. Deterministic baselines are excluded from CRPS tables unless explicitly converted to a one-member degenerate forecast.

\textit{Continuous Ranked Probability Skill Score (CRPSS).}
CRPSS reports the relative improvement in CRPS over a reference forecast. We use a persistence forecast, obtained by holding the latest observed radar field constant over the forecast horizon, as the reference forecast when computing CRPSS. For a method with score $\mathrm{CRPS}_{\mathrm{model}}$ and a reference score $\mathrm{CRPS}_{\mathrm{ref}}$, we compute
\begin{equation}
  \mathrm{CRPSS}=1-\frac{\mathrm{CRPS}_{\mathrm{model}}}{\mathrm{CRPS}_{\mathrm{ref}}}.
\end{equation}
Higher CRPSS values indicate better probabilistic skill relative to the reference forecast; a value of zero indicates equal skill to the reference, and negative values indicate worse skill.

\subsubsection{Relative economic value}
Following the cost--loss decision framework used by DGMR~\cite{dgmr} and standard ensemble verification~\cite{richardson2000}, we evaluate whether probabilistic forecasts can reduce the expense of protective decisions. At a fixed region, forecast lead and rain-rate threshold $\rho$, each grid cell and issue time defines a binary event. For a $K$-member forecast, the predicted event probability is
\begin{equation}
  \widehat{p}
  =
  \frac{1}{K}\sum_{k=1}^{K}\mathbf{1}\!\left\{r^{(k)}\geq\rho\right\},
\end{equation}
where $r^{(k)}$ is the rain rate of member $k$; the verifying event is $y=\mathbf{1}\{r^{\mathrm{obs}}\geq\rho\}$. A user takes protective action when $\widehat{p}\geq\gamma$, where $\gamma\in[0,1]$ is a probability decision threshold. Taking action incurs cost $C$, whereas failing to act when the observed event occurs incurs loss $L$. If $N_{\mathrm{TP}}$, $N_{\mathrm{FP}}$, $N_{\mathrm{TN}}$ and $N_{\mathrm{FN}}$ are the resulting contingency counts, the mean forecast expense is
\begin{equation}
  E_{\mathrm{f}}(\gamma)
  =
  \frac{(N_{\mathrm{TP}}+N_{\mathrm{FP}})C+N_{\mathrm{FN}}L}
  {N_{\mathrm{TP}}+N_{\mathrm{FP}}+N_{\mathrm{TN}}+N_{\mathrm{FN}}}.
\end{equation}
Let $p_{\mathrm{c}}$ be the observed climatological event frequency. The minimum expense of the optimal climatological strategy and the expense of a perfect forecast are, respectively,
\begin{equation}
  E_{\mathrm{c}}=\min(C,p_{\mathrm{c}}L),
  \qquad
  E_{\mathrm{p}}=p_{\mathrm{c}}C.
\end{equation}
Because the value depends on $C$ and $L$ only through the cost--loss ratio $\alpha=C/L$, we set $L=1$ and $C=\alpha$. For each $\alpha\in(0,1)$, we sweep the attainable probability thresholds $\gamma$ and report the maximum relative economic value
\begin{equation}
  \mathrm{REV}(\alpha)
  =
  \max_{\gamma}
  \frac{E_{\mathrm{c}}-E_{\mathrm{f}}(\gamma)}
       {E_{\mathrm{c}}-E_{\mathrm{p}}}.
\end{equation}
This normalisation is algebraically equivalent to that used by DGMR: $\mathrm{REV}=1$ for a perfect forecast, $\mathrm{REV}=0$ for the optimal climatological strategy, and negative values indicate worse expense than climatology. We use $K=4$ members and evaluate rain-rate exceedances at $\rho\in\{16,32\}$ mm h$^{-1}$ for lead times of 2, 4 and 6 h in each region. Counts and climatological frequencies are pooled over all grid cells and samples in the corresponding chronologically held-out compact test set. Deterministic predictions are treated as degenerate one-member probabilities.

\subsubsection{Structural diagnostic}
\textit{Power Spectral Density (PSD).}
PSD \cite{psd} is used as a diagnostic of spatial structure. For each lead time and field $X\in\mathbb{R}^{S\times S}$, we compute the two-dimensional discrete Fourier transform $\mathcal{F}(X)$ and its power spectrum
\begin{equation}
  P(k_x,k_y)=\frac{1}{S^2}\left|\mathcal{F}(X)(k_x,k_y)\right|^2.
\end{equation}
The PSD curve is obtained by radially averaging $P(k_x,k_y)$ over spatial wave number $k=\sqrt{k_x^2+k_y^2}$. In figures, we plot forecast and observation PSD curves together at selected forecast lead times. PSD is used as a structural diagnostic curve rather than collapsed into a single scalar score.

\subsection{{Motivation for the residual formulation}}

All model inputs and outputs are radar reflectivity composites in dBZ on a fixed Cartesian grid after preprocessing. Observed radar sequences combine relatively predictable storm-scale displacement with less predictable local growth, decay, merging, splitting and convective initiation. For the following conceptual continuous-time statement, $\mX(t)$ denotes the continuously indexed reflectivity field corresponding to $\mX_t$, and $\mathbf{v}(t)$ denotes the standard advection velocity. We express the physical motivation as
\begin{equation}
  \frac{\partial \mX(t)}{\partial t}
  =
  \underbrace{-\mathbf{v}(t)\cdot\nabla\mX(t)}_{\mathbf{S}_t\;\text{(transport-dominated)}}
  +
  \underbrace{\mathbf{R}_t^{(k)}}_{\text{source-like residual}},
  \label{eq:transport_balance}
\end{equation}
where the underbraces indicate the modelling correspondence to the previously defined deterministic prediction $\mathbf{S}_t$ and probabilistic residual $\mathbf{R}_t^{(k)}$. Equation~\eqref{eq:transport_balance} is used only as physical motivation: the model does not explicitly estimate $\mathbf{v}(t)$ or impose the transport equation as a training constraint.

Radar reflectivity evolution over several hours contains components with different predictability. Storm-scale displacement and broad precipitation organisation are often partly constrained by the recent radar history, whereas local growth, decay, merging, splitting and convective initiation become increasingly uncertain with lead time. This motivates the decomposition in Eq.~\eqref{eq:forecast_factorization}: the deterministic prediction $\mathbf{S}_{1:T}$ represents the shared forecast structure supported by the observed history, while the probabilistic residual $\mathbf{R}_{1:T}^{(k)}$ represents member-dependent local departures from that shared structure. This decomposition is not imposed by an explicit advection equation; instead, the allocation between deterministic structure and probabilistic residual variability is learned jointly from data.

\section{Data availability}
The US severe-weather event records used in this study were obtained from the NOAA/NCEI Storm Events Database through the official NCEI bulk-data archive (\url{https://www.ncei.noaa.gov/pub/data/swdi/stormevents/csvfiles/}). The corresponding radar observations were obtained from the NOAA Multi-Radar/Multi-Sensor (MRMS) system, for which official data-access information is available at \url{https://www.nssl.noaa.gov/projects/mrms/MRMS_data.php}. The European severe-weather event records were obtained from the European Severe Weather Database (ESWD; \url{https://www.eswd.eu/}), operated by the European Severe Storms Laboratory, and were used in accordance with the ESWD User's Agreement. The European maximum-reflectivity radar mosaics were obtained from EUMETNET OPERA under licence; access to these data requires application to and approval by EUMETNET. The China dataset was constructed from composite reflectivity data provided by the China Meteorological Administration (CMA); access to these data requires application to and approval by CMA.

\section{Code availability}
The model code and trained checkpoints for \ours{} will be made publicly available upon publication at \url{https://github.com/microsoft/MW-Nowcast}. The baseline implementations are based on PySTEPS (\url{https://github.com/pySTEPS/pysteps}), NowcastNet (\url{https://doi.org/10.24433/CO.0832447.v1}) and OpenSTL (\url{https://github.com/chengtan9907/OpenSTL}).

\section{Acknowledgements}
We acknowledge the National Centers for Environmental Information (NCEI) for the NOAA Storm Events Database, the National Oceanic and Atmospheric Administration (NOAA) for the Multi-Radar Multi-Sensor (MRMS) products, the European Severe Storms Laboratory (ESSL) for the European Severe Weather Database (ESWD), EUMETNET OPERA for the European radar mosaics, and the China Meteorological Administration (CMA) for the composite reflectivity data used in this study.

\section{Contributions}
H.D., N.W. and Z.F. conceived the study and designed the overall methodology. Z.F. and Z.L. curated and processed the datasets. Z.F. and N.W. implemented the experimental pipeline, performed the experiments and analysed the results. S.Q., P.Z. and S.X. maintained the operational data pipelines. N.W. drafted the initial manuscript. W.J. substantially revised the manuscript and led its scientific framing, including the central story, claims and interpretation of the results. J.W., R.E.T., K.T., N.W., Z.F. and H.D. contributed to the scientific framing, interpretation of the results and revision of the manuscript. J.B. and H.X. reviewed the manuscript and provided feedback. M.C., H.S., J.K. and  K.T. contributed to product requirements and operational context. N.G., B.Z., L.Z., D.D., Q.Z., H.S., K.T. and S.I. provided scientific and strategic guidance and project oversight. H.D. supervised and coordinated the study. All authors reviewed and approved the final manuscript.

\section{Competing interests}
The authors declare no competing interests.

\clearpage
\setcounter{figure}{0}
\renewcommand{\figurename}{Extended Data Fig.}
\renewcommand{\theHfigure}{extended.\arabic{figure}}
\section*{Extended Data Figures}

\begin{figure*}[!ht]
  \centering
  \includegraphics[width=\textwidth,keepaspectratio]{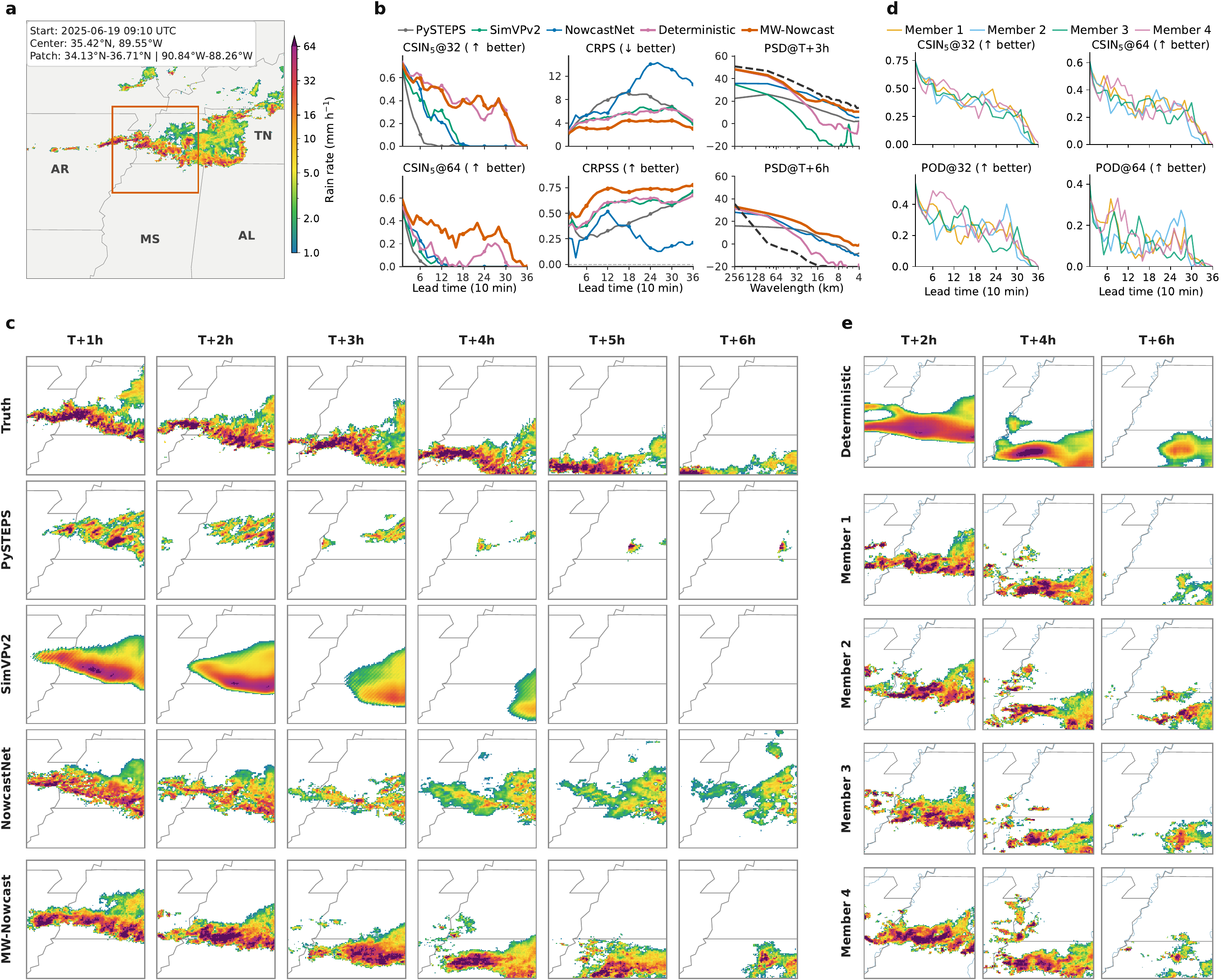}
  \caption{\textbf{US flash-flood precipitation forecast.} The case is located in Tipton County, Tennessee, on 19 June 2025, and the forecast is initialised at 09:10 UTC. The observed sequence shows an organised heavy-precipitation band moving from the northwestern part of the forecast patch towards the southeast. The band remains relatively continuous during the first half of the 6 h window, before the main precipitation shifts towards the southern to southeastern part of the patch and locally weakens at later lead times. Embedded intense cores repeatedly reorganise during this evolution, making the case sensitive to rain-band displacement, core retention and late-lead structural evolution. All Extended Data case figures use the same five-panel layout:
  \textbf{a}, Geographic and radar context, with the forecast domain outlined in orange.
  \textbf{b}, Forecast diagnostics: CSIN in a $5\times5$ grid-cell neighbourhood at rain-rate thresholds of 32 and 64 mm h$^{-1}$, CRPS, CRPSS, and radially averaged PSD at lead times of 3 and 6 h. ``Deterministic'' denotes the output of the deterministic decoder of \ours{}.
  \textbf{c}, Observed and predicted rain-rate fields at hourly lead times from 1 to 6 h. Rows show observations, PySTEPS, SimVPv2, NowcastNet and \ours{}.
  \textbf{d}, CSIN and POD at thresholds of 32 and 64 mm h$^{-1}$ for the four \ours{} ensemble members and their PMM summary.
  \textbf{e}, The deterministic prediction and four ensemble members at lead times of 2, 4 and 6 h.}
  \label{fig:extended-us-event182747}
\end{figure*}

\begin{figure*}[p]
  \centering
  \includegraphics[width=\textwidth,keepaspectratio]{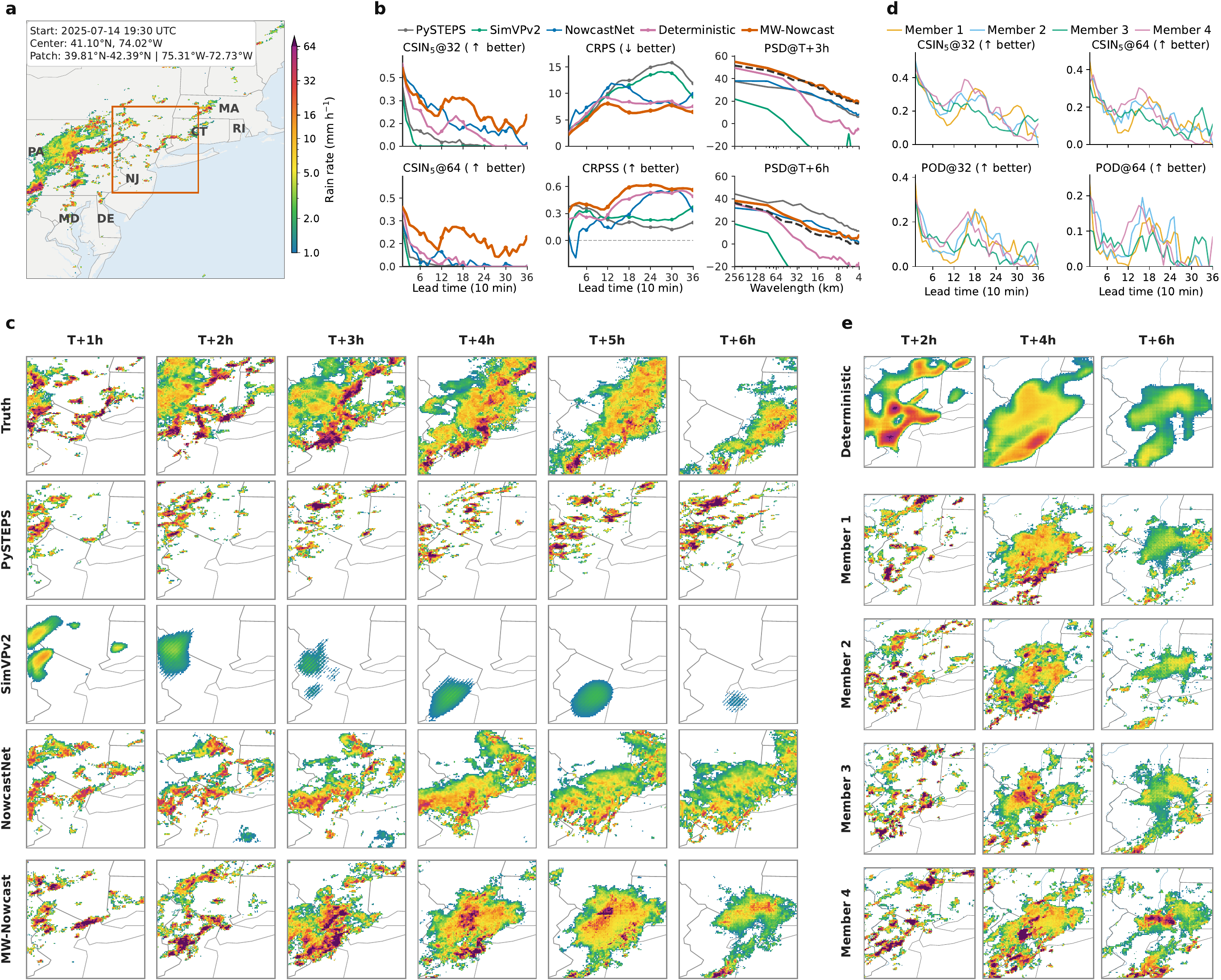}
  \caption{\textbf{US thunderstorm-wind convective precipitation forecast.} The case is located in Rockland County, New York, on 14 July 2025, and the forecast is initialised at 19:30 UTC. The observed sequence shows convective precipitation moving generally eastward and becoming more organised into a coherent rain band during the first half of the forecast window. At later lead times, the precipitation coverage contracts, and the embedded intense cores weaken and become more fragmented. This case tests whether a model can retain rain-band structure during convective organisation while representing the subsequent contraction and decay of high-intensity cores. Panel descriptions are as in Extended Data Fig.~\ref{fig:extended-us-event182747}.}
  \label{fig:extended-us-event189530}
\end{figure*}

\begin{figure*}[p]
  \centering
  \includegraphics[width=\textwidth,keepaspectratio]{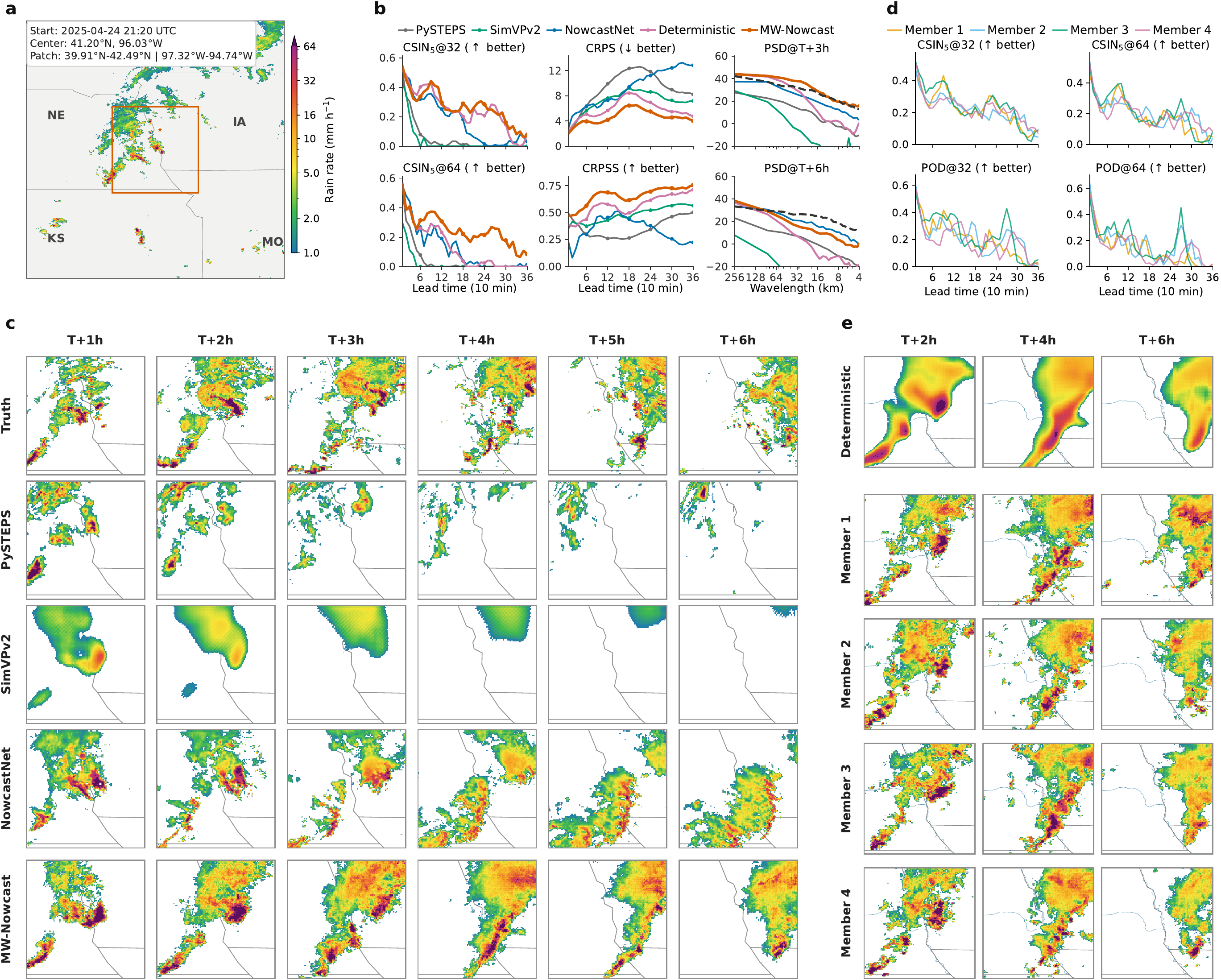}
  \caption{\textbf{US hail-associated convective precipitation forecast.} The case is located in Douglas County, Nebraska, on 24 April 2025, and the forecast is initialised at 21:20 UTC. The observed sequence shows a convective echo field moving generally eastward and expanding across the forecast patch, with intense precipitation organised as several compact local cores rather than a single continuous rain band. These cores repeatedly develop, split and reorganise during the 6 h window, making the case sensitive to convective-core initiation, displacement and multi-core structural retention. Panel descriptions are as in Extended Data Fig.~\ref{fig:extended-us-event182747}.}
  \label{fig:extended-us-event167588}
\end{figure*}

\begin{figure*}[p]
  \centering
  \includegraphics[width=\textwidth,keepaspectratio]{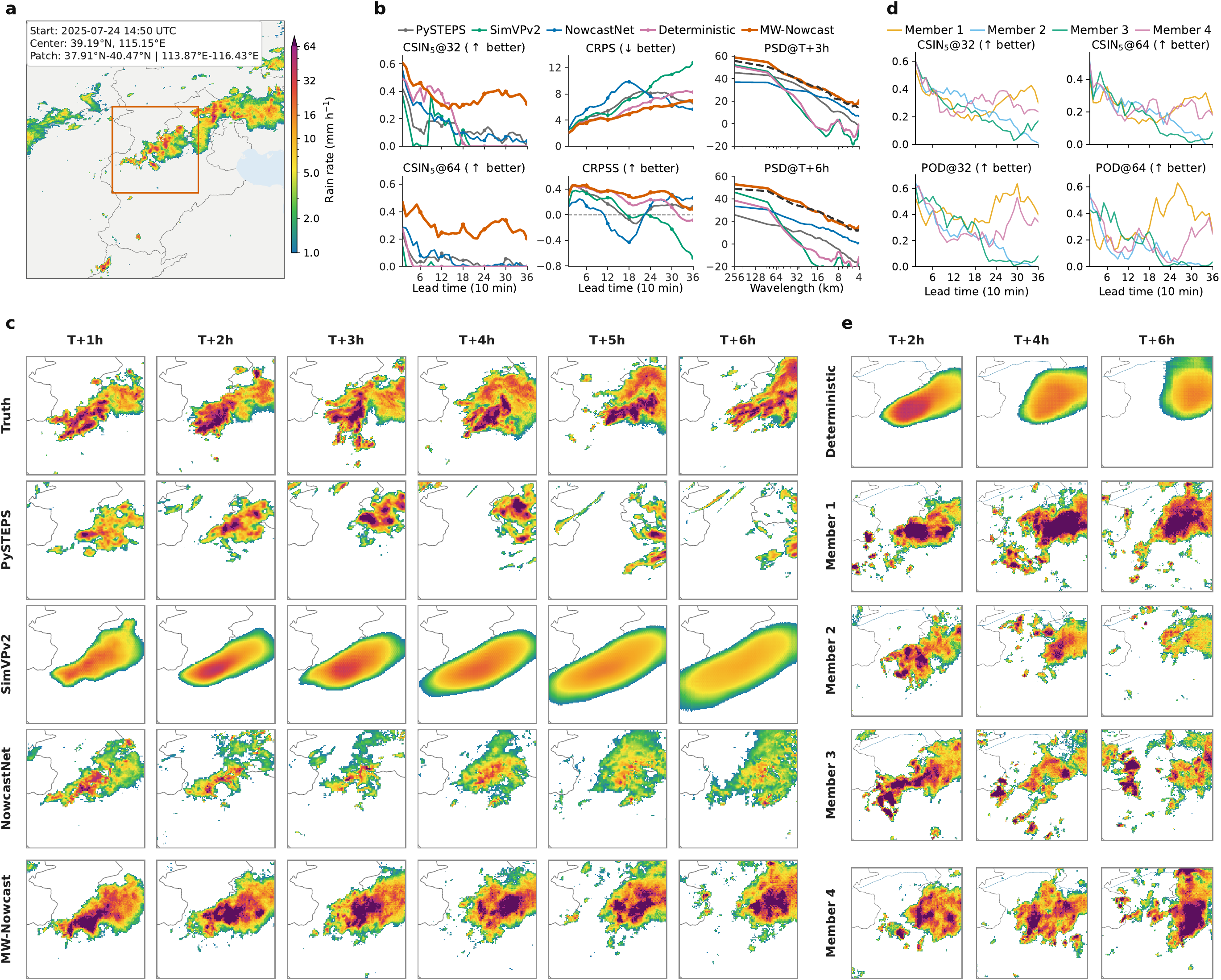}

\caption{\textbf{Mountainous heavy-rainfall forecast over western Hebei.} The forecast patch is centred near Yi County, Baoding, and is initialised at 14:50 UTC on 24 July 2025. The observed sequence shows persistent heavy precipitation that remains active within the forecast patch and gradually expands, while the main rain area retains broad spatial continuity. Peak intensity weakens at later lead times, but embedded local intense cores continue to reorganise. This case tests whether a model can retain rain-area extent and structure in a persistent mountainous rainfall setting while representing the weakening and reorganisation of local intense cores. Panel descriptions are as in Extended Data Fig.~\ref{fig:extended-us-event182747}.}
  \label{fig:extended-cn-180027}
\end{figure*}

\begin{figure*}[p]
  \centering
  \includegraphics[width=\textwidth,keepaspectratio]{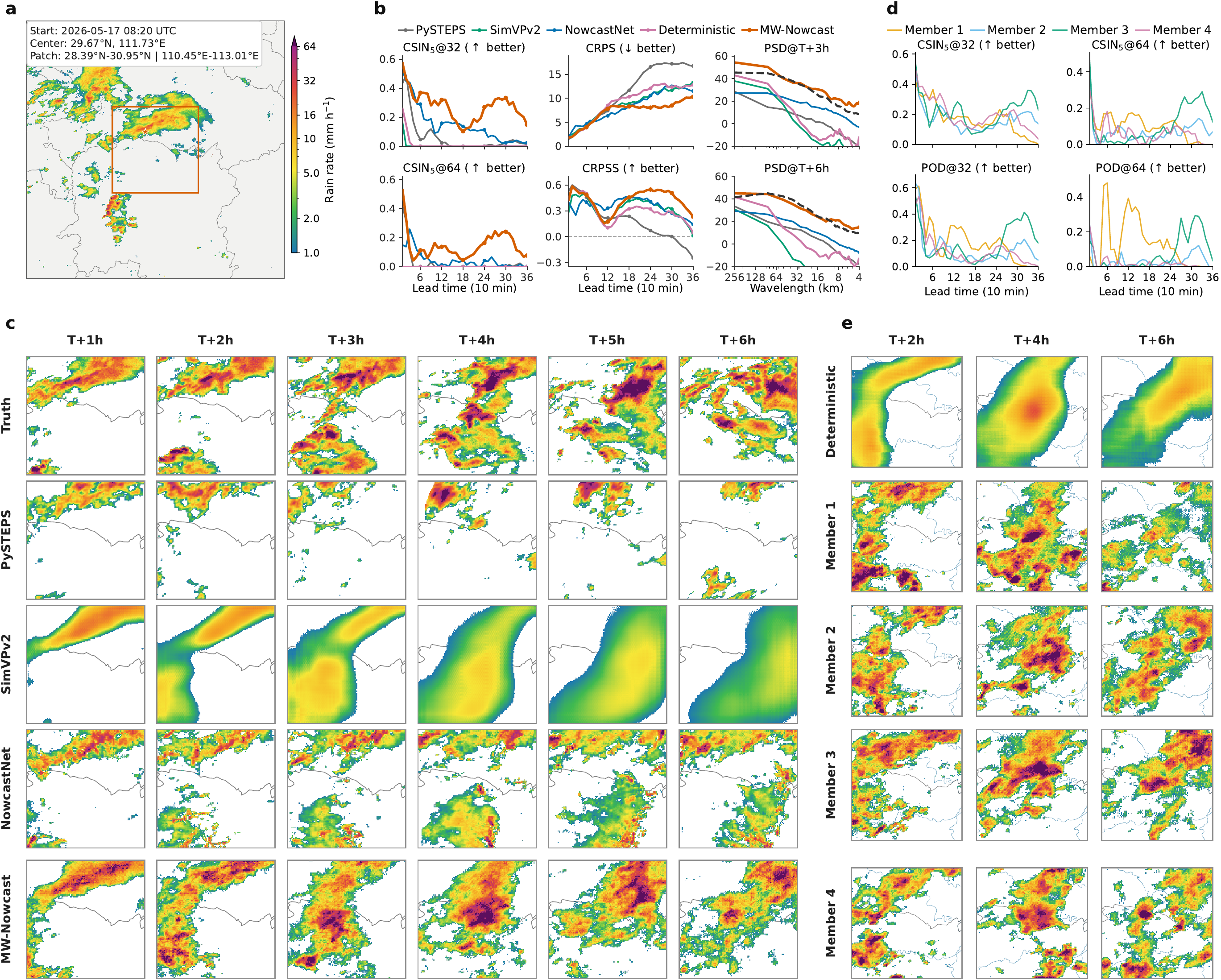}
  \caption{\textbf{Hunan Shimen mountainous heavy-rainfall forecast.} This forecast window forms part of the May 2026 heavy-rainfall episode in Shimen County, Hunan, during which public reports documented six deaths and ten missing people. The forecast is initialised at 08:20 UTC on 17 May 2026. The observed sequence shows a persistent heavy-precipitation system moving generally eastward and expanding across the forecast patch. Embedded intense cores remain strong through the 6 h window and locally reorganise as the rain area expands. This case tests whether a model can retain high-intensity cores and coherent rain-area structure in a moving and expanding mountainous rainfall system. Panel descriptions are as in Extended Data Fig.~\ref{fig:extended-us-event182747}.}
  \label{fig:extended-cn-219270}
\end{figure*}

\begin{figure*}[p]
  \centering
  \includegraphics[width=\textwidth,keepaspectratio]{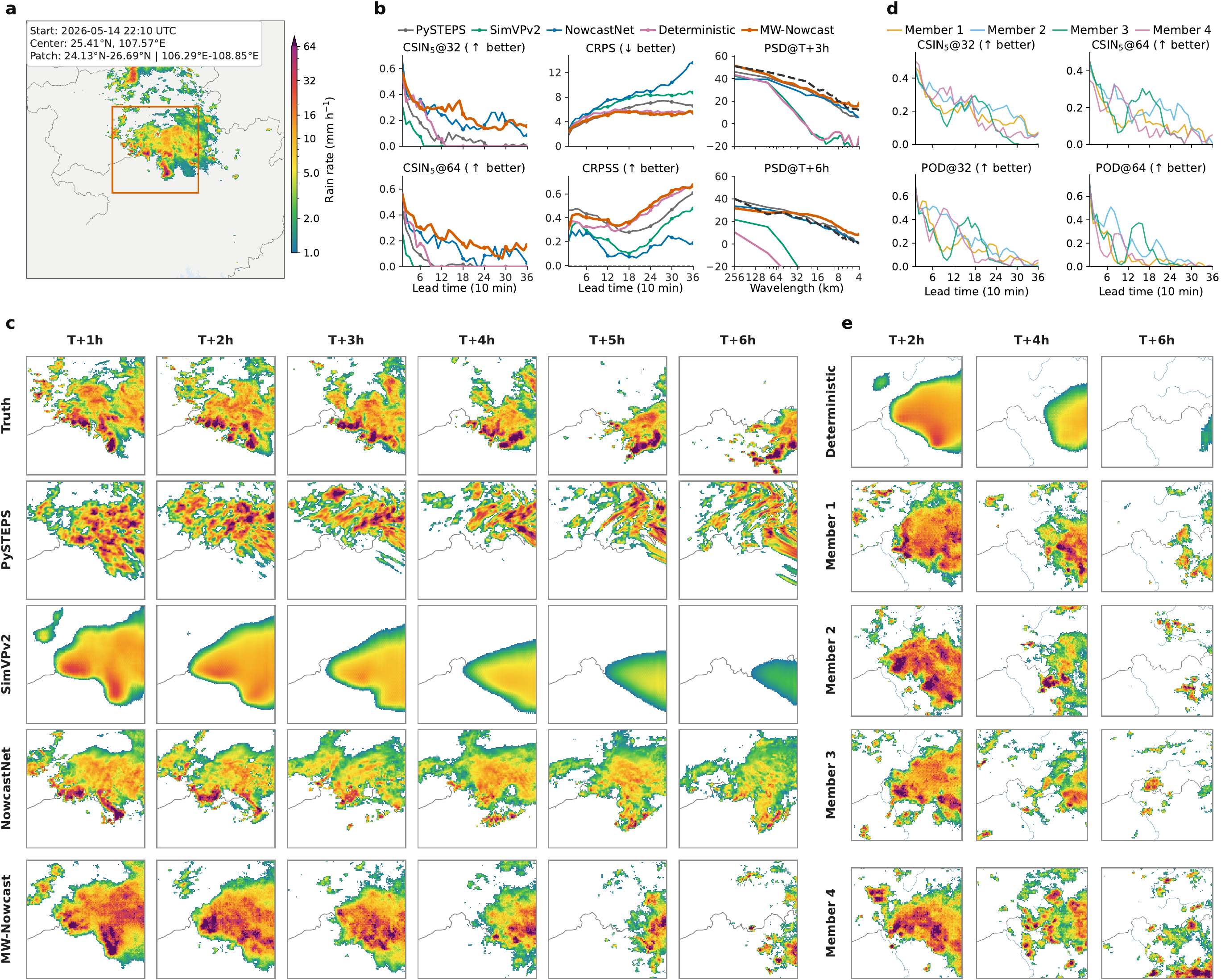}
  \caption{\textbf{Guizhou Guiding--Majiang heavy-rainfall forecast.} This forecast window forms part of the 15--20 May 2026 rainstorm episode in Guizhou; an official disaster summary reported 19 people dead or missing across Guiding, Majiang and several other counties. The forecast is initialised at 22:10 UTC on 14 May 2026. The observed sequence shows a heavy-precipitation system moving generally eastward and gradually contracting after the middle of the forecast window. Although precipitation coverage decreases, embedded intense cores persist and locally reorganise. This case tests whether a model can retain the placement, intense cores and late-lead organisation of a moving heavy-rainfall system during contraction and structural adjustment. Panel descriptions are as in Extended Data Fig.~\ref{fig:extended-us-event182747}.}
  \label{fig:extended-cn-218780}
\end{figure*}

\begin{figure*}[p]
  \centering
  \includegraphics[width=\textwidth,keepaspectratio]{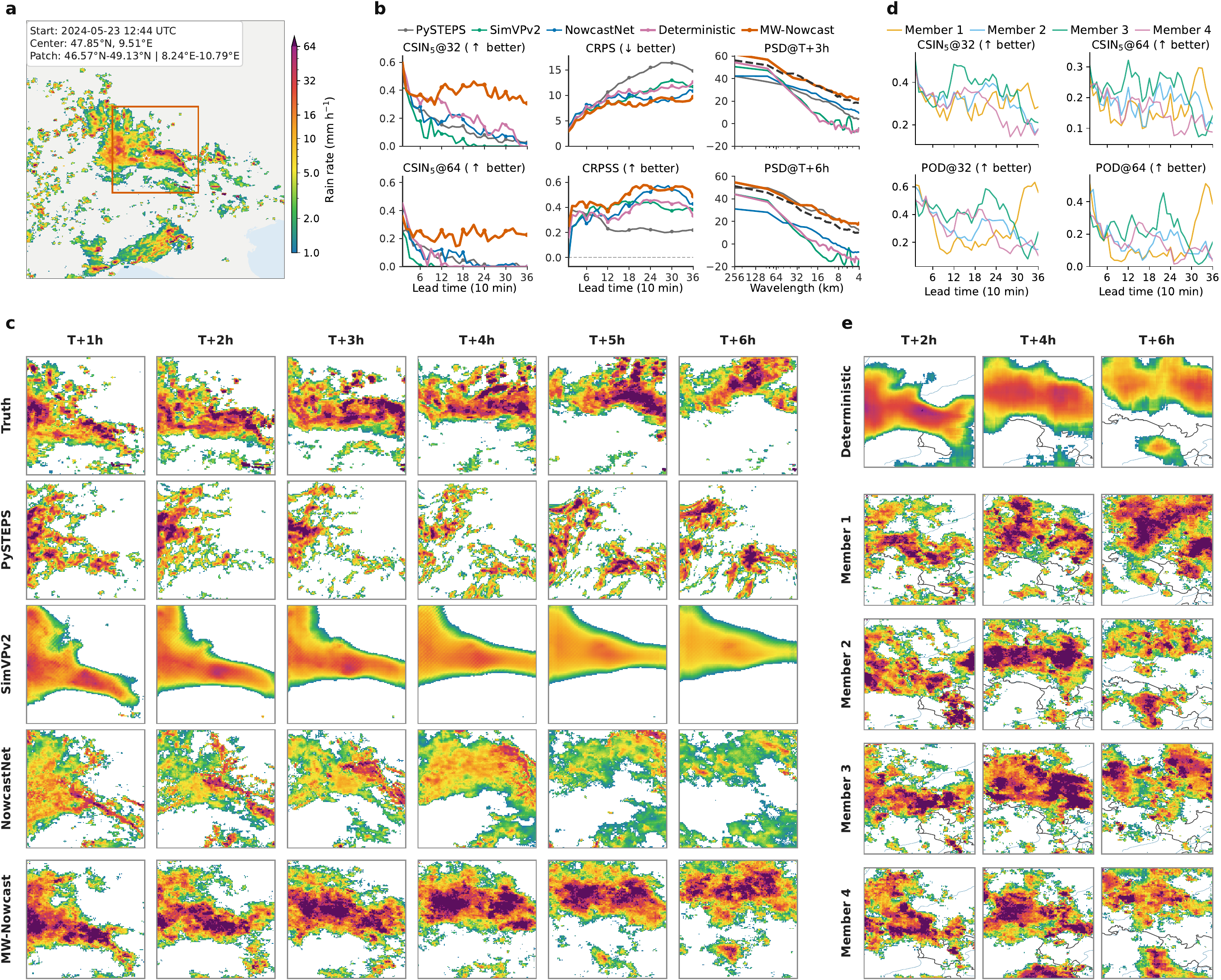}
  \caption{\textbf{European lightning-associated convective precipitation forecast.} The case is located in southern Germany on 23 May 2024, and the forecast is initialised at 12:44 UTC. The observed sequence shows organised convective precipitation moving generally eastward and expanding through the first two-thirds of the forecast window. The precipitation area then contracts modestly, while embedded intense cores persist and locally reorganise along the main rain area. This case tests whether a model can retain rain-area continuity and intense-core structure in a moving and expanding convective system while representing the later modest contraction. Panel descriptions are as in Extended Data Fig.~\ref{fig:extended-us-event182747}.}
  \label{fig:extended-eu-172467}
\end{figure*}

\begin{figure*}[p]
  \centering
  \includegraphics[width=\textwidth,keepaspectratio]{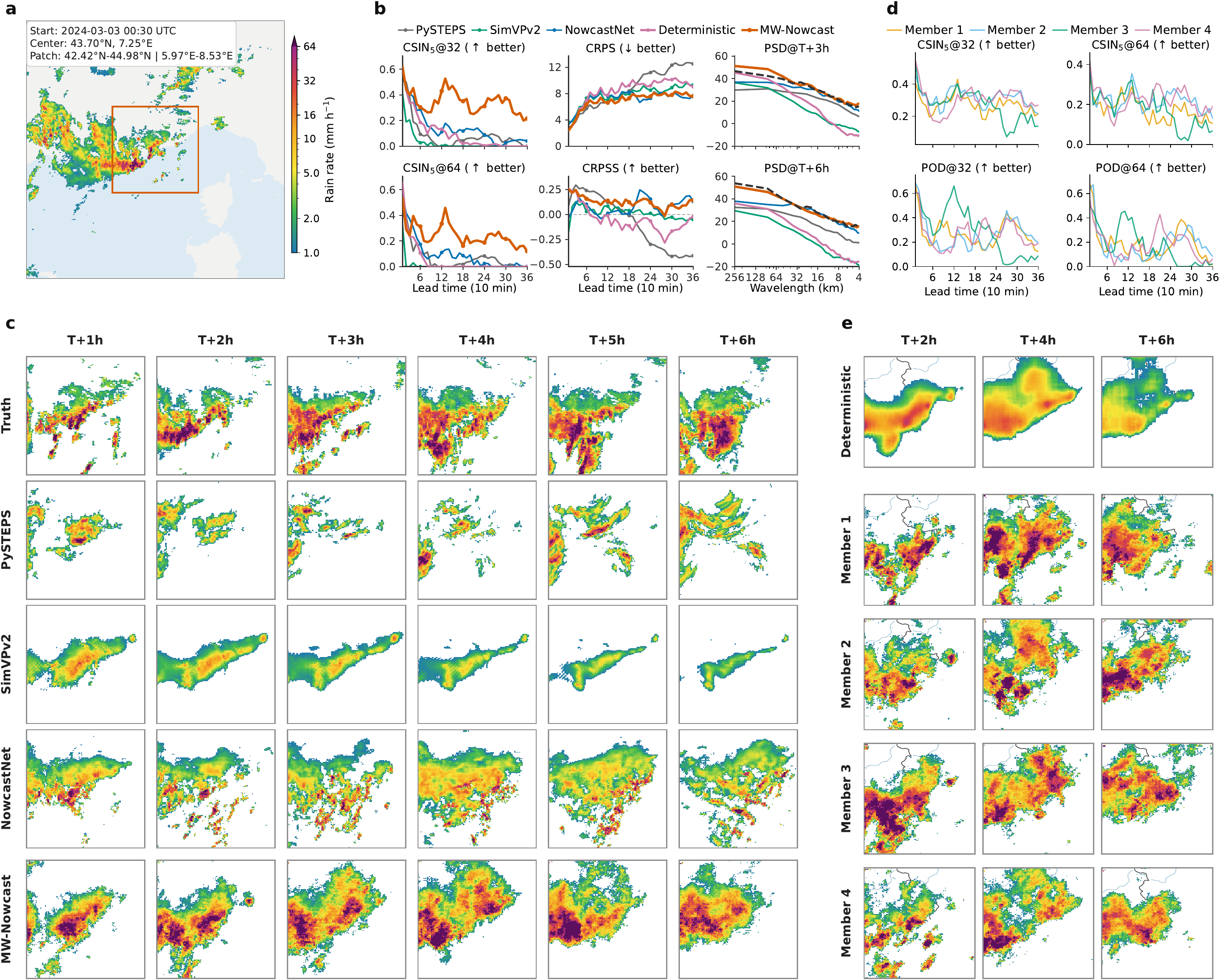}
  \caption{\textbf{European tornado-associated convective precipitation forecast.} The case is located in southern France on 3 March 2024, and the forecast is initialised at 00:30 UTC. The observed sequence shows a slowly moving precipitation system that expands through the 6 h window. Intense precipitation is organised mainly as multiple compact cores that repeatedly develop, split and reorganise, rather than as a stable single rain band. This case tests whether a model can represent local core initiation, displacement and morphology changes within a slowly moving but expanding convective system. Panel descriptions are as in Extended Data Fig.~\ref{fig:extended-us-event182747}.}
  \label{fig:extended-eu-162245}
\end{figure*}

\begin{figure*}[p]
  \centering
  \includegraphics[width=\textwidth,keepaspectratio]{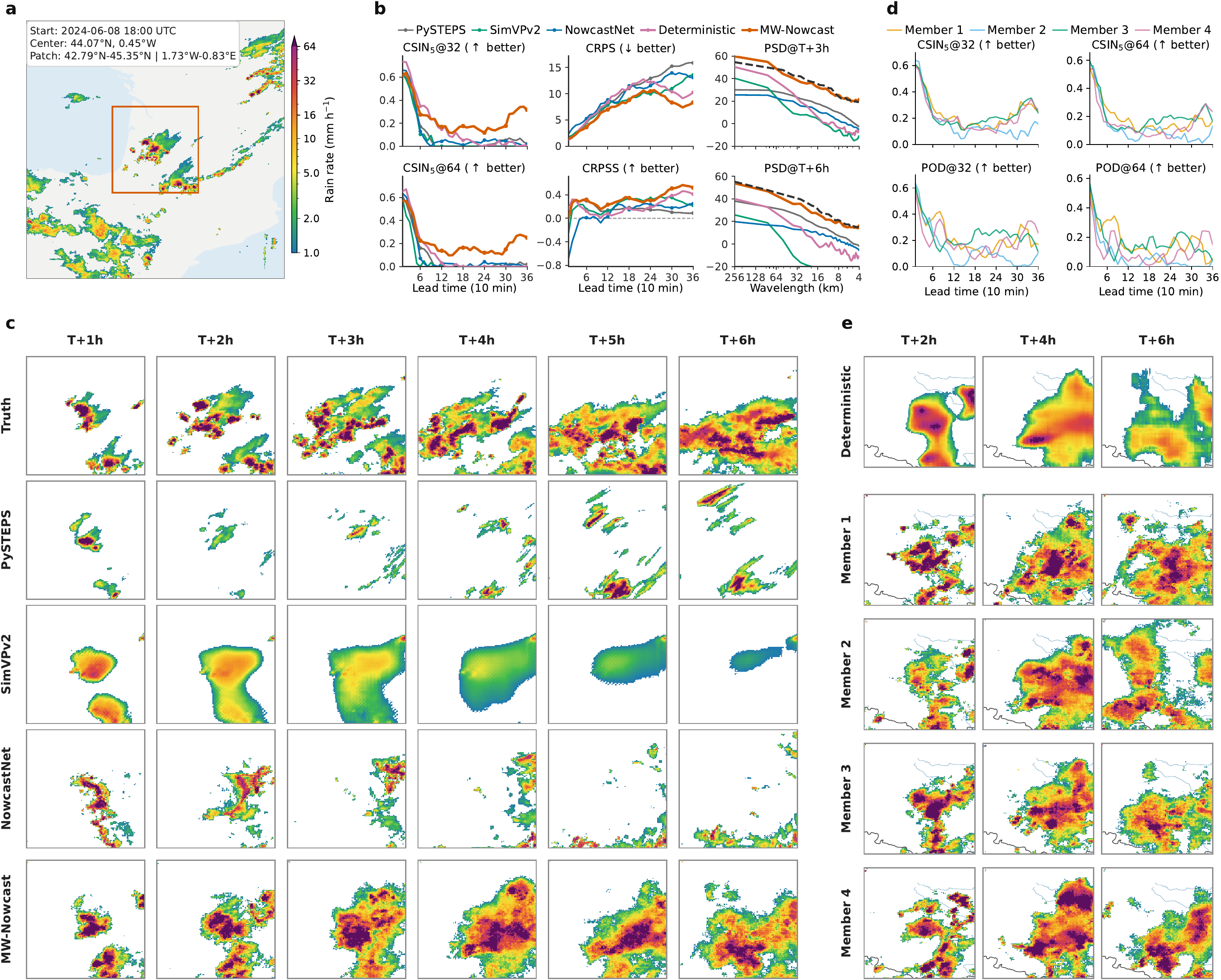}
  \caption{\textbf{European heavy-precipitation forecast over southwestern France.} The case is located in southwestern France on 8 June 2024, and the forecast is initialised at 18:00 UTC. The observed sequence shows a slowly moving heavy-precipitation area that expands substantially through the 6 h window, gradually forming a broader and more continuous rain region. Embedded intense cores continue to reorganise within the main rain area rather than rapidly moving out of the forecast patch. This case tests whether a model can retain rain-area extent, spatial continuity and local intense-core evolution in a slowly moving and steadily expanding heavy-precipitation system. Panel descriptions are as in Extended Data Fig.~\ref{fig:extended-us-event182747}.}
  \label{fig:extended-eu-175810}
\end{figure*}

\begin{figure*}[p]
  \centering
  \includegraphics[width=\textwidth,keepaspectratio]{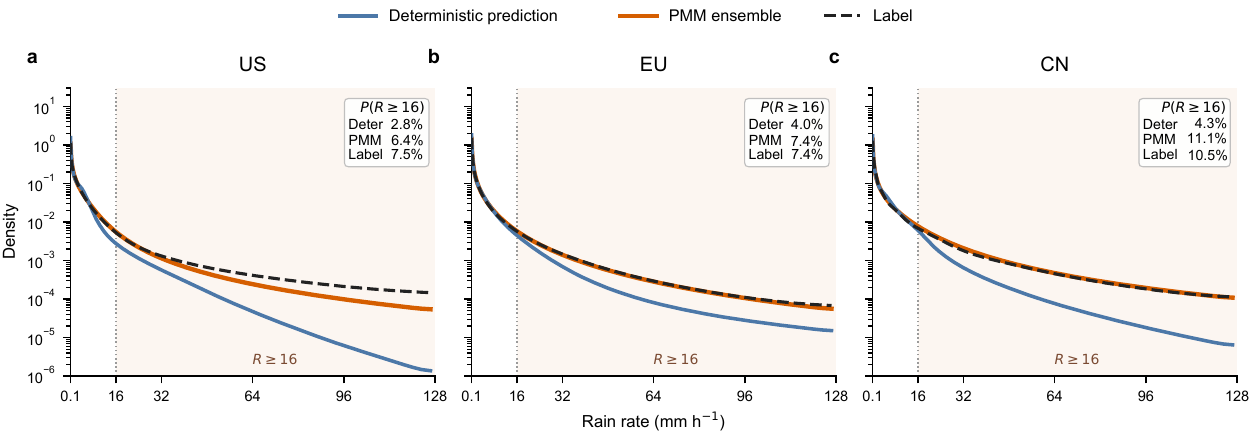}
  \caption{\textbf{Regional rain-rate distributions and high-intensity-tail recovery.} \textbf{a--c}, Conditional rain-rate densities pooled over 4,000 held-out events in the United States (US), Europe (EU) and China (CN), respectively. Densities use effective-rain pixels with $0.1\leq R<128$~mm~h$^{-1}$. Blue solid curves show the deterministic branch, orange solid curves show the Probability Matched Mean (PMM) ensemble summary, and black dashed curves show ground truth. The shaded region marks the high-intensity tail, $R\geq16$~mm~h$^{-1}$, corresponding to the lower high-rain threshold used in the regional quantitative evaluation. Insets report pooled pixel--time exceedance probability conditional on the displayed rain-rate range. Across all three regions, PMM recovers substantially more high-intensity probability mass than the deterministic branch and closely approaches the observed tail.}
  \label{fig:extended-regional-rain-density}
\end{figure*}

\FloatBarrier

\clearpage
\newgeometry{margin=1in,bindingoffset=0pt}
\setcounter{section}{0}
\setcounter{figure}{0}
\setcounter{table}{0}
\setcounter{equation}{0}
\renewcommand{\theHsection}{supplementary.\arabic{section}}
\renewcommand{\theHfigure}{supplementary.\arabic{figure}}
\renewcommand{\theHtable}{supplementary.\arabic{table}}
\renewcommand{\theequation}{S\arabic{equation}}
\renewcommand{\theHequation}{supplementary.\arabic{equation}}
\renewcommand{\figurename}{Supplementary Fig.}
\renewcommand{\tablename}{Supplementary Table}
\pdfbookmark[0]{Supplementary Information}{supplementary-information}

\section*{Supplementary Information}

\section{Related works}

\subsection{Historical perspective}

Precipitation nowcasting has developed along two complementary lines: extrapolating radar-echo motion and learning nonlinear evolution from radar sequences. Classical radar nowcasting starts from the observation that short-lead precipitation fields often contain a strong advective component. Echo motion is estimated from recent radar frames and then extrapolated into the future. STEPS and its PySTEPS implementation extend this idea by combining scale-dependent extrapolation with stochastic perturbations, yielding fast and interpretable ensemble forecasts \cite{steps,pysteps}. Related optical-flow and radar-tracking methods also exploit motion information in recent radar observations and remain valuable in rapidly updated operational settings \cite{casa}. Their limitation is that severe convective precipitation is not a passive tracer. As lead time extends towards 3--6 h, local growth, decay, merging, splitting and initiation become increasingly important, making pure extrapolation insufficient for maintaining intense cores and organised precipitation structure \cite{guidelines,nwp_challenge}.

Deep learning broadened the modelling capacity of radar nowcasting by learning nonlinear spatiotemporal evolution from large radar archives. ConvLSTM formulated precipitation nowcasting as convolutional recurrent sequence prediction, and later models such as PredRNN, EarthFormer and SimVPv2 further improved the representation of spatiotemporal dependencies \cite{convlstm,predrnn,earthformer,simvpv2}. These deterministic models can capture morphology beyond linear extrapolation, but point-wise training losses tend to average over multiple plausible futures, weakening intense echoes at longer lead times. Generative radar nowcasting models address this limitation by sampling multiple future fields. DGMR showed that deep generative models can produce useful probabilistic precipitation nowcasts, and NowcastNet further demonstrated the value of incorporating physical structure for extreme-precipitation nowcasting \cite{dgmr,nowcastnet}. Together, these studies show that long-lead extreme-precipitation nowcasting requires both the retention of predictable structure supported by the recent radar history and the representation of uncertain local evolution.

\subsection{Diffusion-based and flow-based nowcasting}

Recent diffusion and flow-based nowcasting studies extend generative radar forecasting along complementary directions. PreDiff and LDCast use latent diffusion to make probabilistic precipitation nowcasting tractable while representing forecast uncertainty \cite{prediff,ldcast}. StormDiT brings latent rectified-flow transformers into the 2--6 h precipitation nowcasting grey zone, and FlowCast explores conditional flow matching for precipitation nowcasting \cite{stormdit,flowcast}. CasCast and DiffCast use cascaded or residual formulations to connect deterministic evolution with stochastic refinement \cite{cascast,diffcast}. Together, these studies suggest a useful principle for longer-range nowcasting: forecast uncertainty should be represented explicitly, while the more predictable component of storm evolution should remain structurally anchored.

MW-Nowcast follows this principle but makes a different set of modelling choices. Several existing diffusion nowcasting systems operate in latent space, which improves tractability but can make the recovery of compact high-intensity radar structures more dependent on the decoder. Other cascaded or residual systems separate deterministic prediction from stochastic refinement, so the residual model is trained relative to a fixed or separately trained forecast. In contrast, MW-Nowcast is a pixel-level, end-to-end flow-matching ensemble nowcaster. It pairs a deterministic decoder, which provides the shared predictable forecast structure, with a probabilistic residual decoder, which samples local growth, decay, reorganisation and initiation around that shared forecast. Because the residual target is recomputed as the deterministic prediction changes during joint training, the two branches learn the decomposition together rather than forming a fixed post-processing cascade.

This distinction motivates the ablation experiments reported below. We compare pixel- and latent-space configurations, and use controlled comparisons to test an explicit deterministic-prediction--probabilistic-residual decomposition against a single generative head, joint end-to-end training against a two-stage residual cascade, and flow matching against an EDM diffusion objective.

\section{Ablation study}\label{sec:supp-ablation}

The ablation study evaluates which modelling choices contribute to the skill of MW-Nowcast. All comparison variants are evaluated on the held-out US test set described in the main text, using the same 9-frame input, 36-frame output and 6 h nowcasting protocol as the main quantitative comparison. We compare MW-Nowcast with five variants: \texttt{Latent}, \texttt{OneStage}, \texttt{TwoStage}, \texttt{EDM-ODE} and \texttt{EDM-SDE}. These comparisons examine generation space, decoder decomposition, training strategy and probabilistic objective.

\subsection{Decoder coupling and training strategy}

The \texttt{OneStage} and \texttt{TwoStage} variants examine the design hypotheses underlying the deterministic-prediction--probabilistic-residual formulation and the way this decomposition is learned. \texttt{OneStage} removes the explicit deterministic decoder and uses a single flow-matching generative head to produce the complete future sequence directly. Its lower CSIN$_5$ across all reported thresholds and lead times suggests that directly generating the full forecast sequence provides a less effective basis for detecting intense precipitation. One possible explanation is that a single generator must represent both the large-scale, relatively predictable evolution and the smaller-scale uncertain departures, whereas the explicit decomposition allows these components to be handled by specialised decoders.

\texttt{TwoStage} retains the deterministic decoder and residual decoder but trains them sequentially: the deterministic decoder is trained first, and the residual decoder is then trained against a fixed residual target. Compared with \texttt{OneStage}, this configuration recovers part of the skill at the 32 and 64 mm h$^{-1}$ thresholds, supporting the value of an explicit deterministic forecast anchor. However, it remains weaker than MW-Nowcast at all representative lead times and in the mean over the full forecast horizon reported in Supplementary Table~\ref{tab:supp-ablation-metrics}. This remaining gap may arise because the residual decoder is learned from predictions produced by a fixed deterministic model. Joint optimisation in MW-Nowcast instead allows the shared representation, deterministic prediction and probabilistic residual target to adapt together during training.

\subsection{Generation space and diffusion objective}

The \texttt{Latent} variant represents an alternative generative design that moves forecasting from radar-pixel space to a compact VAE latent space. Its consistently lower CSIN$_5$ suggests that this configuration retains less information relevant to intense-precipitation detection. A plausible contributing factor is the compression and reconstruction process, which may smooth compact precipitation cores or attenuate local extremes, causing fewer locations to exceed the high rain-rate thresholds after decoding. This effect is particularly relevant at high thresholds, where even modest attenuation can change whether an intense precipitation feature is detected.

The \texttt{EDM-ODE} and \texttt{EDM-SDE} variants retain the two-branch architecture and joint training setting, but replace the flow-matching residual objective used in MW-Nowcast with an EDM diffusion objective. They differ only in the sampling procedure: \texttt{EDM-ODE} uses deterministic ODE sampling, whereas \texttt{EDM-SDE} uses stochastic sampling with added noise. MW-Nowcast achieves higher CSIN$_5$ than both EDM variants in every reported column, indicating a consistent advantage for flow matching under the evaluated setting. The absolute differences are nevertheless modest, particularly at the highest threshold and longest lead times, and are consistent with flow matching being well suited to learning the residual transformation rather than identifying the probabilistic objective as the dominant source of improvement.

Taken together, the ablation results provide consistent empirical support for the main design hypotheses of MW-Nowcast. The clearest gains are associated with combining an explicit deterministic forecast anchor with probabilistic residual generation and optimising the two components jointly. The full pixel-space configuration performs best among the evaluated generation designs, while flow matching provides a further, more modest improvement over the EDM alternatives.

Supplementary Table~\ref{tab:supp-ablation-metrics} summarises the quantitative comparison using CSIN$_5$ at representative lead times and as a mean over the complete 6 h forecast horizon.

\begin{figure}[p]
  \centering
  \includegraphics[width=0.95\linewidth]{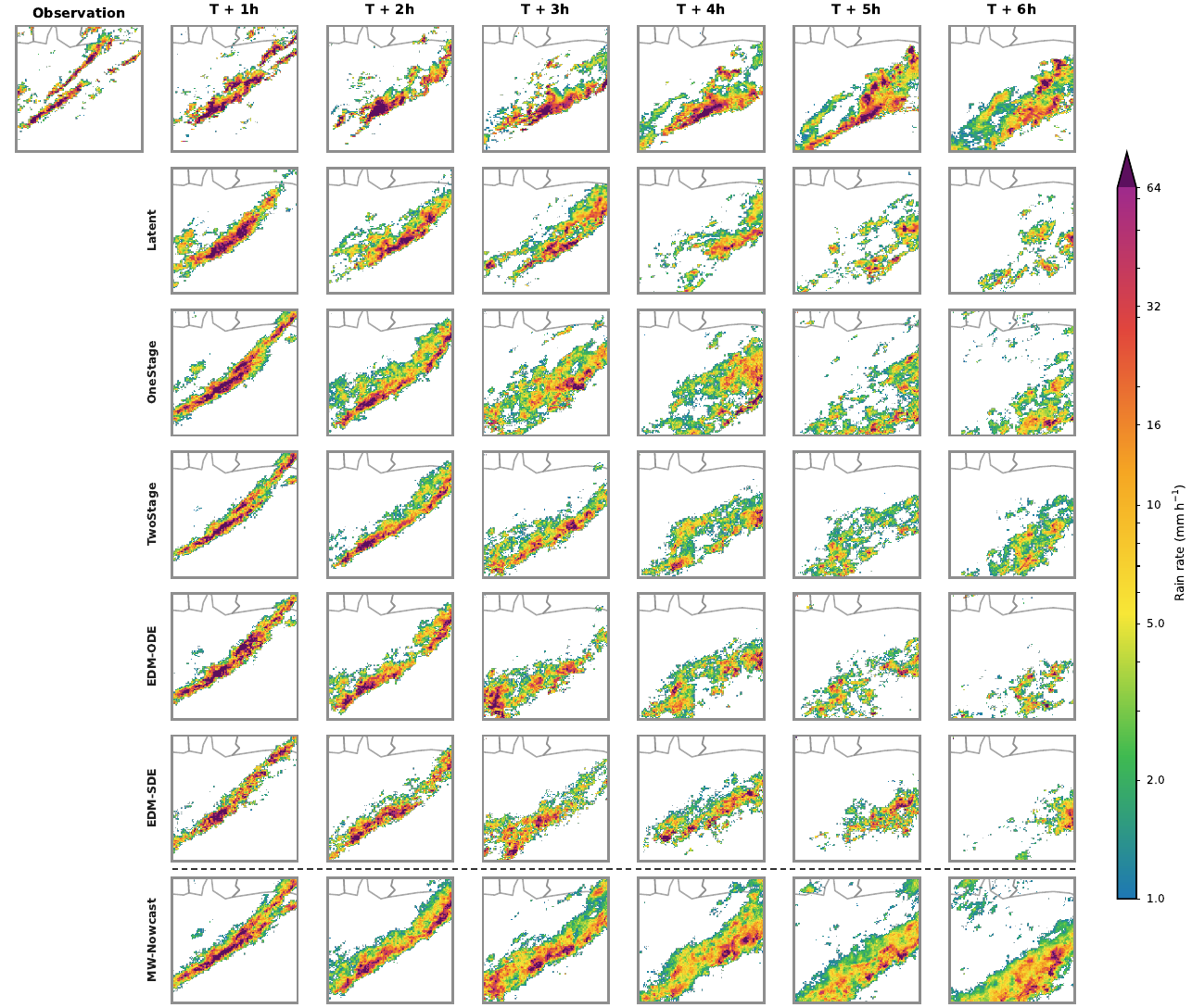}
  \captionsetup{font=small}
  \caption{\textbf{Visual comparison of MW-Nowcast ablation variants for a held-out US strong-precipitation case.} The forecast is initialised at 09:00 UTC on 13 February 2025 for a patch centred near 29.4$^\circ$N, 87.5$^\circ$W. The figure includes the input context and the subsequent 6 h forecast evolution. Rows compare the observed fields, MW-Nowcast and the comparison variants \texttt{Latent}, \texttt{OneStage}, \texttt{TwoStage}, \texttt{EDM-ODE} and \texttt{EDM-SDE}. The case illustrates how the ablated design choices affect the retention of compact intense precipitation cores and organised storm structure over the forecast horizon.}
  \label{fig:supp-ablation-case-us}
\end{figure}

Supplementary Fig.~\ref{fig:supp-ablation-case-us} complements these aggregate results with a qualitative comparison for a representative strong-precipitation case. The example illustrates how the alternative configurations differ in their ability to retain compact intense precipitation cores and organised storm structure as lead time increases.

\clearpage
\begin{table}[p]
  \centering
  \captionsetup{font=small}
  \caption{\textbf{CSIN$_5$ comparison of MW-Nowcast ablation variants on the held-out US test set.} Neighbourhood Critical Success Index with a $5\times5$ neighbourhood is reported at rain-rate thresholds of 16, 32 and 64~mm~h$^{-1}$ for representative lead times T+2 h, T+4 h and T+6 h, together with the arithmetic mean of the unsmoothed per-lead scores over all 36 lead times from T+10 min to T+6 h. Higher is better. The best displayed value in each column is shown in bold.}
  \label{tab:supp-ablation-metrics}
  \scriptsize
  \setlength{\tabcolsep}{2.6pt}
  \renewcommand{\arraystretch}{1.08}
  \begin{tabular}{lcccccccccccc}
    \toprule
    Model & \multicolumn{3}{c}{T+2 h} & \multicolumn{3}{c}{T+4 h} & \multicolumn{3}{c}{T+6 h} & \multicolumn{3}{c}{Mean} \\
    \cmidrule(lr){2-4} \cmidrule(lr){5-7} \cmidrule(lr){8-10} \cmidrule(lr){11-13}
    & 16 & 32 & 64 & 16 & 32 & 64 & 16 & 32 & 64 & 16 & 32 & 64 \\
    \midrule
    \textbf{MW-Nowcast} & \textbf{0.284} & \textbf{0.214} & \textbf{0.160} & \textbf{0.197} & \textbf{0.129} & \textbf{0.085} & \textbf{0.095} & \textbf{0.055} & \textbf{0.032} & \textbf{0.256} & \textbf{0.194} & \textbf{0.147} \\
    \texttt{Latent}   & 0.216 & 0.147 & 0.099 & 0.155 & 0.100 & 0.062 & 0.072 & 0.038 & 0.021 & 0.204 & 0.144 & 0.103 \\
    \texttt{OneStage} & 0.239 & 0.168 & 0.118 & 0.164 & 0.096 & 0.058 & 0.078 & 0.039 & 0.021 & 0.225 & 0.164 & 0.121 \\
    \texttt{TwoStage} & 0.233 & 0.174 & 0.129 & 0.163 & 0.114 & 0.079 & 0.073 & 0.044 & 0.027 & 0.218 & 0.169 & 0.130 \\
    \texttt{EDM-ODE}  & 0.247 & 0.181 & 0.129 & 0.187 & 0.126 & 0.085 & 0.075 & 0.046 & 0.027 & 0.235 & 0.178 & 0.134 \\
    \texttt{EDM-SDE}  & 0.244 & 0.178 & 0.127 & 0.186 & 0.126 & 0.085 & 0.068 & 0.042 & 0.025 & 0.234 & 0.177 & 0.133 \\
    \bottomrule
  \end{tabular}
\end{table}

\clearpage
\clearpage

\section{Flow-matching visualisation}

This section visualises how the residual decoder in MW-Nowcast constructs ensemble members. Supplementary Figs.~\ref{fig:supp-fm-us-150035} and~\ref{fig:supp-fm-eu-21477} show single-member residual trajectories, illustrating how Gaussian noise is transformed into a structured probabilistic residual and then added to the deterministic prediction. Supplementary Figs.~\ref{fig:supp-fm-ens-us-150035} and~\ref{fig:supp-fm-ens-eu-21477} show how multiple members diverge along the same learned flow, revealing where ensemble uncertainty is concentrated.

\subsection{Single-member residual construction}\label{sec:supp-single-member}

\begin{figure}[p]
  \centering
  \begin{subfigure}{\linewidth}
    \includegraphics[width=\linewidth]{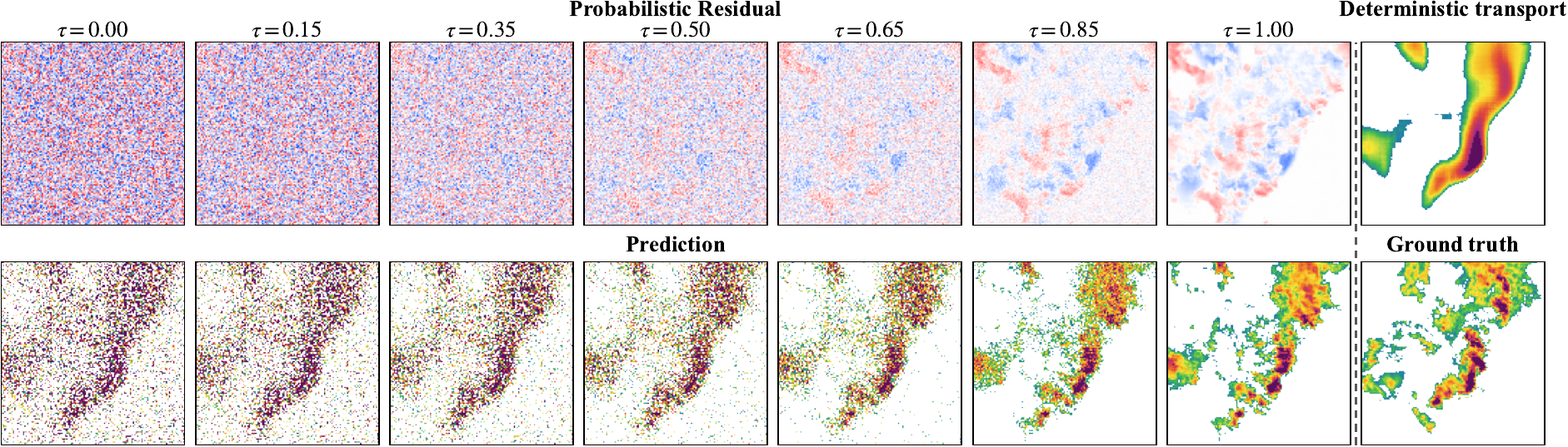}
    \caption{Lead time T+4 h.}
    \label{fig:supp-fm-us-150035-t4h}
  \end{subfigure}

  \vspace{0.5em}
  \begin{subfigure}{\linewidth}
    \includegraphics[width=\linewidth]{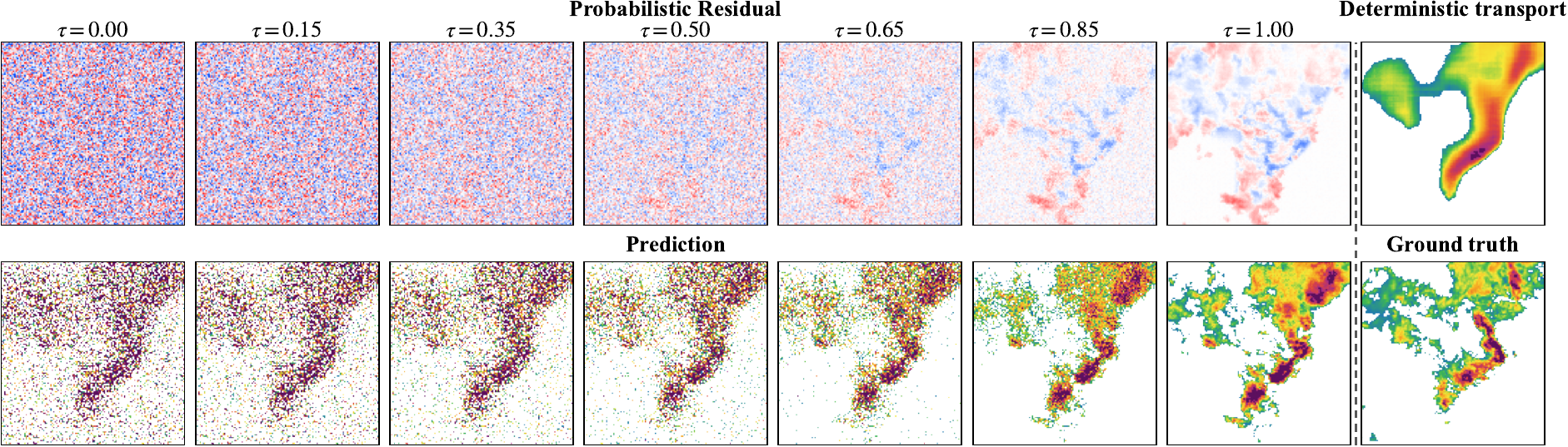}
    \caption{Lead time T+5 h.}
    \label{fig:supp-fm-us-150035-t5h}
  \end{subfigure}

  \vspace{0.5em}
  \begin{subfigure}{\linewidth}
    \includegraphics[width=\linewidth]{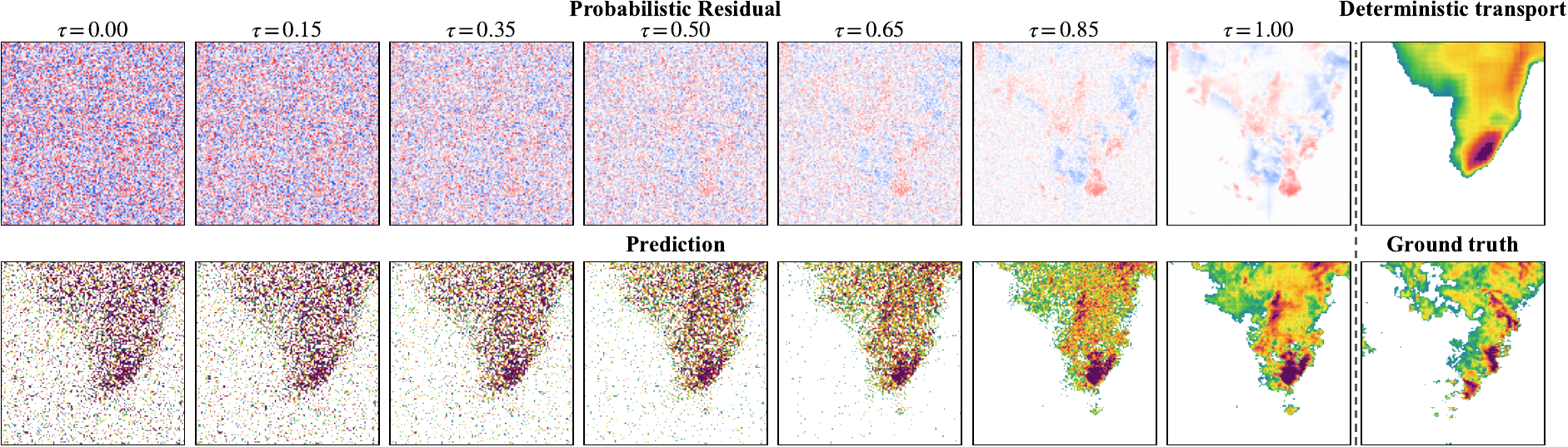}
    \caption{Lead time T+6 h.}
    \label{fig:supp-fm-us-150035-t6h}
  \end{subfigure}
  \captionsetup{font=small}
  \caption{\textbf{Flow-matching construction of the probabilistic residual for a US hail case.} The case was reported at 23:07 UTC on 19 October 2024 near 33.87$^\circ$N, 104.59$^\circ$W, with the radar clip spanning 20:07 UTC on 19 October to 05:07 UTC on 20 October over a patch of approximately 31.29--36.45$^\circ$N and 107.17--102.01$^\circ$W. For each lead time, the top row shows the probabilistic residual evolving along the flow-matching trajectory from a Gaussian sample at $\tau=0.00$ to its terminal state at $\tau=1.00$, and the bottom row shows the prediction formed by adding the evolving probabilistic residual to the deterministic prediction. The right-hand column gives the deterministic prediction (top) and the ground-truth observation (bottom).}
  \label{fig:supp-fm-us-150035}
\end{figure}

\begin{figure}[p]
  \centering
  \begin{subfigure}{\linewidth}
    \includegraphics[width=\linewidth]{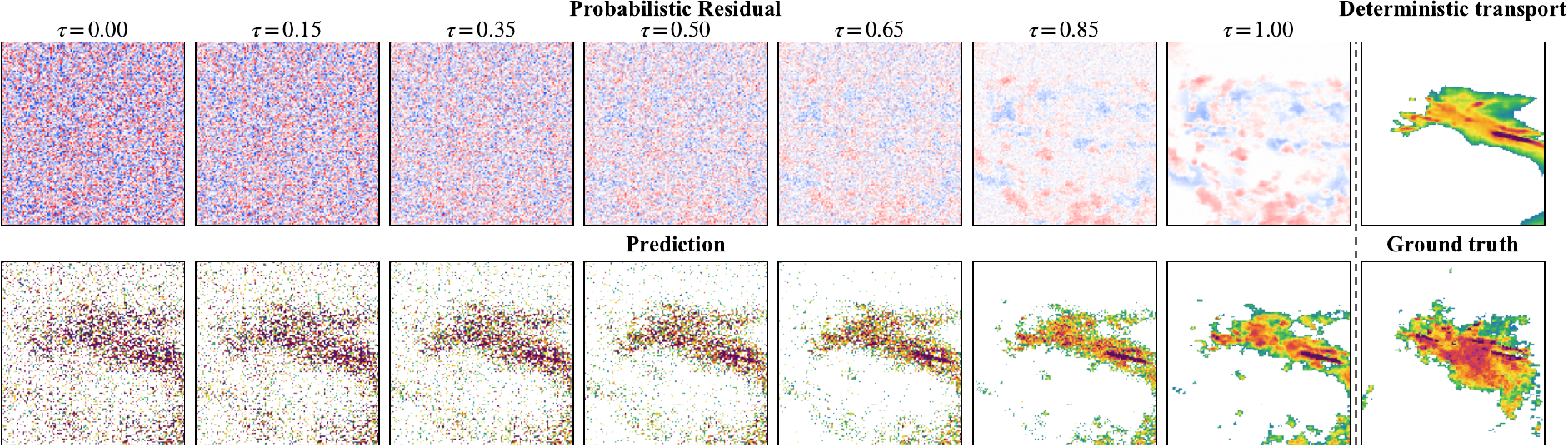}
    \caption{Lead time T+4 h.}
    \label{fig:supp-fm-eu-21477-t4h}
  \end{subfigure}

  \vspace{0.5em}
  \begin{subfigure}{\linewidth}
    \includegraphics[width=\linewidth]{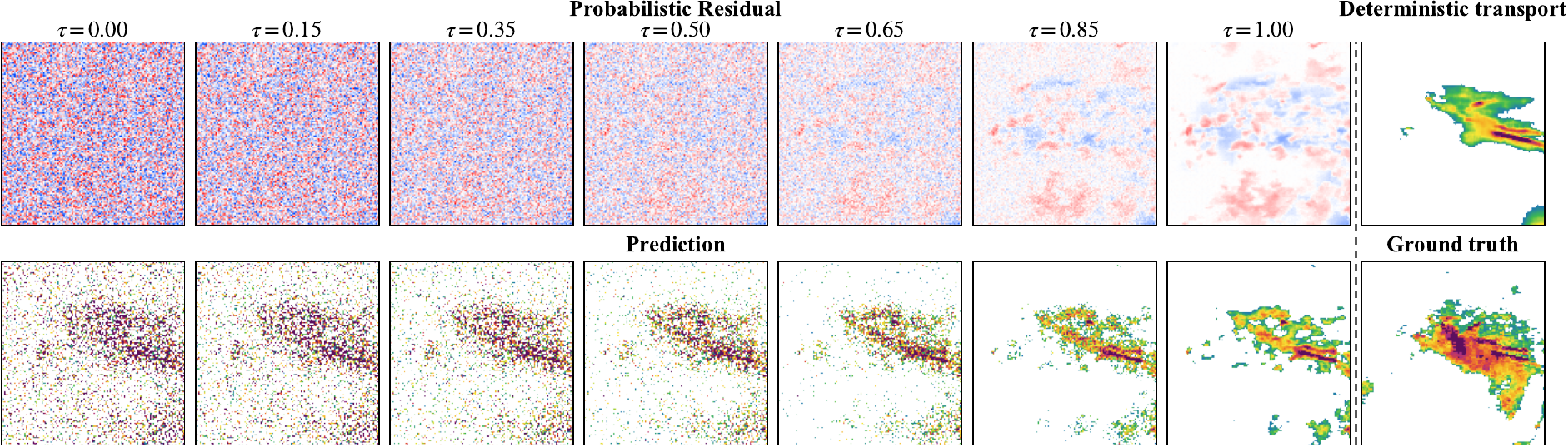}
    \caption{Lead time T+5 h.}
    \label{fig:supp-fm-eu-21477-t5h}
  \end{subfigure}

  \vspace{0.5em}
  \begin{subfigure}{\linewidth}
    \includegraphics[width=\linewidth]{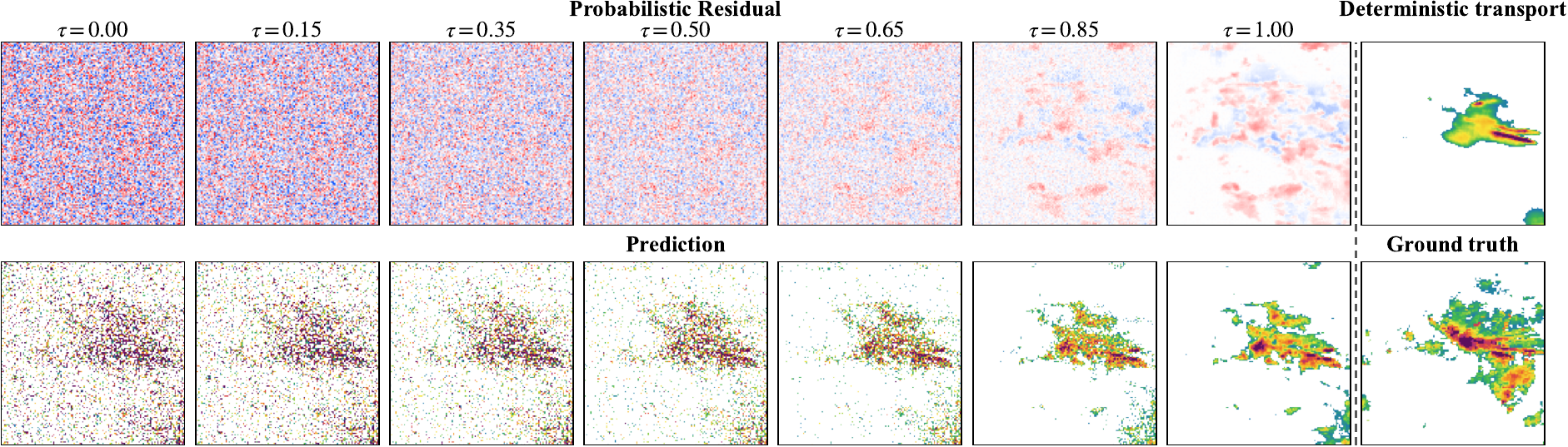}
    \caption{Lead time T+6 h.}
    \label{fig:supp-fm-eu-21477-t6h}
  \end{subfigure}
  \captionsetup{font=small}
  \caption{\textbf{Flow-matching construction of the probabilistic residual for a European thunderstorm case.} The radar clip spans 07:50--15:50 UTC on 15 May 2024 over the Alpine region, with the patch centred near 46.04$^\circ$N, 8.67$^\circ$E and covering approximately 43.38--48.50$^\circ$N and 6.31--11.43$^\circ$E. Layout follows Supplementary Fig.~\ref{fig:supp-fm-us-150035}: the top row of each panel traces the probabilistic residual along the flow-matching trajectory and the bottom row shows the resulting prediction, with the deterministic prediction and ground-truth observation in the right-hand column.}
  \label{fig:supp-fm-eu-21477}
\end{figure}

For a fixed forecast lead time, the residual flow evolves from a Gaussian sample at $\tau=0$ to a probabilistic residual at $\tau=1$, using the same Heun ODE solver as in deployment. Adding this probabilistic residual to the deterministic prediction yields the final prediction for that ensemble member. Supplementary Figs.~\ref{fig:supp-fm-us-150035} and~\ref{fig:supp-fm-eu-21477} show this construction for a US hail case and a European thunderstorm case, each at lead times T+4 h, T+5 h and T+6 h.

In every panel, the top row traces the probabilistic residual as it evolves from noise at $\tau=0.00$ to a structured field at $\tau=1.00$, and the bottom row shows the corresponding prediction obtained by adding the evolving probabilistic residual to the deterministic prediction. The right-hand column reports the deterministic prediction and the ground-truth observation for reference. The trajectories show that the probabilistic residual is not unstructured noise, but a spatially organised correction that modifies precipitation cores and fine-scale structure around the deterministic forecast.

\clearpage
\clearpage

\clearpage

\subsection{Ensemble divergence along the flow}\label{sec:supp-ensemble}

The single-member trajectories above show how one realisation of the probabilistic residual is constructed, but they do not show how ensemble members separate from one another. Because all members share the same deterministic prediction and differ only in the Gaussian sample drawn at $\tau=0$, member-to-member spread is a direct visualisation of the conditional uncertainty represented by the residual decoder.

\begin{figure}[p]
  \centering
  \begin{subfigure}{\linewidth}
    \centering
    \includegraphics[width=0.9\linewidth]{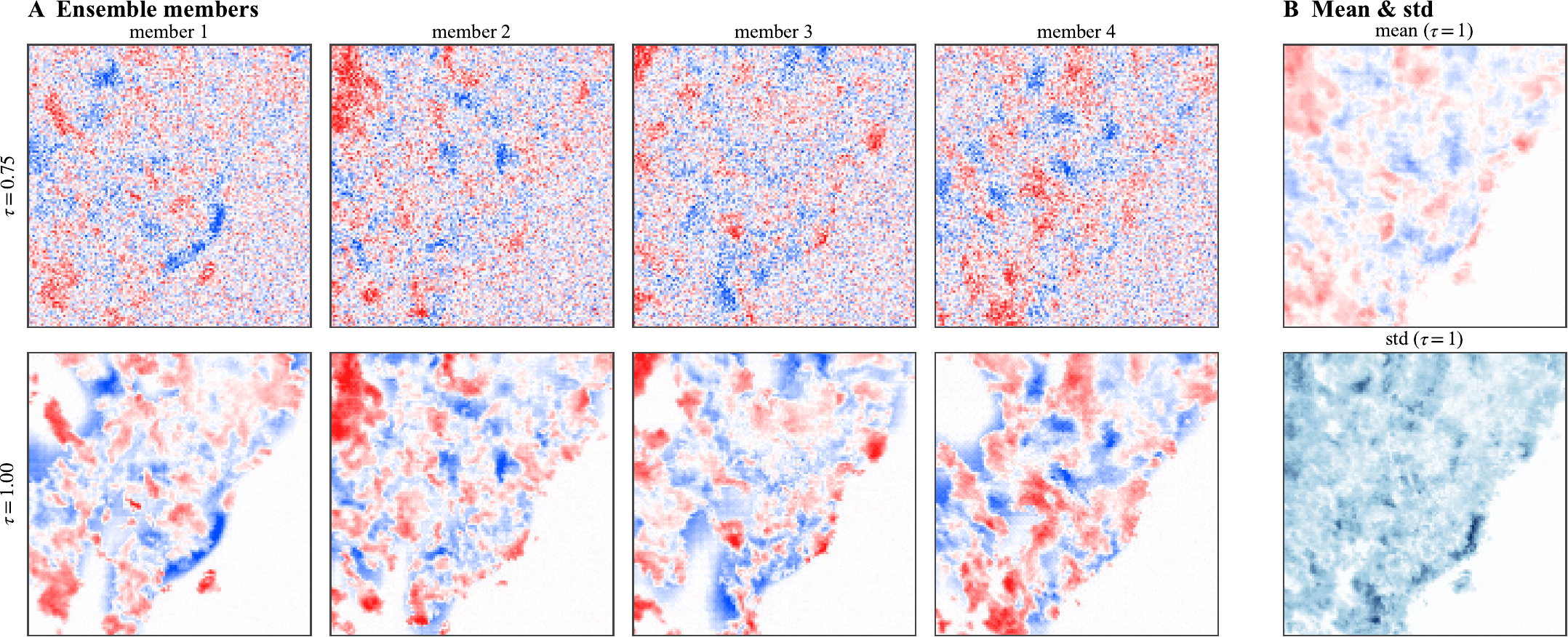}
    \caption{Lead time T+4 h.}
    \label{fig:supp-fm-ens-us-150035-t4h}
  \end{subfigure}

  \vspace{0.5em}
  \begin{subfigure}{\linewidth}
    \centering
    \includegraphics[width=0.9\linewidth]{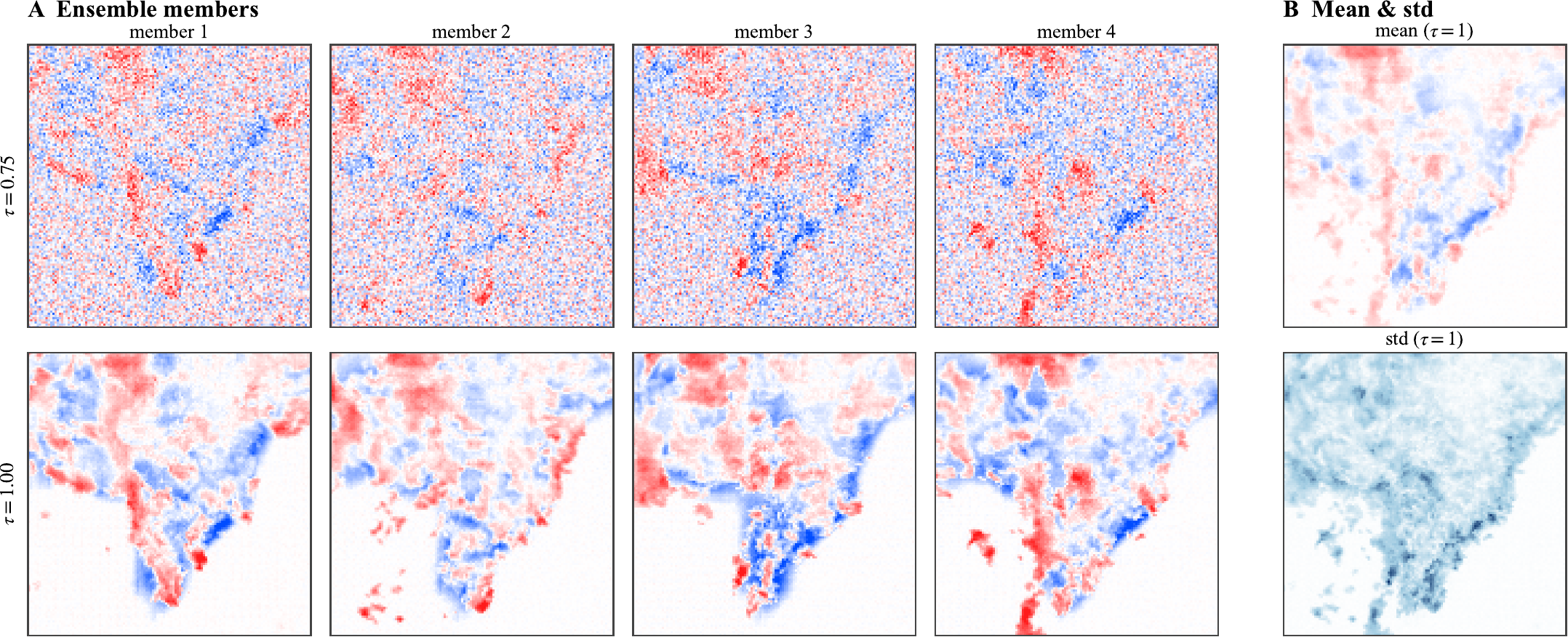}
    \caption{Lead time T+5 h.}
    \label{fig:supp-fm-ens-us-150035-t5h}
  \end{subfigure}

  \vspace{0.5em}
  \begin{subfigure}{\linewidth}
    \centering
    \includegraphics[width=0.9\linewidth]{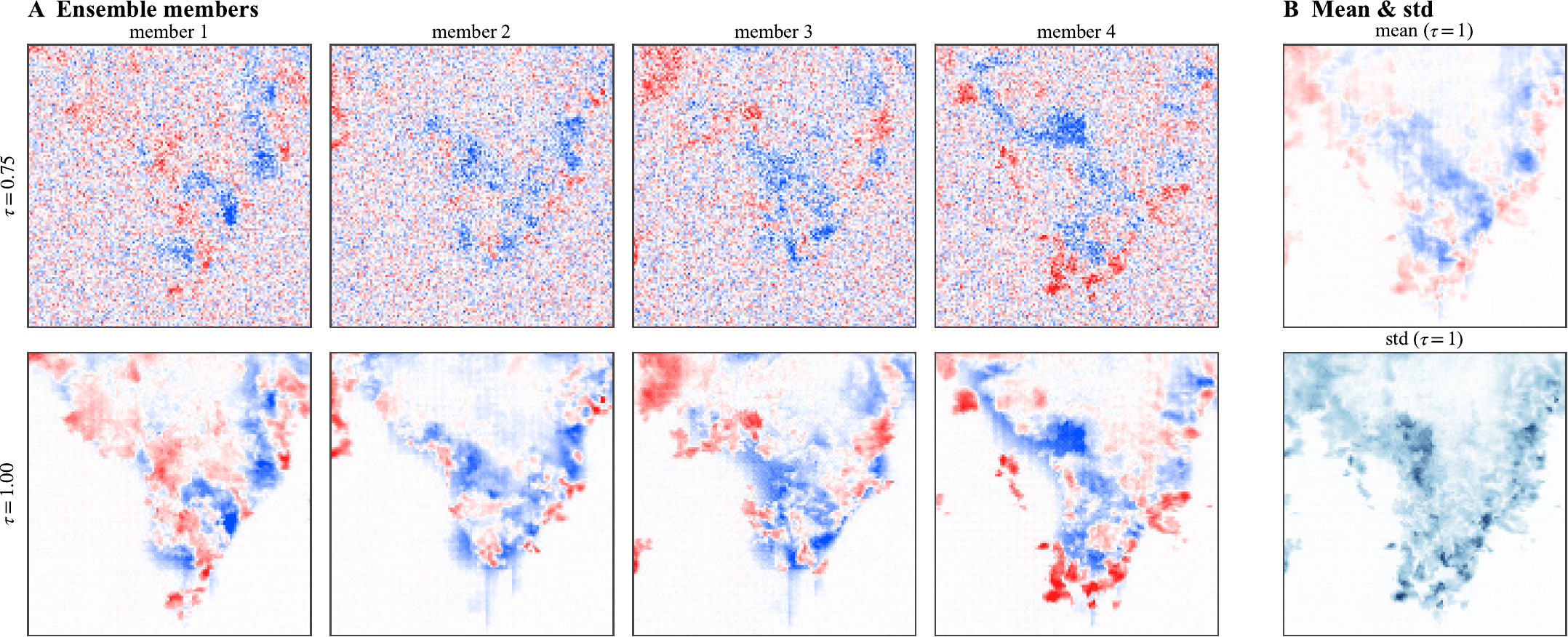}
    \caption{Lead time T+6 h.}
    \label{fig:supp-fm-ens-us-150035-t6h}
  \end{subfigure}
  \captionsetup{font=small}
  \caption{\textbf{Ensemble divergence of the probabilistic residual for a US hail case.} The case was reported at 23:07 UTC on 19 October 2024 near 33.87$^\circ$N, 104.59$^\circ$W. Four members (seeds 1--4) are integrated along the flow from independent Gaussian samples. \textbf{Section A}: the probabilistic residual for each member at flow positions $\tau=0.75$ (top row) and $\tau=1.00$ (bottom row). \textbf{Section B}: the per-pixel ensemble mean (top) and standard deviation (bottom) of the probabilistic residual at $\tau=1.00$. The standard deviation localises along precipitation cores and edges, showing where the probabilistic residual contributes the most uncertainty.}
  \label{fig:supp-fm-ens-us-150035}
\end{figure}

\begin{figure}[p]
  \centering
  \begin{subfigure}{\linewidth}
    \centering
    \includegraphics[width=0.9\linewidth]{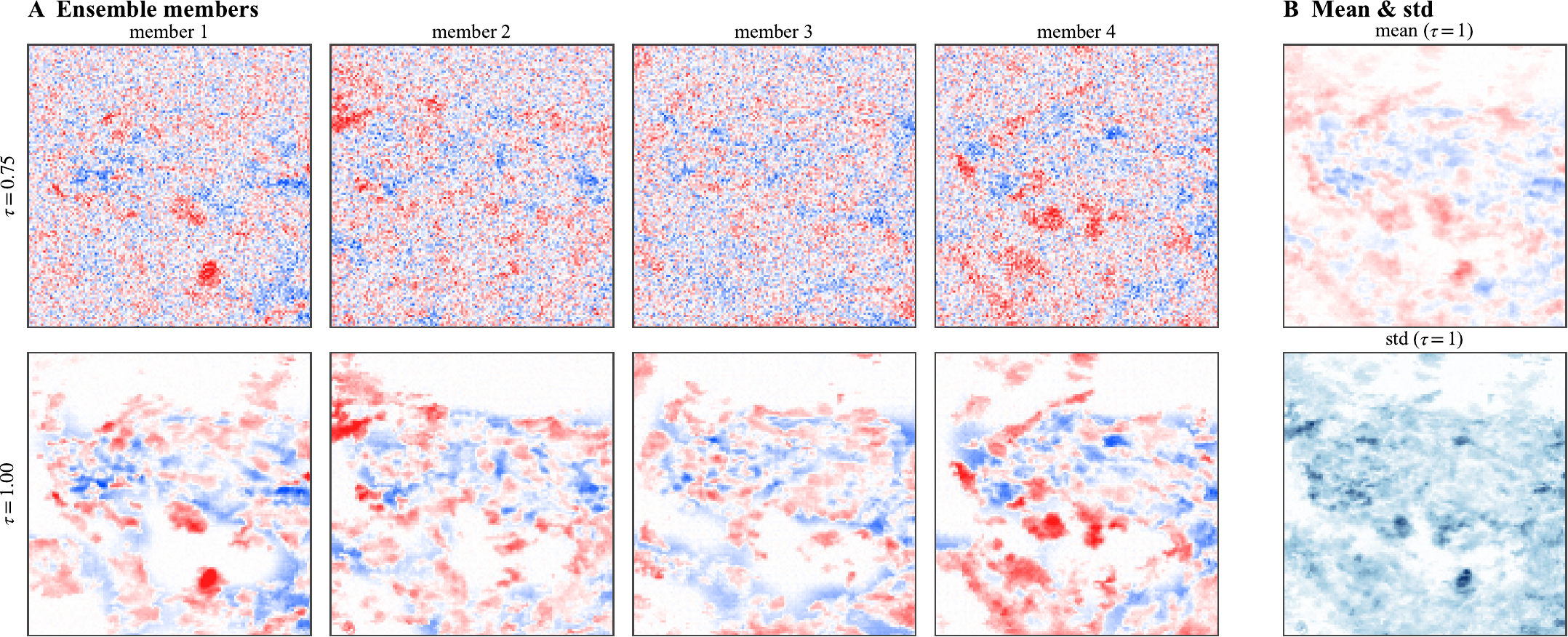}
    \caption{Lead time T+4 h.}
    \label{fig:supp-fm-ens-eu-21477-t4h}
  \end{subfigure}

  \vspace{0.5em}
  \begin{subfigure}{\linewidth}
    \centering
    \includegraphics[width=0.9\linewidth]{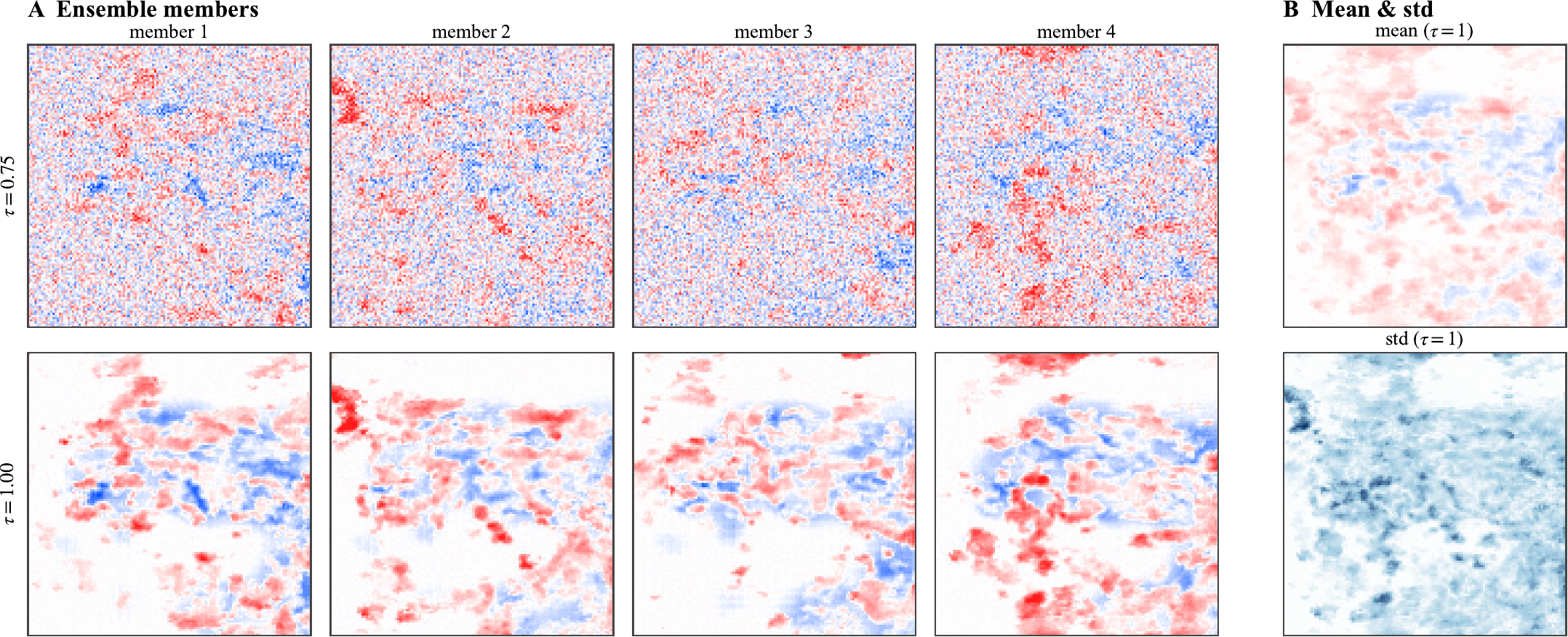}
    \caption{Lead time T+5 h.}
    \label{fig:supp-fm-ens-eu-21477-t5h}
  \end{subfigure}

  \vspace{0.5em}
  \begin{subfigure}{\linewidth}
    \centering
    \includegraphics[width=0.9\linewidth]{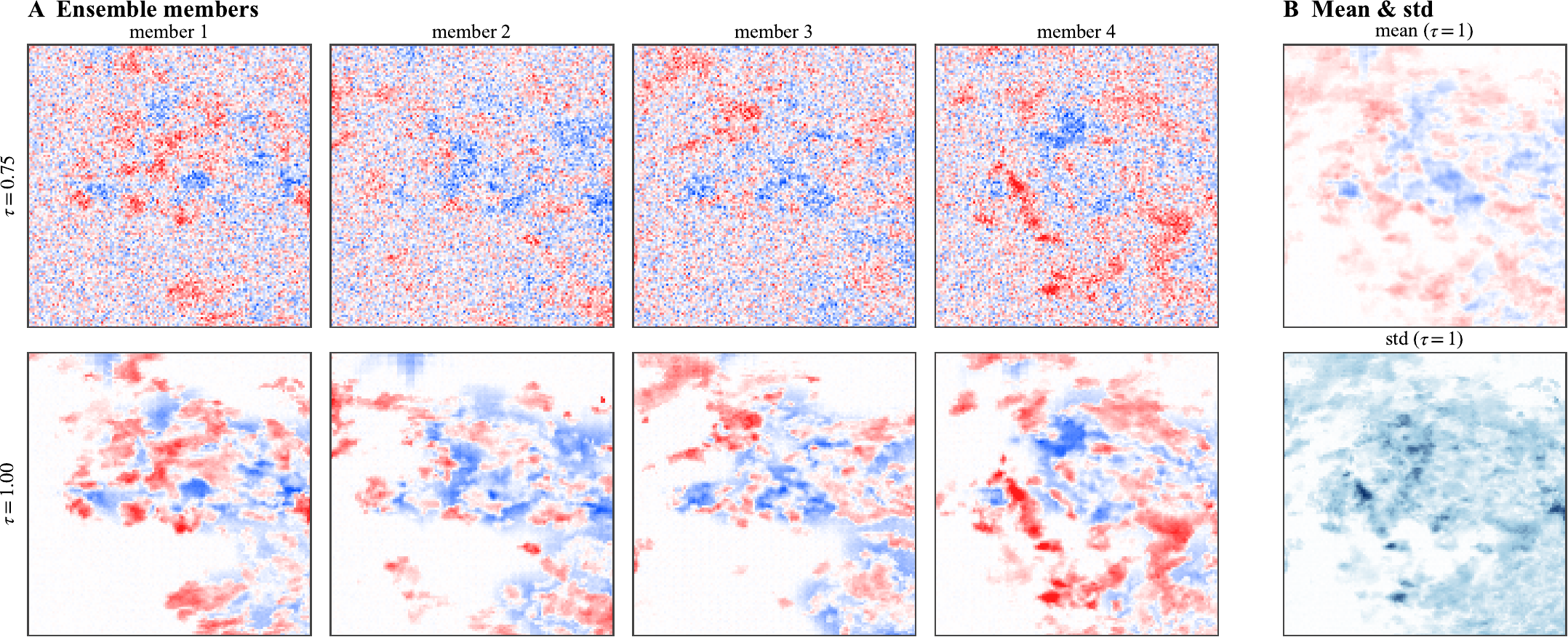}
    \caption{Lead time T+6 h.}
    \label{fig:supp-fm-ens-eu-21477-t6h}
  \end{subfigure}
  \captionsetup{font=small}
  \caption{\textbf{Ensemble divergence of the probabilistic residual for a European thunderstorm case.} The case occurred on 15 May 2024 over the Alpine region, with the patch centred near 46.04$^\circ$N, 8.67$^\circ$E. Layout follows Supplementary Fig.~\ref{fig:supp-fm-ens-us-150035}: Section A shows four members at $\tau=0.75$ and $\tau=1.00$, and Section B shows the ensemble mean and standard deviation of the probabilistic residual at $\tau=1.00$.}
  \label{fig:supp-fm-ens-eu-21477}
\end{figure}

Supplementary Figs.~\ref{fig:supp-fm-ens-us-150035} and~\ref{fig:supp-fm-ens-eu-21477} show four independently sampled members for the same two cases and lead times. Section A shows residual fields at $\tau=0.75$ and $\tau=1.00$, when the residual has developed structured precipitation-related features. Section B summarises the converged residual ensemble using the per-pixel ensemble mean and standard deviation at $\tau=1.00$. The standard deviation concentrates near precipitation cores and edges rather than spreading uniformly across the domain, indicating that the residual decoder expresses uncertainty around meteorologically active structures.

\clearpage
\clearpage

\clearpage

\section{Additional quantitative results}

\subsection{Supplementary verification metrics}
\label{sec:supp-verification-metrics}

The main text defines the core metrics used in its figures and quantitative results, including CSIN, POD, CRPS, CRPSS, REV and PSD. Here we give additional details for the supplementary diagnostic metrics reported below.

\textit{False Alarm Ratio (FAR).}
FAR measures the fraction of forecast threshold exceedances that are not observed. Using the pooled threshold-contingency counts, it is defined as
\begin{equation}
  \mathrm{FAR}(\rho)=\frac{\mathrm{FP}}{\mathrm{TP}+\mathrm{FP}},
\end{equation}
where lower values are better.

\textit{Neighbourhood Critical Success Index (CSIN).}
CSIN allows a spatial tolerance around threshold exceedances. Let $O_{\rho}(p)=\mathbf{1}\{y(p)\geq\rho\}$ and $F_{\rho}(p)=\mathbf{1}\{x(p)\geq\rho\}$ be the observed and forecast binary fields at grid cell $p$, and let $\mathcal{N}_{n}(p)$ denote the centred $n\times n$ neighbourhood of $p$. We dilate both binary fields using
\begin{equation}
  O^{\max}_{\rho,n}(p)=\max_{q\in\mathcal{N}_{n}(p)}O_{\rho}(q),
  \qquad
  F^{\max}_{\rho,n}(p)=\max_{q\in\mathcal{N}_{n}(p)}F_{\rho}(q).
\end{equation}
After pooling the contingency counts $\mathrm{TP}_{n}$, $\mathrm{FP}_{n}$ and $\mathrm{FN}_{n}$ from these dilated fields over the test set, we compute
\begin{equation}
  \mathrm{CSIN}_{n}(\rho)
  =\frac{\mathrm{TP}_{n}}
  {\mathrm{TP}_{n}+\mathrm{FP}_{n}+\mathrm{FN}_{n}}.
\end{equation}
Values outside the domain are treated as non-events. In the tables and figures below, the subscript denotes the centred neighbourhood width in grid cells.

\textit{Fractions Skill Score (FSS).}
FSS \cite{roberts2008fss} compares neighbourhood event fractions. For the same binary fields, define
\begin{equation}
  o_{\rho,n}(p)=\frac{1}{n^2}
  \sum_{q\in\mathcal{N}_{n}(p)}O_{\rho}(q),
  \qquad
  f_{\rho,n}(p)=\frac{1}{n^2}
  \sum_{q\in\mathcal{N}_{n}(p)}F_{\rho}(q).
\end{equation}
The score for one forecast field is
\begin{equation}
  \mathrm{FSS}_{n}(\rho)
  =1-\frac{\sum_p\left[f_{\rho,n}(p)-o_{\rho,n}(p)\right]^2}
  {\sum_p f_{\rho,n}(p)^2+\sum_p o_{\rho,n}(p)^2}.
\end{equation}
Higher values are better and a perfect forecast has value 1. We compute FSS per sample and lead time and then average the valid scores for each reported threshold and neighbourhood window.

\subsection{Regional quantitative diagnostics}\label{sec:supp-regional}

To complement the lead-time curves in the main text, Supplementary Table~\ref{tab:csin5-regional} reports neighbourhood Critical Success Index with a $5\times5$ neighbourhood (CSIN$_5$) at the three high rain-rate thresholds (16, 32 and 64 mm h$^{-1}$). The table gives both representative lead-time values at T+2 h, T+4 h and T+6 h and the mean over all 36 lead times from T+10 min to T+6 h. These values are computed from the same regional evaluation data as the main-text high-threshold comparison and provide a compact numerical reading of the regional CSIN$_5$ curves. Across all regions and thresholds, MW-Nowcast retains the highest CSIN$_5$, with the advantage remaining clear at later lead times and heavier precipitation thresholds.

\begin{figure}[p]
  \centering
  \includegraphics[width=\linewidth]{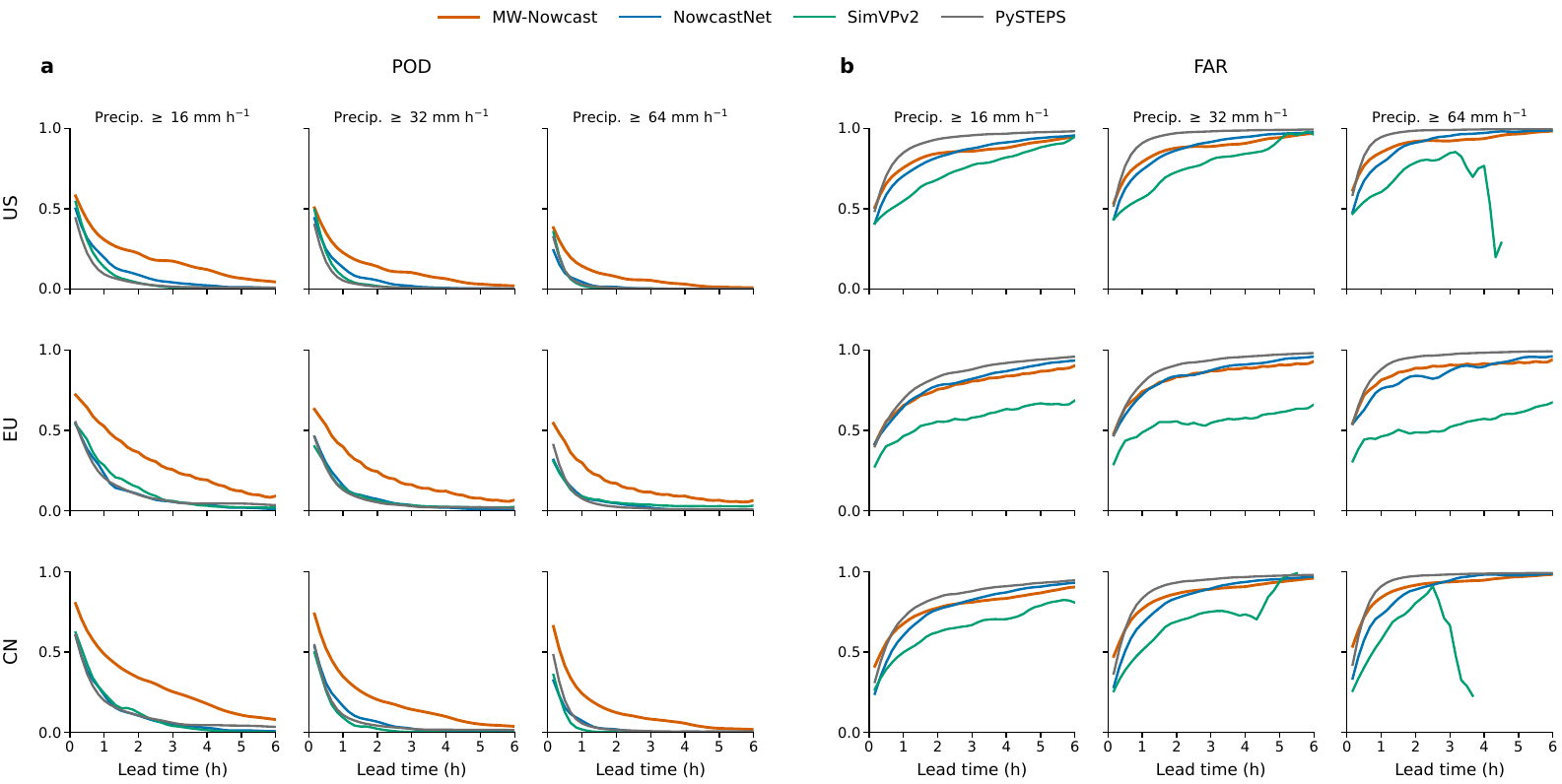}
  \captionsetup{font=small}
  \caption{\textbf{Pointwise detection and false-alarm characteristics across regions.} Rows show the US, EU and CN test sets, and columns show rain-rate thresholds of 16, 32 and 64~mm~h$^{-1}$. \textbf{a}, Probability of detection (POD). \textbf{b}, False alarm ratio (FAR). Both metrics are computed from pointwise threshold exceedances over lead times from T+10~min to T+6~h. Curves compare MW-Nowcast with NowcastNet, SimVPv2 and PySTEPS. Higher values are better for POD, whereas lower values are better for FAR.}
  \label{fig:supp-pod-far}
\end{figure}

\begin{figure}[p]
  \centering
  \includegraphics[width=\linewidth]{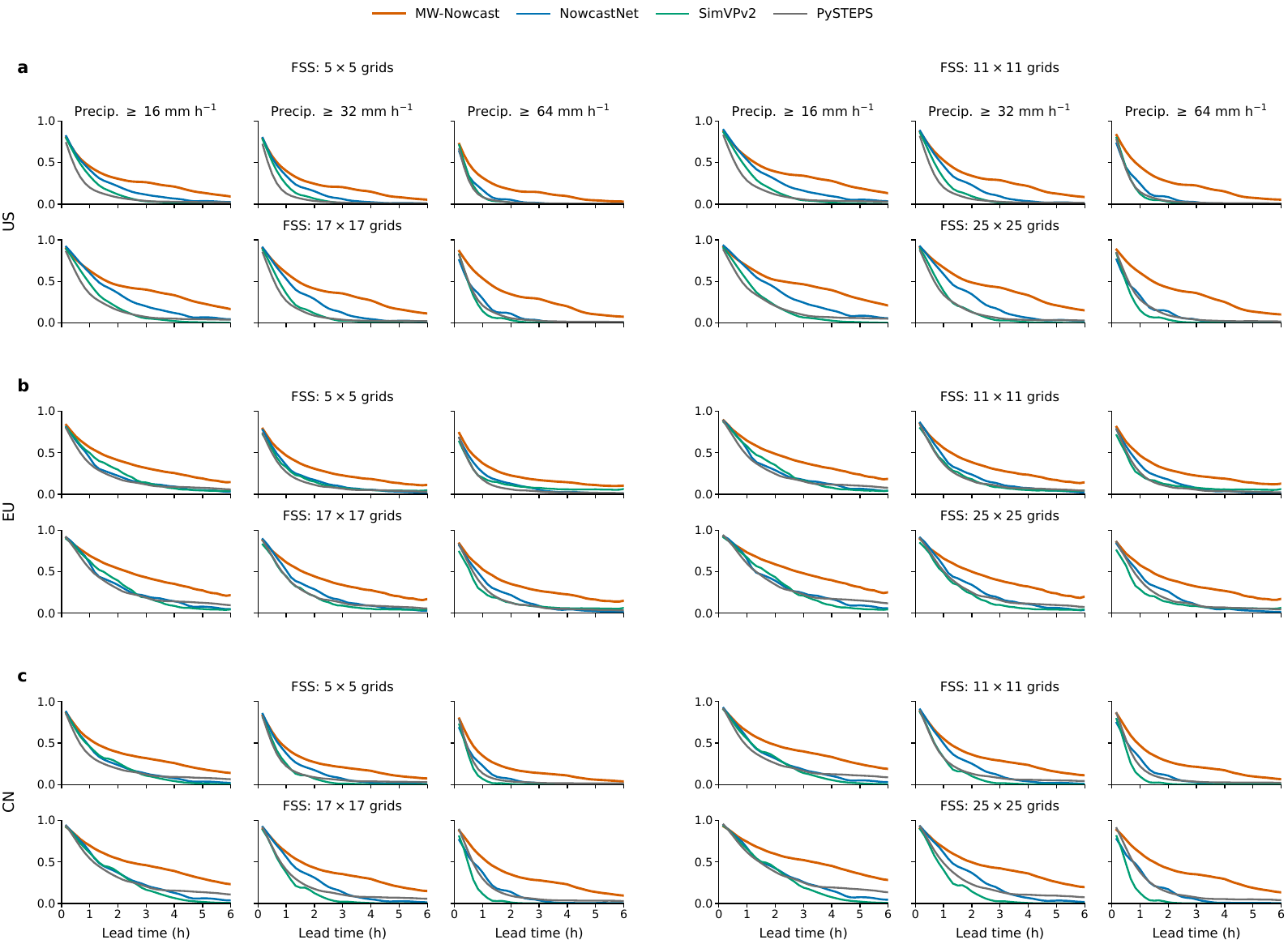}
  \captionsetup{font=small}
  \caption{\textbf{FSS sensitivity to neighbourhood size across regions.} \textbf{a}, US; \textbf{b}, EU; \textbf{c}, CN. Within each regional block, the upper row reports FSS for $5\times5$ and $11\times11$ grid-cell neighbourhoods, and the lower row reports FSS for $17\times17$ and $25\times25$ neighbourhoods. For each neighbourhood size, columns show rain-rate thresholds of 16, 32 and 64~mm~h$^{-1}$. Curves compare MW-Nowcast with NowcastNet, SimVPv2 and PySTEPS over lead times from T+10~min to T+6~h. Higher values indicate better neighbourhood-scale forecast skill.}
  \label{fig:supp-fss-windows}
\end{figure}

\begin{figure}[p]
  \centering
  \includegraphics[width=\linewidth]{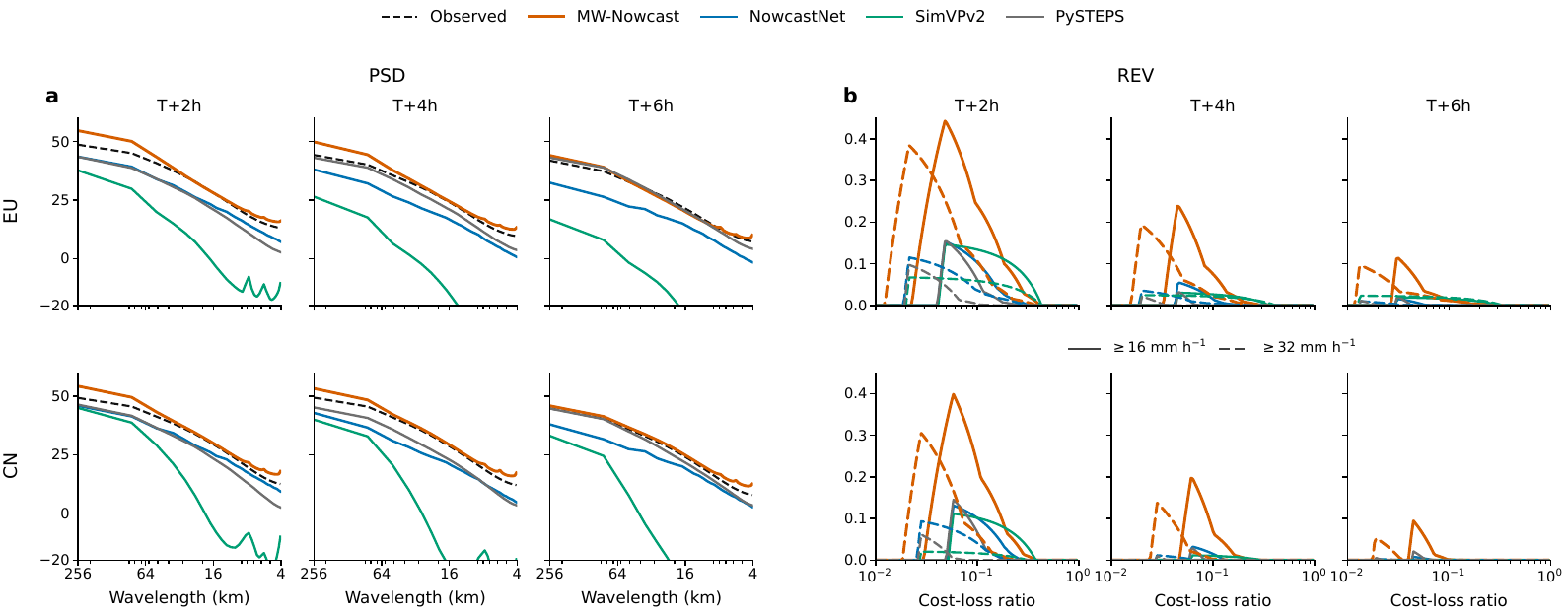}
  \captionsetup{font=small}
  \caption{\textbf{Structural and relative economic value diagnostics for the EU and CN regions.} The upper row shows the EU test set and the lower row shows the CN test set. \textbf{a}, Radially averaged power spectral density (PSD) at T+2 h, T+4 h and T+6 h. The observed spectrum is shown together with forecast spectra from MW-Nowcast, NowcastNet, SimVPv2 and PySTEPS; closer agreement with the observed spectrum indicates better preservation of multiscale precipitation structure. \textbf{b}, Standard pixel-wise relative economic value (REV) at the same lead times as a function of cost--loss ratio. Solid and dashed curves denote rain-rate thresholds of 16 and 32~mm~h$^{-1}$, respectively. Positive REV indicates lower expected expense than the optimal climatological strategy, and higher values indicate greater economic value under the corresponding cost--loss setting.}
  \label{fig:supp-psd-rev-eu-cn}
\end{figure}

Supplementary Figs.~\ref{fig:supp-pod-far}--\ref{fig:supp-psd-rev-eu-cn} provide diagnostic views that complement the main-text CSIN$_5$ and CRPS results. Supplementary Fig.~\ref{fig:supp-pod-far} shows that MW-Nowcast maintains substantially higher POD across regions and thresholds, while its FAR is generally lower than that of PySTEPS and comparable to that of NowcastNet. Although SimVPv2 often produces a lower FAR, this is accompanied by markedly lower POD, indicating more frequent missed events.

Supplementary Fig.~\ref{fig:supp-fss-windows} shows that the advantage of MW-Nowcast persists as the FSS neighbourhood broadens from $5\times5$ to $25\times25$ grid cells. From T+1 h onwards, it achieves the highest FSS for every evaluated region--threshold--neighbourhood combination, indicating that the improvement is not specific to the neighbourhood used in the main comparison.

Supplementary Fig.~\ref{fig:supp-psd-rev-eu-cn} extends the PSD and REV diagnostics to EU and CN. MW-Nowcast remains closer to the observed spectra at T+2 h, T+4 h and T+6 h, while retaining positive economic value over a broader range of cost--loss ratios than the baselines, particularly at the longer lead times.

\clearpage
\clearpage

\clearpage

\clearpage

\begin{table}[p]
  \centering
  \captionsetup{font=small}
  \caption{\textbf{Regional CSIN$_5$ comparison of baseline models.} Neighbourhood Critical Success Index with a $5\times5$ neighbourhood is reported at rain-rate thresholds of 16, 32 and 64~mm~h$^{-1}$ for representative lead times T+2 h, T+4 h and T+6 h, together with the mean over all 36 lead times from T+10 min to T+6 h. Values are computed from the same regional evaluation data as the main-text quantitative comparison. Higher is better; the best value within each region and column is shown in bold.}
  \label{tab:csin5-regional}
  \scriptsize
  \setlength{\tabcolsep}{2.5pt}
  \renewcommand{\arraystretch}{1.08}
  \begin{tabular}{llcccccccccccc}
    \toprule
    Region & Model & \multicolumn{3}{c}{T+2 h} & \multicolumn{3}{c}{T+4 h} & \multicolumn{3}{c}{T+6 h} & \multicolumn{3}{c}{Mean} \\
    \cmidrule(lr){3-5} \cmidrule(lr){6-8} \cmidrule(lr){9-11} \cmidrule(lr){12-14}
    & & 16 & 32 & 64 & 16 & 32 & 64 & 16 & 32 & 64 & 16 & 32 & 64 \\
    \midrule
    \multirow{4}{*}{US}
      & NowcastNet & 0.192 & 0.144 & 0.070 & 0.075 & 0.037 & 0.004 & 0.038 & 0.016 & 0.002 & 0.171 & 0.129 & 0.064 \\
      & SimVPv2 & 0.043 & 0.023 & 0.006 & 0.005 & 0.001 & 0.000 & 0.000 & 0.000 & 0.000 & 0.065 & 0.047 & 0.025 \\
      & PySTEPS & 0.089 & 0.043 & 0.022 & 0.037 & 0.017 & 0.009 & 0.026 & 0.014 & 0.007 & 0.099 & 0.064 & 0.045 \\
      & \textbf{MW-Nowcast} & \textbf{0.284} & \textbf{0.214} & \textbf{0.160} & \textbf{0.197} & \textbf{0.129} & \textbf{0.085} & \textbf{0.095} & \textbf{0.055} & \textbf{0.032} & \textbf{0.256} & \textbf{0.194} & \textbf{0.147} \\
    \midrule
    \multirow{4}{*}{EU}
      & NowcastNet & 0.171 & 0.123 & 0.087 & 0.081 & 0.051 & 0.028 & 0.036 & 0.022 & 0.011 & 0.158 & 0.118 & 0.086 \\
      & SimVPv2 & 0.147 & 0.083 & 0.070 & 0.046 & 0.039 & 0.043 & 0.034 & 0.038 & 0.045 & 0.122 & 0.086 & 0.075 \\
      & PySTEPS & 0.122 & 0.072 & 0.039 & 0.069 & 0.037 & 0.020 & 0.048 & 0.026 & 0.014 & 0.130 & 0.085 & 0.055 \\
      & \textbf{MW-Nowcast} & \textbf{0.288} & \textbf{0.213} & \textbf{0.150} & \textbf{0.188} & \textbf{0.132} & \textbf{0.095} & \textbf{0.111} & \textbf{0.080} & \textbf{0.063} & \textbf{0.254} & \textbf{0.192} & \textbf{0.144} \\
    \midrule
    \multirow{4}{*}{CN}
      & NowcastNet & 0.184 & 0.128 & 0.067 & 0.074 & 0.034 & 0.004 & 0.034 & 0.012 & 0.002 & 0.166 & 0.118 & 0.067 \\
      & SimVPv2 & 0.109 & 0.022 & 0.002 & 0.014 & 0.001 & 0.000 & 0.001 & 0.000 & 0.000 & 0.095 & 0.042 & 0.020 \\
      & PySTEPS & 0.134 & 0.067 & 0.028 & 0.071 & 0.034 & 0.014 & 0.052 & 0.027 & 0.013 & 0.140 & 0.086 & 0.053 \\
      & \textbf{MW-Nowcast} & \textbf{0.293} & \textbf{0.197} & \textbf{0.129} & \textbf{0.195} & \textbf{0.123} & \textbf{0.077} & \textbf{0.112} & \textbf{0.063} & \textbf{0.033} & \textbf{0.264} & \textbf{0.184} & \textbf{0.130} \\
    \bottomrule
  \end{tabular}
\end{table}

\section{Additional extended precipitation cases}

The main text and appendix show representative examples from the US, EU and CN regions. Here we add two additional cases for each region and two additional US ablation examples. The regional cases broaden the qualitative coverage across storm types and geographic settings, whereas the ablation cases complement Supplementary Fig.~\ref{fig:supp-ablation-case-us} by showing how the comparison variants behave in additional US strong-precipitation settings.

\subsection{Additional regional forecast examples}

\begin{figure}[p]
  \centering
  \includegraphics[width=\linewidth]{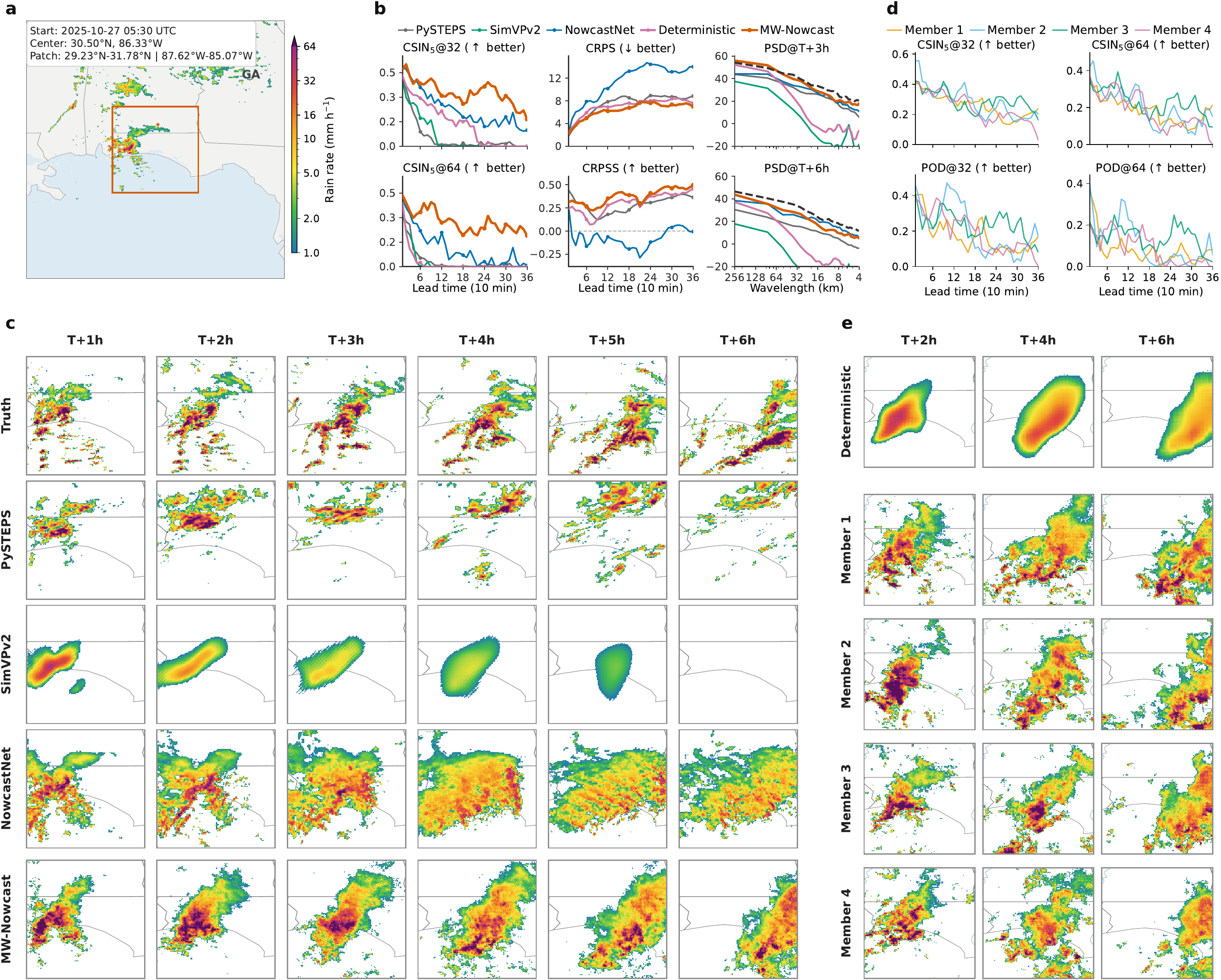}
  \captionsetup{font=small}
  \caption{\textbf{Convective heavy-precipitation forecast over the US Gulf Coast.} The forecast is initialised at 05:30 UTC on 27 October 2025 for a patch centred near 30.50$^\circ$N, 86.33$^\circ$W. The observed sequence shows convective precipitation developing over the Gulf Coast region and generally expanding eastward to southeastward across the forecast patch. Rainfall intensity increases during the first half of the 6 h window, while the strongest precipitation is organised as multiple compact, discrete cores that merge, split and locally reorganise during the evolution.}
  \label{fig:supp-case-us-gulf}
\end{figure}

\begin{figure}[p]
  \centering
  \includegraphics[width=\linewidth]{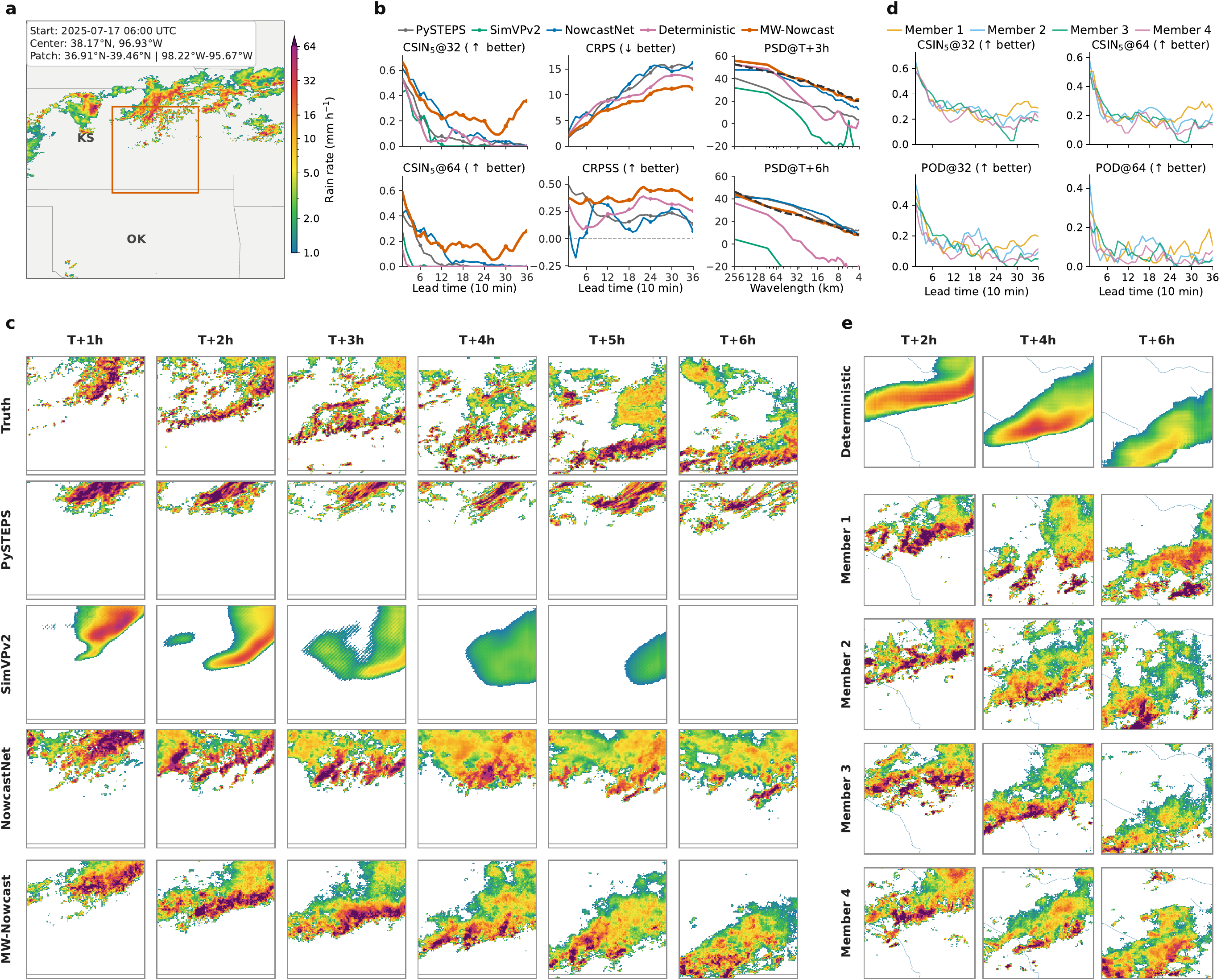}
  \captionsetup{font=small}
  \caption{\textbf{Convective heavy-precipitation forecast over the central US.} The forecast is initialised at 06:00 UTC on 17 July 2025 for a patch centred near 38.17$^\circ$N, 96.93$^\circ$W. The observed sequence shows organised convective precipitation over the central US, with the rain area first advancing eastward to southeastward and then extending mainly towards the southern part of the patch. The high-intensity area broadens through the forecast window, but its embedded cores become increasingly fragmented, producing a multi-core structure rather than a single continuous rain band.}
  \label{fig:supp-case-us-plains}
\end{figure}

\begin{figure}[p]
  \centering
  \includegraphics[width=\linewidth]{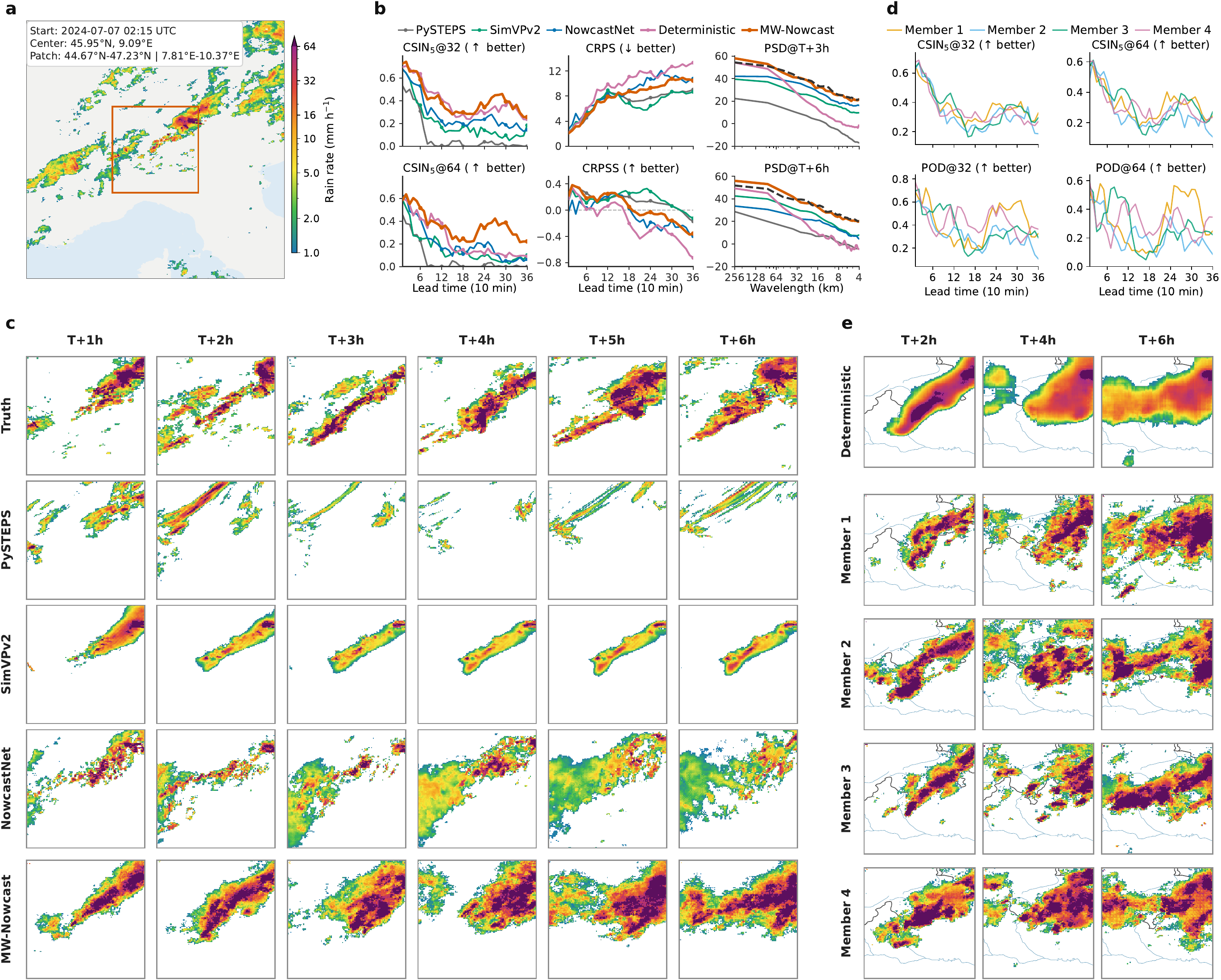}
  \captionsetup{font=small}
  \caption{\textbf{Heavy-precipitation forecast over northern Italy.} The forecast is initialised at 02:15 UTC on 7 July 2024 and is associated with a precipitation report near 45.95$^\circ$N, 9.09$^\circ$E in northern Italy. The observed sequence shows Alpine convective precipitation that remains concentrated near the main rain area rather than translating as a single coherent band. Localised intense echoes are repeatedly renewed and reorganised during the 6 h window, giving the case a compact but intermittently multi-core structure.}
  \label{fig:supp-case-eu-italy}
\end{figure}

\begin{figure}[p]
  \centering
  \includegraphics[width=\linewidth]{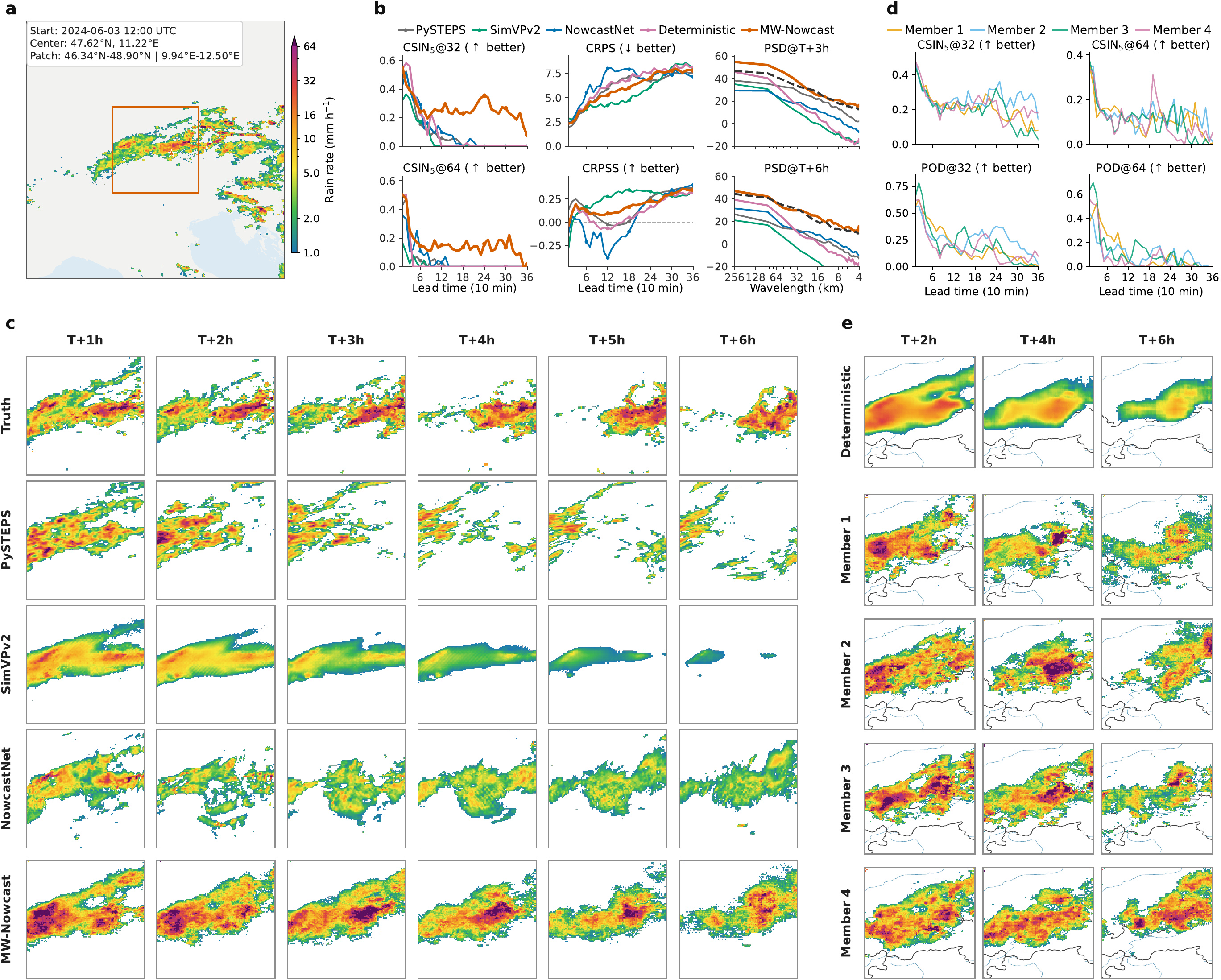}
  \captionsetup{font=small}
  \caption{\textbf{Heavy-precipitation forecast over southern Germany.} The forecast is initialised at 12:00 UTC on 3 June 2024 and is associated with a precipitation report near 47.62$^\circ$N, 11.22$^\circ$E in southern Germany. The observed sequence shows organised heavy precipitation along the Alpine region, with the main precipitation area remaining relatively coherent through the forecast window. Intense echoes persist within this broader system, while local cores are renewed and rearranged along the rain area instead of separating into fully isolated cells.}
  \label{fig:supp-case-eu-precip}
\end{figure}

\begin{figure}[p]
  \centering
  \includegraphics[width=\linewidth]{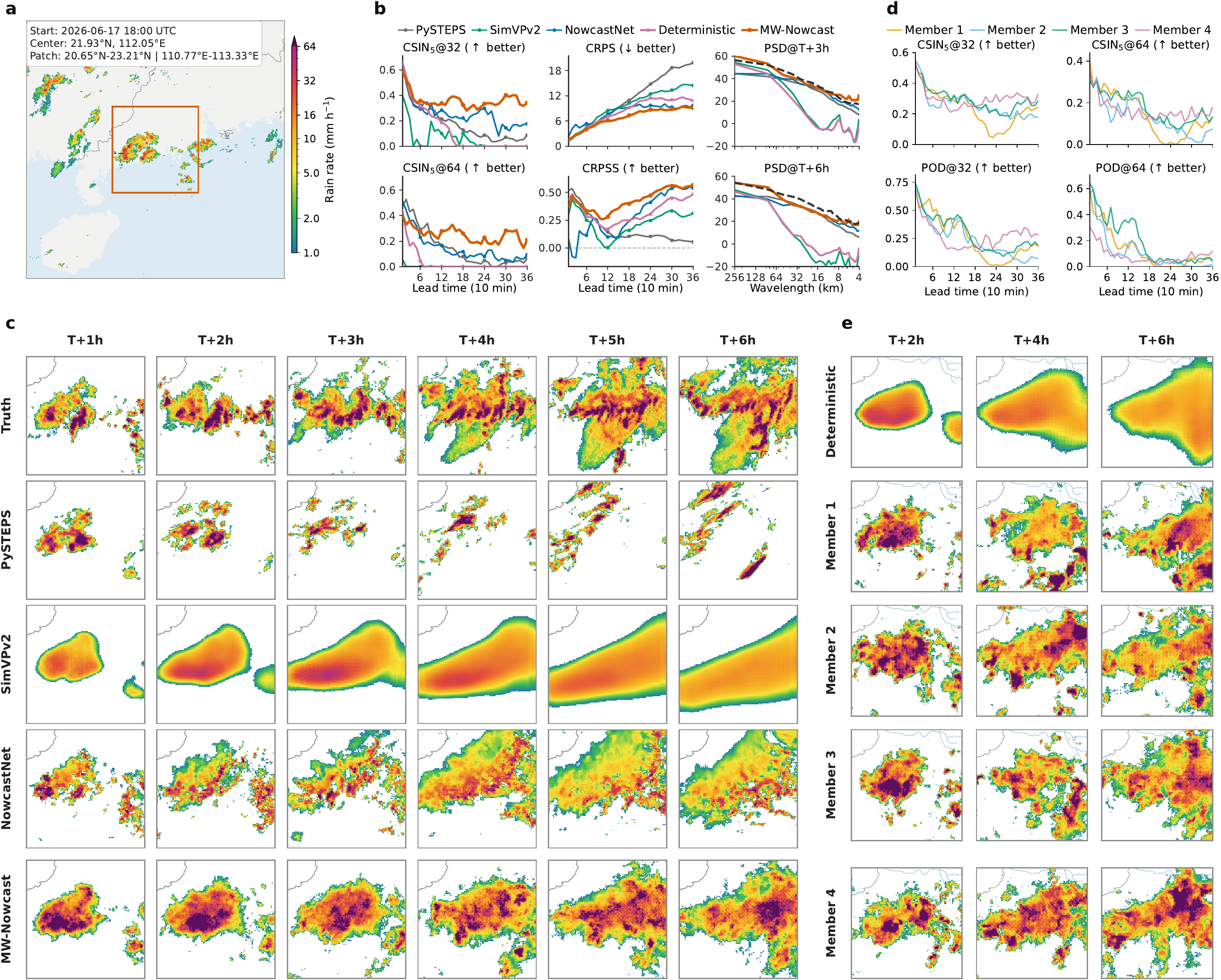}
  \captionsetup{font=small}
  \caption{\textbf{Heavy-precipitation forecast over southern China.} The forecast is initialised at 18:00 UTC on 17 June 2026 for a patch centred near 21.93$^\circ$N, 112.05$^\circ$E. The observed sequence shows developing heavy precipitation over southern China, with the rain area expanding mainly towards the eastern to northeastern part of the patch. The intense precipitation area strengthens through the first half of the forecast window and then forms a broader, semi-continuous core region with smaller embedded maxima that reorganise locally.}
  \label{fig:supp-case-cn-south}
\end{figure}

\begin{figure}[p]
  \centering
  \includegraphics[width=\linewidth]{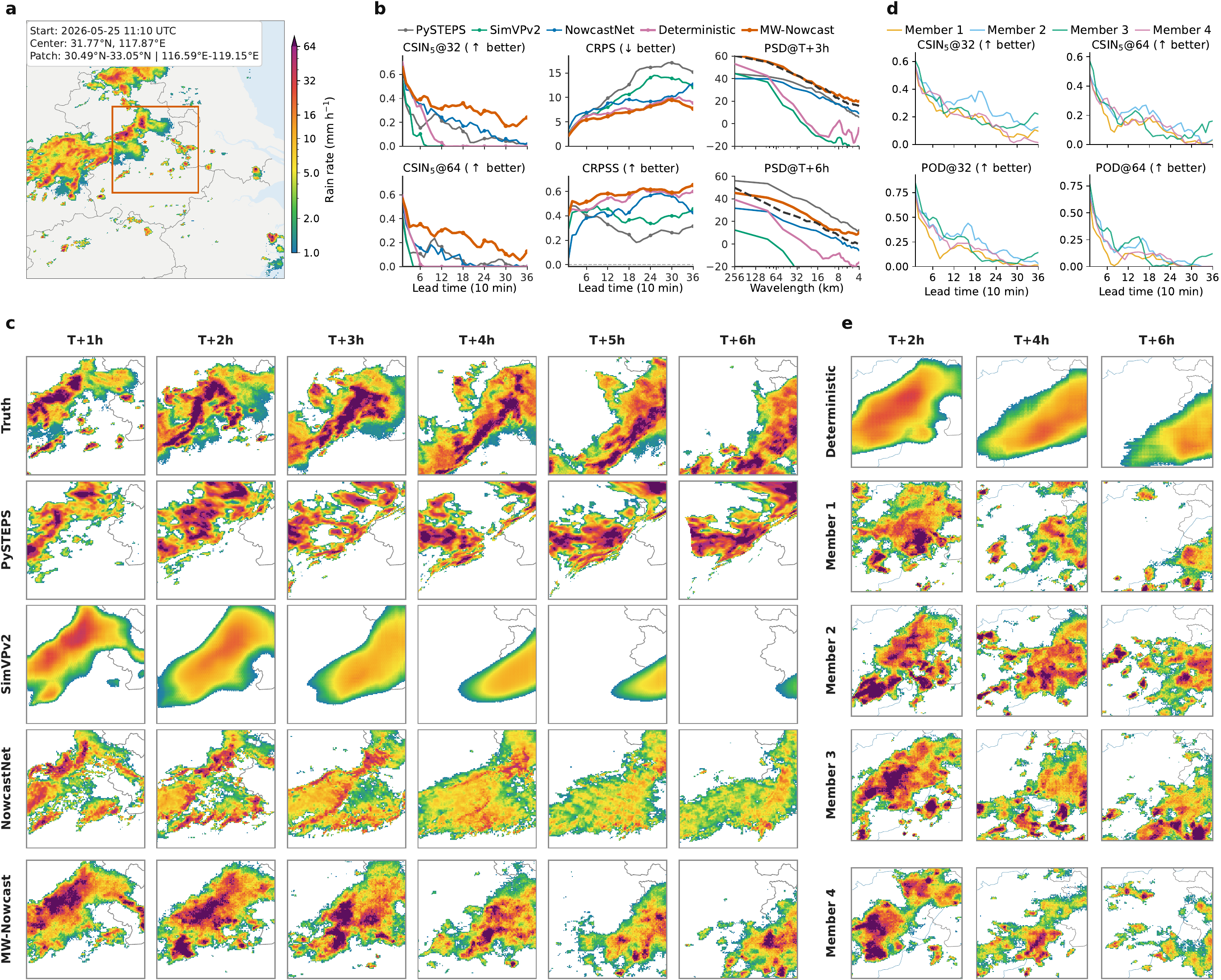}
  \captionsetup{font=small}
  \caption{\textbf{Heavy-precipitation forecast over eastern China.} The forecast is initialised at 11:10 UTC on 25 May 2026 for a patch centred near 31.77$^\circ$N, 117.87$^\circ$E. The observed sequence shows an organised precipitation system over eastern China that progresses eastward to southeastward across the forecast patch. The high-intensity precipitation expands and becomes more connected during the first half of the window, before the strongest region shifts downstream and breaks into several embedded cores at later lead times.}
  \label{fig:supp-case-cn-east}
\end{figure}

Supplementary Figs.~\ref{fig:supp-case-us-gulf}--\ref{fig:supp-case-cn-east} provide additional 6 h forecast examples. Each case follows the same five-panel layout as the case figures in the main text. \textbf{a}, Geographic and radar context, with the $256\,\mathrm{km}\times256\,\mathrm{km}$ forecast domain outlined. \textbf{b}, Lead-time diagnostics comprising CSIN$_5$ at rain-rate thresholds of 32 and 64 mm h$^{-1}$, CRPS, CRPSS and radially averaged PSD at lead times of 3 and 6 h. \textbf{c}, Observed and predicted rain-rate fields from PySTEPS, SimVPv2, NowcastNet and MW-Nowcast at hourly lead times from 1 to 6 h. \textbf{d}, CSIN$_5$ and POD at thresholds of 32 and 64 mm h$^{-1}$ for the four MW-Nowcast ensemble members and their PMM summary. \textbf{e}, The MW-Nowcast deterministic prediction and four ensemble members at lead times of 2, 4 and 6 h.

\clearpage
\clearpage

\clearpage

\clearpage

\clearpage

\clearpage

\clearpage

\subsection{Additional ablation examples}

\begin{figure}[p]
  \centering
  \includegraphics[width=\linewidth]{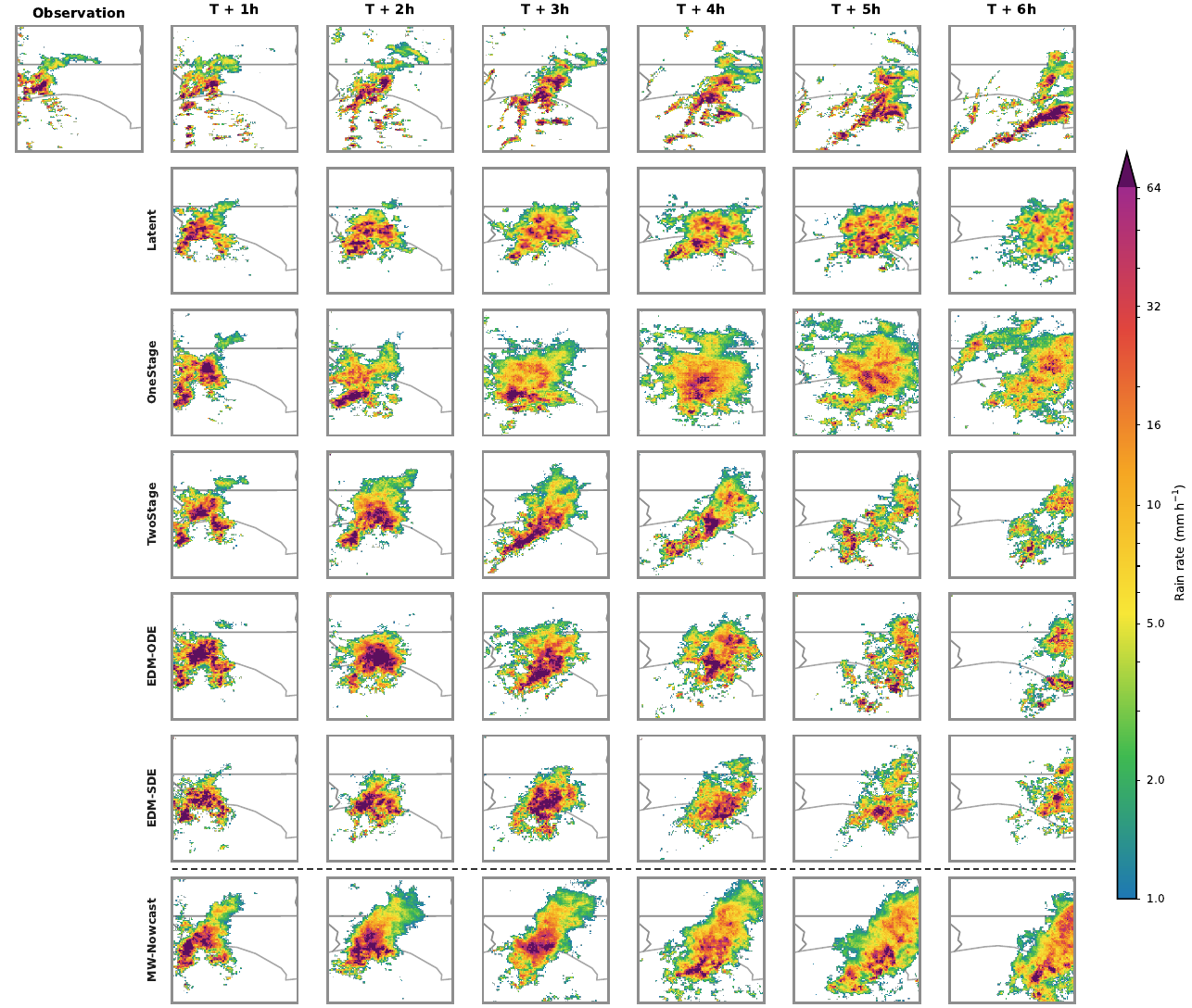}
  \captionsetup{font=small}
  \caption{\textbf{Ablation forecast grid for the US strong-precipitation case initialised at 05:30 UTC on 27 October 2025.} The patch is centred near 30.50$^\circ$N, 86.33$^\circ$W. Rows show the observation and the MW-Nowcast, Latent, OneStage, TwoStage, EDM-ODE and EDM-SDE variants; columns show the input context and forecast lead times through T+6\,h.}
  \label{fig:supp-ablation-us-gulf}
\end{figure}

\begin{figure}[p]
  \centering
  \includegraphics[width=\linewidth]{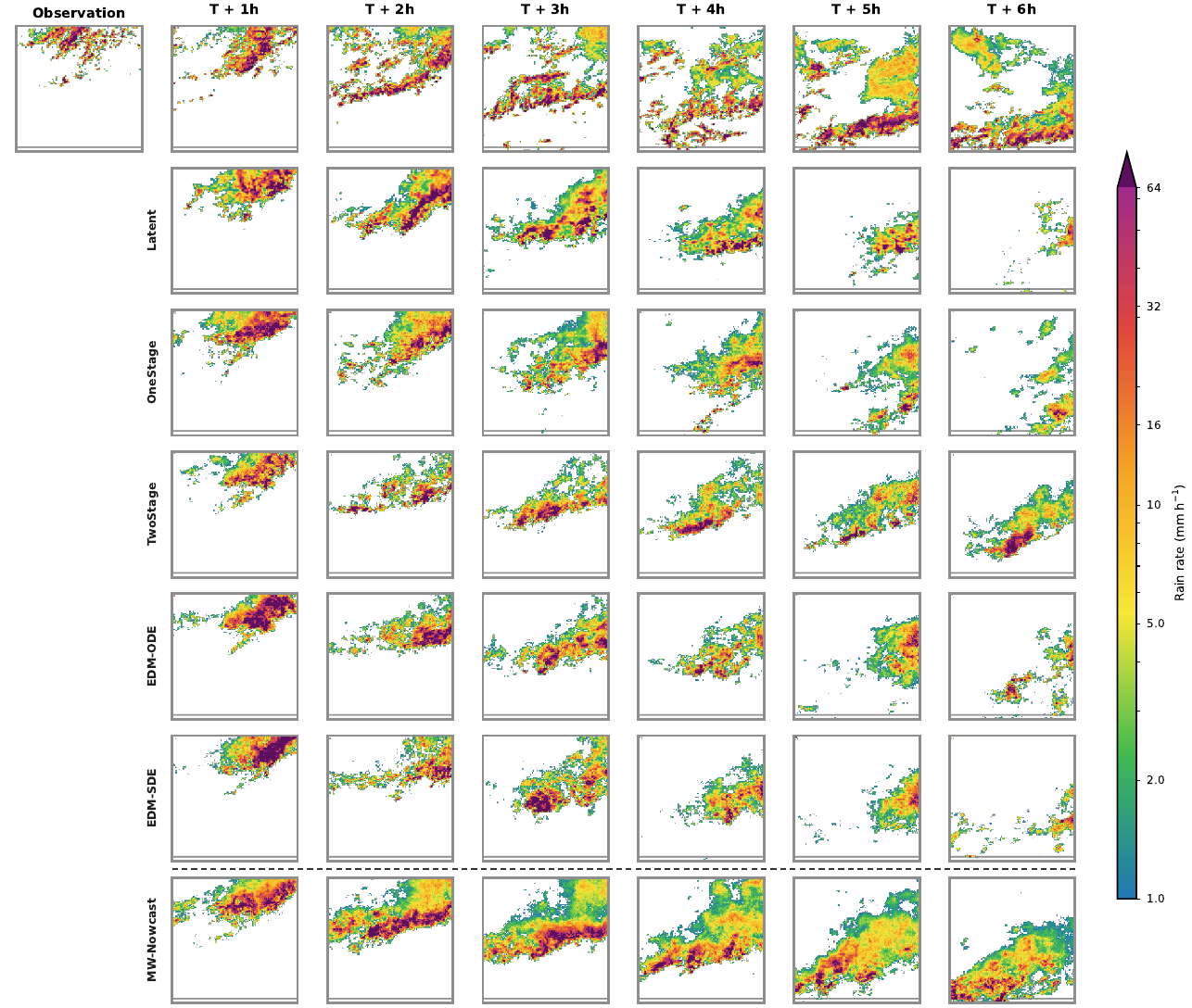}
  \captionsetup{font=small}
  \caption{\textbf{Ablation forecast grid for the US strong-precipitation case initialised at 06:00 UTC on 17 July 2025.} The patch is centred near 38.17$^\circ$N, 96.93$^\circ$W. Layout follows Supplementary Fig.~\ref{fig:supp-ablation-us-gulf}: rows give the observation and the MW-Nowcast, Latent, OneStage, TwoStage, EDM-ODE and EDM-SDE variants, and columns give the input context and forecast lead times through T+6\,h.}
  \label{fig:supp-ablation-us-plains}
\end{figure}

To extend the single ablation example in Supplementary Fig.~\ref{fig:supp-ablation-case-us}, Supplementary Figs.~\ref{fig:supp-ablation-us-gulf} and~\ref{fig:supp-ablation-us-plains} show the same two US cases under the ablation variants. Each grid follows the same visualisation style as Supplementary Fig.~\ref{fig:supp-ablation-case-us}: rows give the observation and the ablation variants, and columns give the input context and forecast lead times through T+6\,h.

\clearpage
\clearpage

\clearpage

\end{document}